\documentclass[a4paper,11pt]{article}
\pdfoutput=1 

\usepackage{jinstpub} 

\usepackage{lineno}

\collaboration{The MicroBooNE Collaboration}

\title{Design and Construction of the MicroBooNE Detector}

\author[g]{R.~Acciarri}
\author[aa]{C.~Adams}
\author[h]{R.~An}
\author[g]{A.~Aparicio}
\author[g]{S.~Aponte}
\author[x]{J.~Asaadi}
\author[a]{M.~Auger}
\author[f]{N.~Ayoub}
\author[g]{L.~Bagby}
\author[g]{B.~Baller}
% ...

\dedicated{Dedicated to W.~J.~Willis}

\affiliation[a]{Universit{\"a}t Bern, Bern CH-3012, Switzerland}
% ...
\affiliation[f]{Columbia University, New York, NY, 10027, USA}
\affiliation[g]{Fermi National Accelerator Laboratory (FNAL), Batavia, IL 60510, USA}
\affiliation[h]{Illinois Institute of Technology (IIT), Chicago, IL 60616, USA}
% ...
\affiliation[x]{University of Texas, Arlington, TX, 76019, USA}
% ...
\affiliation[aa]{Yale University, New Haven, CT, 06520, USA}

\abstract{
This paper describes the design and construction of the MicroBooNE liquid argon time projection chamber and associated systems. MicroBooNE is the first phase of the Short Baseline Neutrino program, located at Fermilab, and will utilize the capabilities of liquid argon detectors to examine a rich assortment of physics topics.  In this document details of design specifications, assembly procedures, and acceptance tests are reported.
}

\keywords{Time projection chambers; Noble liquid detectors; Neutrino detectors}

\begin{document}
\maketitle
\flushbottom


\newcommand{\lartpc}{LArTPC~}
\newcommand{\lartpcs}{LArTPCs~}


\section{Introduction and Physics Motivation}

The {\bf{Micro}} {\bf{Boo}}ster {\bf{N}}eutrino {\bf{E}}xperiment (MicroBooNE) employes a large ($\sim$100 tonnes) Liquid Argon Time Projection Chamber (LArTPC) detector designed for precision neutrino physics measurements.   MicroBooNE is the latest among a family of detectors that exploit the potential of liquified noble gases as the detection medium for neutrino interactions.   These detectors combine the advantages of high spatial resolution and calorimetry for excellent particle identification with the potential to scale to very large volumes. 

Large calorimeters using cryogenic noble liquids combined with active components were recognized in the 1970s as having use for particle physics applications~\cite{Willis:1974}.  Specifically, much of the liquid argon based technology was developed within the ICARUS program ~\cite{Benetti:1993-3ton,Cennini:1994-3ton,Arneodo:1999-50l} culminating in the realization of the ICARUS T600 detector~\cite{Amerio:2004-T600}.  On a much smaller scale than the ICARUS detector, the ArgoNeuT (Argon Neutrino Test) experiment operated a $\sim$0.25 tonne \lartpc from 2009-2010 in the NuMI neutrino beam at Fermilab.   ArgoNeuT performed a series of detailed studies on the interaction of medium-energy neutrinos \cite{Acciarri:2013-argoneut-recomb}  producing the first published neutrino cross section measurements on argon~\cite{Anderson:2012-argoneut-CCincl,Acciarri:2014-argoneut-CCxsec,Acciarri:2014eit}.   Next generation \lartpcs for the Short Baseline Neutrino Detector (SBND) experiment and the Deep Underground Neutrino Experiment (DUNE) are now being designed and constructed.   
   
MicroBooNE 's principal physics goal is to address short baseline neutrino oscillations, primarily the MiniBooNE observation of an excess of electron-like events at low energy~\cite{AguilarArevalo:2008rc}, at the Fermi National Accelerator Laboratory (Fermilab).  MicroBooNE will be exposed to the  0.5-2 GeV on-axis Booster Neutrino Beam (BNB) at a $\sim$500~m baseline, the same as was employed for MiniBoonE.  The MicroBooNE experiment is exploiting the \lartpc technology because of its superior capability for separation of signal electrons from the background of photon conversions.   While the mass of MicroBooNE is significantly less than the mass of MiniBooNE, this superior discrimination is expected to address the MiniBooNE result at the 5$\sigma$ level.   


In addition to MicroBooNE's signature oscillation analyses, a suite of precision cross-section measurements will be performed, critical both for future \lartpc oscillation experiments and for understanding neutrino interactions in general.   In the BNB, multiple interaction processes (quasi-elastic, resonances, deep inelastic scattering) are possible, and complicated nuclear effects in neutrino interactions on argon result in a variety of final states. These can range from the emission of several nucleons to more complex topologies with multiple pions, all in addition to the leading lepton in charged-current events. The \lartpc technology employed by MicroBooNE is particularly well suited for complicated topologies because of its excellent particle identification capability and calorimetric energy reconstruction down to very low detection thresholds. MicroBooNE's physics program also encompasses searches for supernova and proton decay.  The detector is capable of recording neutrinos from a galactic supernova which would result in $\sim$30 charged current neutrino interactions in MicroBooNE's active volume.   The detector will measure proton decay-like signatures and backgrounds and develop the analysis for this search; though its target mass is insufficiently large to enable a competitive sensitivity, the analysis will provide an important proof-of-principle for future searches in more massive detectors.  


MicroBooNE began operations in late 2015 for an initial anticipated $\sim$3 year run.   In 2018, MicroBooNE will continue operations as part of an expanded Short Baseline Neutrino (SBN) program~\cite{Adams:2013-lar1nd} at Fermilab that includes continued operation of MicroBooNE (at 470~m) along with the SBND (at 110~m)  and ICARUS (at 600~m) detectors.  The SBND and ICARUS experimental halls and detectors are presently under construction.  MicroBooNE will definitively address whether or not the MiniBooNE low energy excess in neutrino mode is due to electrons or photons in its initial run.  SBND will look for this low energy excess at the near location and ICARUS, with its larger mass, will enable the three detector program to cover the entire LSND-allowed region in neutrino parameter space with 5$\sigma$ sensitivity in the $\nu_e$ appearance channel.


This document describes the design, construction, and technical details of the MicroBooNE detector.  Section \ref{sec:overview} gives a brief review of the \lartpc technique and its implementation in MicroBooNE.  Section~\ref{sec:cryostat} describes the cryogenic and purification systems which are required for maintaining a stable volume of highly purified liquid argon.  The \lartpc described in section~\ref{sec:tpc-all} is the centerpiece of the experiment, providing fine-grained images of neutrino interactions.  A light collection system, described in section~\ref{sec:light-collection}, provides timing information, used primarily for triggering beam events, from the prompt scintillation light that is produced in the detector volume. Signals from the light collection system and from the \lartpc are amplified, sampled, and recorded by a custom-designed electronic and readout system, as described in section~\ref{sec:electronics}.  Section \ref{sec:slow-control} describes the auxiliary instrumentation that monitor and control the detector and all of its associated systems, as well as provide an electrically quiet environment for the experiment to operate. Finally, one of the main calibration sources for the experiment is an ultraviolet laser system, described in section~\ref{sec:laser}, that provides the capability to map out geometric track distortions, as induced, for example, by space charge.  A cosmic ray tagger system, under construction at the time of the writing of this paper, will surround the detector to improve cosmic ray identification and rejection.  This system will be described in a subsequent publication.  

More information on the \lartpc technology can be found in existing reviews (see, e.g.,~\cite{Marchionni:2013} and references therein).

\newpage
\section{Experiment Overview}
\label{sec:overview}

\begin{figure}
\centering 
\includegraphics[width=0.99\textwidth]{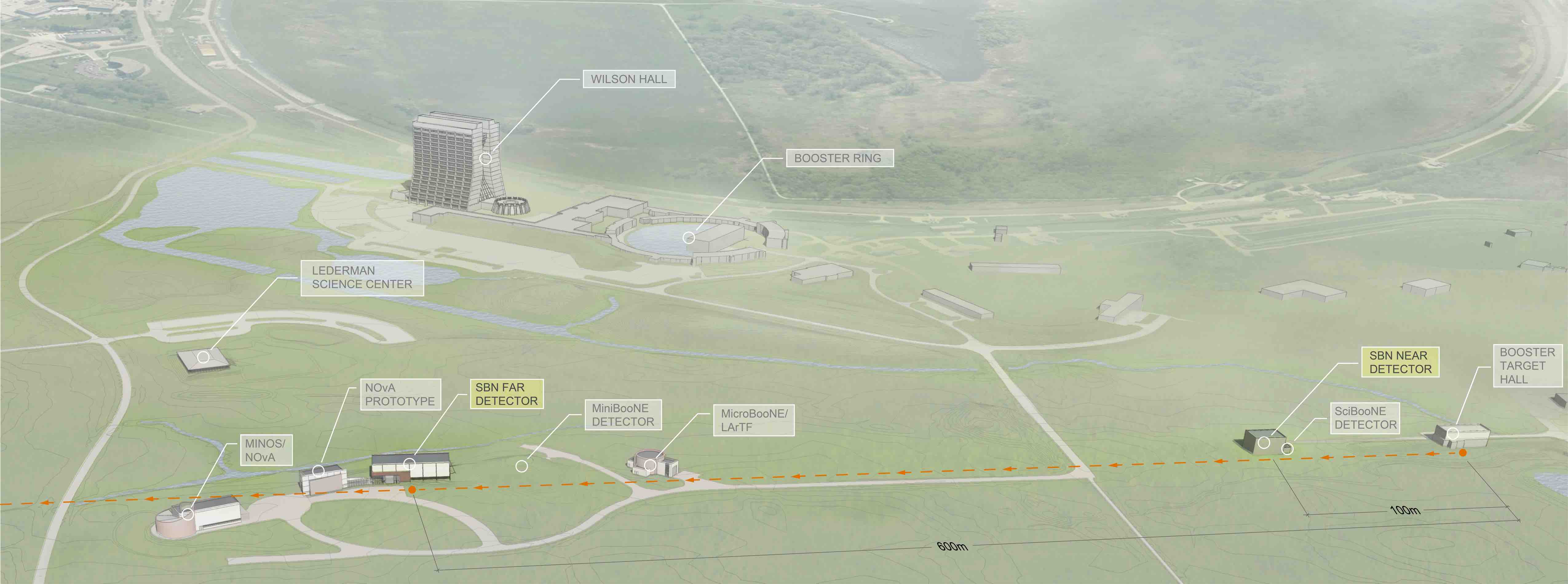}
\caption{Aerial diagram showing location of MicroBooNE along the BNB (orange dashed line) at Fermilab.}
\label{fnalmap}
\end{figure}

The MicroBooNE detector at Fermilab in Batavia, Illinois is sited in the Liquid Argon Test Facility (LArTF) on axis in the BNB, 470 m downstream from the neutrino production target.   The BNB delivers a beam of predominantly muon neutrinos produced primarily from pion decays, with energies peaking at 700~MeV \cite{AguilarArevalo:2008yp}. MicroBooNE is also exposed to an off-axis component of the NuMI beam~\cite{Adamson:2015dkw} produced from pion and kaon decays with average neutrino energies of about 0.25 GeV and 2 GeV respectively.   MicroBooNE is located about 600 m downstream from the NuMI neutrino production target.  The characteristics of the BNB beamline are well measured and understood from many years of data taking and analysis from the MiniBooNE experiment~\cite{AguilarArevalo:2008yp}, which operated directly downstream of the MicroBooNE location.  Figure \ref{fnalmap} shows the arrangement of MicroBooNE with respect to the BNB beamline at Fermilab.  The physics program of MicroBooNE will utilize both BNB and NuMI samples.  MicroBooNE will also collect data that is out-of-time with either beam, which will be useful for developing non-accelerator neutrino-based analyses (e.g. proton decay and supernovae burst neutrino searches) relevant for next-generation detectors.  


\subsection{The MicroBooNE \lartpc}

Charged particles traversing a volume of liquid argon leave trails of ionization electrons in their wake and also create prompt vacuum ultraviolet (VUV) scintillation photons.  In a LArTPC, the liquid argon is highly purified so that the ionization trails can be transported with minimal attenuation over distances of the order of meters~\cite{Aprile:1985} under the influence of a uniform electric field in the detector volume, until they reach sense planes located along one side of the active volume.   The electric field is created by introducing voltage onto a cathode plane and gradually stepping that voltage down in magnitude across a field cage, which is formed from a series of equipotential rings surrounding the drift volume.   Non-uniformities in the electric field, diffusion, recombination, and space charge effects modify the tracks as they are transported.  Calibration of these effects is critical to reconstruction of the initial ionization trails.   

\begin{figure}
\centering 
\includegraphics[width=0.95\textwidth]{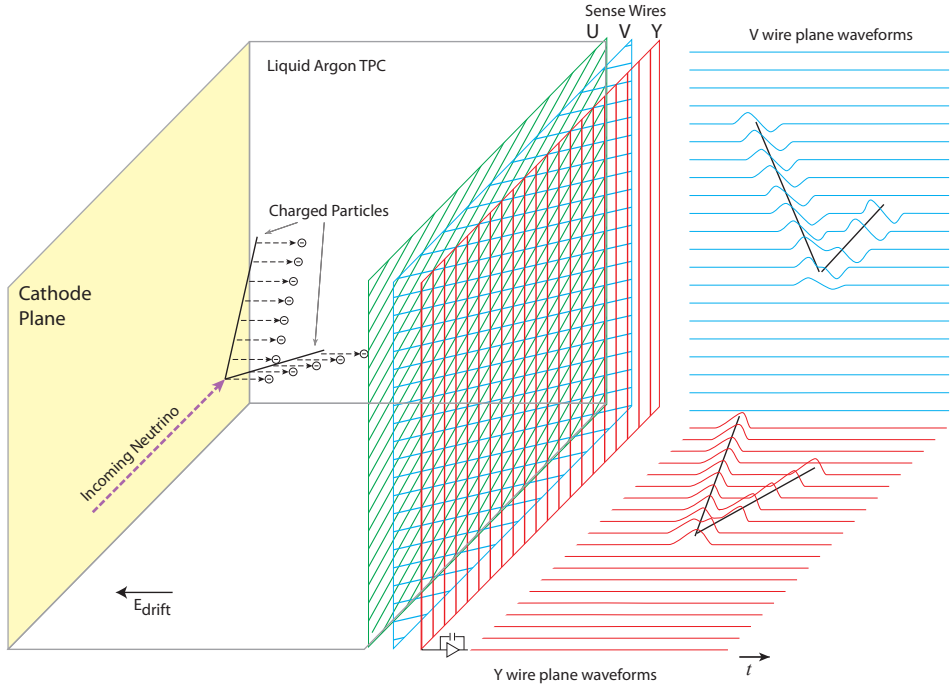}
\caption{Operational principle of the MicroBooNE LArTPC.}
\label{fig:lartpc}
\end{figure}

The anode plane is arranged parallel to the cathode plane, and in MicroBooNE, parallel to the beam direction.   There are three planes comprised of sense wires with a characteristic pitch, held at a predetermined bias voltage, that continuously sense the signals induced by the ionization electrons drifting towards them~\cite{Gatti:1979fba}. The electrostatic potentials of the sequence of anode planes allow ionization electrons to pass undisturbed by the first two planes before ultimately ending their trajectory on a wire in the last plane. The drifting ionization thus induces signals on the first planes (referred to as induction planes) and directly contributes to the signals in the final plane (referred to as the collection plane).  Figure \ref{fig:lartpc} depicts the arrangement of the MicroBooNE LArTPC and its operational principle.

The charged particle trajectory is reconstructed using the known positions of the anode plane wires and the recorded drift time of the ionization.  The drift time is the difference between the arrival times of ionization signals on the wires and the time the interaction took place in the detector ($t_0$) which is provided by an accelerator clock synchronized to the beam (BNB or NuMI) or from a trigger provided by the light collection system.  The characteristics of the waveforms observed by each wire provide a measure of the energy deposition of the traversing particles near that wire, and, when taken as a whole for each contained particle's trajectory, allow for determination of momentum and particle identity.

The scintillation photons are detected by a light collection system that is immersed in the liquid argon and faces into the detector volume.  This system provides signals that can establish the event $t_0$ and supplies trigger information to an electronic readout system.  The light collection system signals are vital in distinguishing detector activity that is in-time with the beam (and therefore possibly originating from beam interactions) from activity which is out-of-time (and therefore probably not associated with the beam), benefiting triggering and event reconstruction.  Information on the $z$ and $y$ position of an interaction can also be inferred from the light system, further aiding in the reconstruction.


Liquid argon as a target for neutrinos is attractive due to its density, allowing a more compact detector with a substantial boost in event rate over a comparable detector using less dense media.  A tradeoff to this aspect is the fact that the complicated structure of the argon nucleus (relative to hydrogen or helium, for example) will introduce nuclear effects that the data analysis must take into account.  The cryogenic temperatures at which the noble elements are in the liquid phase also introduces the need for additional design considerations to ensure stable and safe operations.

Table \ref{tab:nobleparam} lists some of the properties of liquid argon that are salient for \lartpc design.  
The noble liquids produce copious numbers of UV photons for every traversing charged particle, as well as large amounts of ionization.  The electrons from this ionization can be drifted for distances of meters under a modest electric field ($\sim$500~V/cm).  Finally, building LArTPCs on increasingly large scales for neutrino detection becomes economically possible, given the abundance (1\% of atmosphere) and low cost of argon and the convenient feature that it can be maintained as a liquid through refrigeration using liquid nitrogen which is plentiful and cheap.    

\begin{table}[!htb]
   \centering
    \caption{Selected properties of liquid argon.} 
      \begin{tabular}{lcr} 
      \hline
      Property & Value & Reference\\
    \hline
   Atomic number & 18 &\\
   Atomic weight [g/mol] & 39.95 &\\
   Boiling point [K] @ 1 atm & 87.3 & \cite{ArProperties}\\
   Density [g/cm$^3$] @ 1 atm & 1.394 & \cite{ArProperties} \\
   Dielectric constant & 1.505 & \cite{DielectricPaper} \\
   Radiation length [cm] & 14.0 & \cite{Amsler20081} \\
   Moli\`{e}re radius [cm] & 10.0 &\cite{Amsler20081} \\
   W-value for ionization [eV/pair] & 23.6 & \cite{1975NucIM.131..249S,PhysRevA.9.1438} \\
   Minimum specific energy loss [MeV/cm] & 2.12 & \cite{Amsler20081} \\
   Electron transverse diffusion coef. [cm$^2$/s] & 13 & \cite{PhysRevA.20.2547,Derenzo,Cennini:1994-3ton} \\
   Electron longitudinal diffusion coef. [cm$^2$/s] & 5 & \cite{Cennini:1994-3ton, Atrazhev}\\
   \hline
   \end{tabular}

   \label{tab:nobleparam}
\end{table}

The successful implementation of the \lartpc technique depends critically on a number of factors.   The liquid argon must be purified of any electronegative contaminants, such as water or oxygen, to accommodate the very long drift path of ionization through a MicroBooNE-sized \lartpc without significant charge loss.  The signals that the ionization electrons create on the anode wires are very small, requiring low-noise electronics to discriminate between signal pulses and background noise.  The MicroBooNE collaboration has designed and constructed an experiment that addresses these considerations, providing critical technological development upon which the next generation of LArTPCs may capitalize.

\subsection{MicroBooNE \lartpc Implementation}

MicroBooNE's \lartpc active volume, which is defined as the volume immediately within the confines of the \lartpc field cage, is a rectangular liquid argon volume with dimensions as given in table~\ref{tab:detectorparam}. This is the maximum volume that can be used for physics analyses.  The cathode and the anode planes define the beam-left and beam-right sides of the active volume.  The end of the \lartpc that the beam first encounters is referred to as the ``upstream'' end, while the opposite end is referred to as ``downstream.''  Anode plane-to-plane spacing is 3~mm, and each plane has 3~mm wire pitch. The induction plane wires are oriented at $\pm60^{\circ}$ relative to vertical, and the collection plane wires are oriented vertically. Field cage loops are employed to maintain uniformity of the electric field across the entire width of the detector, and these loops also act to define the top, bottom, upstream, and downstream sides of the active volume.  

MicroBooNE uses a right-handed Cartesian coordinate system, with the origin defined to be located on the upstream face of the LArTPC, centered halfway up the vertical height of the active volume and horizontally centered on the anode plane closest to the cathode (the innermost anode plane).  In this system, $x$ ranges from 0.0 m at the innermost anode plane to $+2.6$~m at the cathode, $y$ ranges from $-1.15$~m on the bottom of the active volume to $+1.15$~m at the top of the active volume, and $z$ ranges from 0.0~m at the upstream end of the active volume to $+10.4$~m at the downstream end.  

\begin{figure}
\centering 
\includegraphics[width=0.65\textwidth]{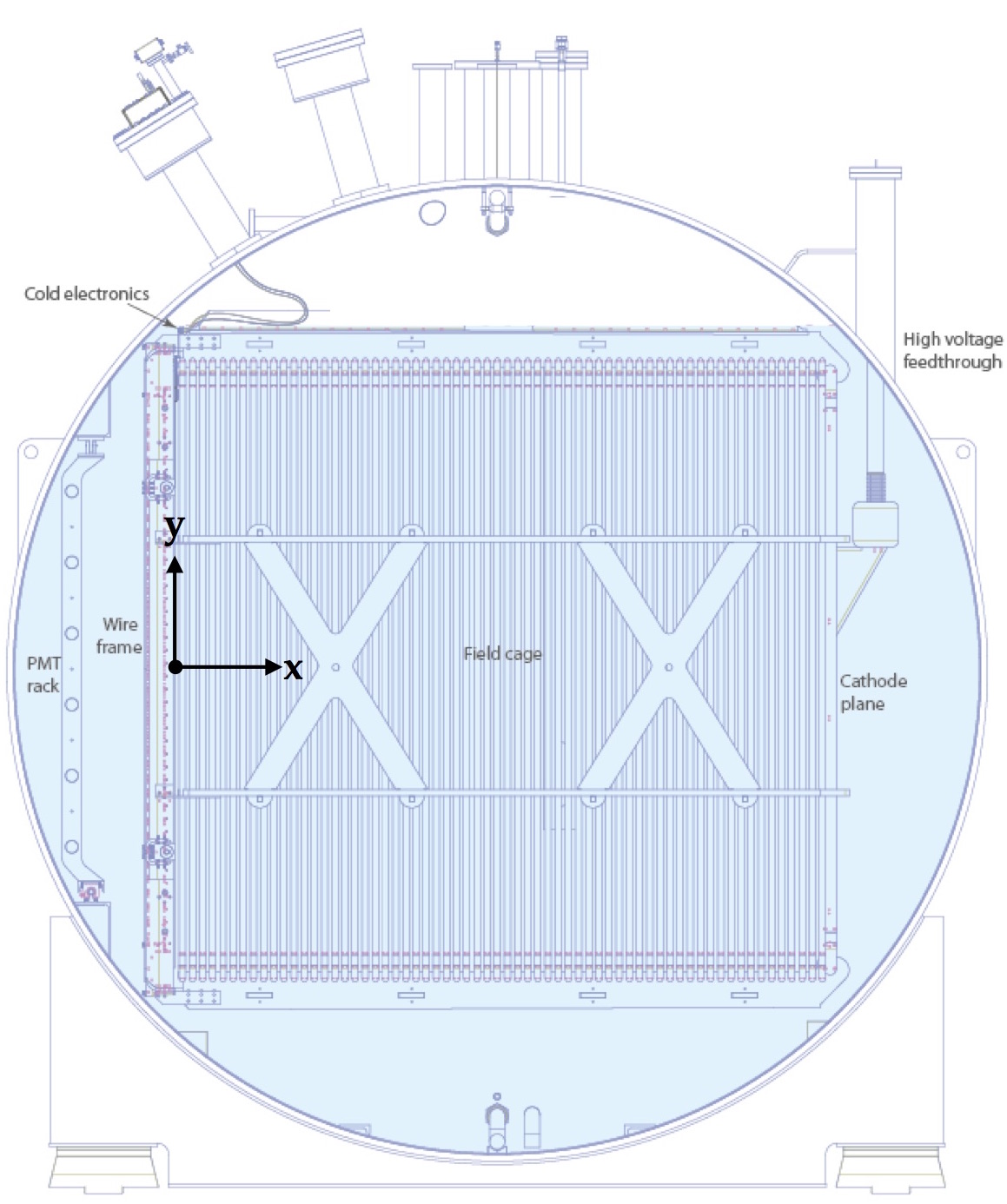}
\caption{Schematic of the cross section of the MicroBooNE LArTPC.  In this view, the beam would be directed out of the page (in the $z$ direction).}
\label{fig:microboonetpc}
\end{figure}

The light collection system, an array of photomultiplier tubes (PMTs) and lightguide paddles, is located directly behind the anode planes on beam-right, facing the detector volume through the anode planes.  The \lartpc and light collection system are immersed in liquid argon contained within a single-walled cryostat with a 170 tonne capacity.  Analog front-end electronics mounted directly on the \lartpc amplify the signals on the wires; these signals are then passed out of the cryostat for further processing and storage on disk.  Table \ref{tab:detectorparam} lists the primary detector design parameters of MicroBooNE, and figure \ref{fig:microboonetpc} shows a schematic of the cross section of the detector. Details of these design parameters and construction of all detector systems will be provided in the subsequent sections.

\begin{table}[!htb]
   \centering
    \caption{Primary detector design parameters for MicroBooNE.} 
    \begin{tabular}{lr} 
    \hline
    Parameter & Value \\
    \hline
     \lartpc Dimensions & 2.325 m vertically \\
     & 2.560 m horizontally \\
     & 10.368 m longitudinally  \\	
     \lartpc argon mass & 90 tonnes \\
     Total Number of Wires & 8256 \\
     Drift field & 500 V/cm\\
      Light collection & 32 200 mm (8 in) diameter PMTs \\
      & 4 lightguide paddles \\
      Total liquid argon mass & 170 tonnes  \\
      Operating temperature & 87 K\\
      Operating pressure & 1.24 bar\\
    \hline
   \end{tabular}
   \label{tab:detectorparam}
\end{table}

\newpage
\section{Cryogenic System}
\label{sec:cryostat}

The use of large quantities of highly-purified liquid argon as a detector medium in MicroBooNE requires a sophisticated cryogenic infrastructure that can maintain stable operations for many years with minimal downtime.  Not only must the purity of the liquid argon be maintained, but the pressure and temperature gradients within the \lartpc active volume must be tightly controlled as the drift velocity of electrons is dependent on these quantities.  A customized cryogenic system that serves these purposes has been built, and the requirements for this system are shown in table \ref{tab:cryoreq}.

\begin{table}[!htb]
   \centering
    \caption{Primary design requirements for MicroBooNE cryogenic and purification systems.} 
    \begin{tabular}{lll} 
    \hline
    Parameter & Value & Motivation\\
    \hline
      Argon purity    & $<$100 ppt O$_2$ equivalent & MIP identification at longest drift\\
      Argon purity    & $<$2 ppm N$_2$ & Scintillation light output\\
      LAr temperature gradient & $<$0.1 K throughout volume & Drift-velocity uniformity\\
      LAr recirculation rate & 1 volume change/2.5 days & Maintain purity\\
      Cryostat heat load & $<$15 W/m$^2$ & Minimize convection and bubbles\\
      Cryogenic cooling capacity & 10 kW & Capacity for expected heat load\\
      Cryostat max. operating pressure & 2.1 bar & Determines relief sizing\\
            \hline
   \end{tabular}
   \label{tab:cryoreq}
\end{table} 

\begin{figure}
\centering 
\includegraphics[width=0.95\textwidth]{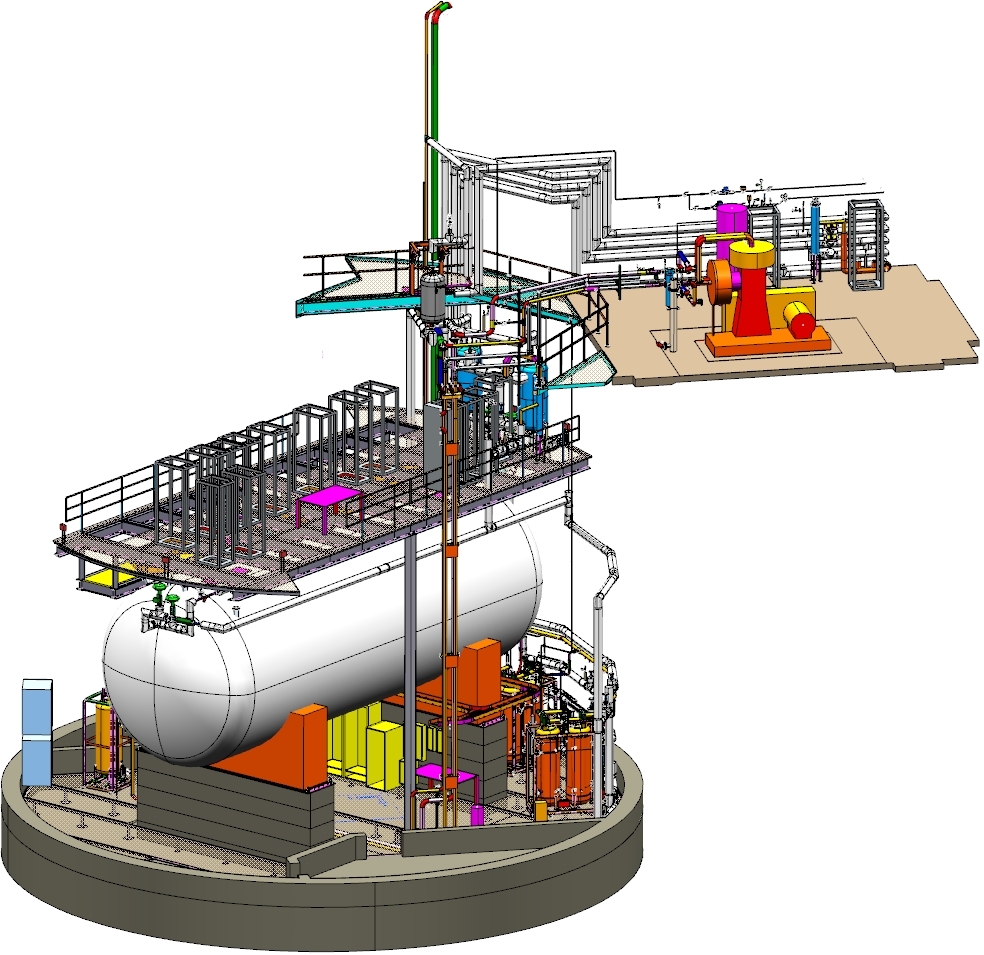}
\caption{A rendering showing the MicroBooNE cryostat and cryogenic system, and the platform for the electronics racks as installed in LArTF.}
\label{fig:cryogenics}
\end{figure}

The MicroBooNE cryogenic system is represented in figure \ref{fig:cryogenics}.  The central component of the system is a cryostat that houses the complete \lartpc and light-collection detector systems.  The cryostat is supported by three major subsystems: the argon purification system, the nitrogen refrigeration system, and the controls and monitoring system.  These systems each represent the next generation of \lartpc cryogenic system after the Liquid Argon Purity Demonstrator (LAPD)~\cite{Adamowski:2014-LAPD} and make considerable use of the expertise gained during the design and implementation of that apparatus.  

\subsection{Cryostat Design Overview}

Three major components make up the MicroBooNE cryostat: a type 304 stainless steel vessel to contain the liquid argon and all the active detector elements, front and rear supports to carry the weight of the fully loaded cryostat, and foam insulation covering the cryostat outer surfaces.  The foam insulation serves to reduce heat input from the ambient environment to a sufficiently low level to prevent large temperature gradients and boiling of the liquid argon. The cryostat and the cryogenic systems are designed to achieve the requisite high purity of liquid argon needed to allow ionization electrons to drift to the anode wires with low probability of capture, and the high degree of thermal homogeneity needed to avoid introducing non-constant drift velocities for the ionization electrons. Finally, the outer diameter of the vessel is designed to be the maximum standard size for highway transport.

\begin{figure}[htb]
\centering	
\includegraphics[width=\linewidth]{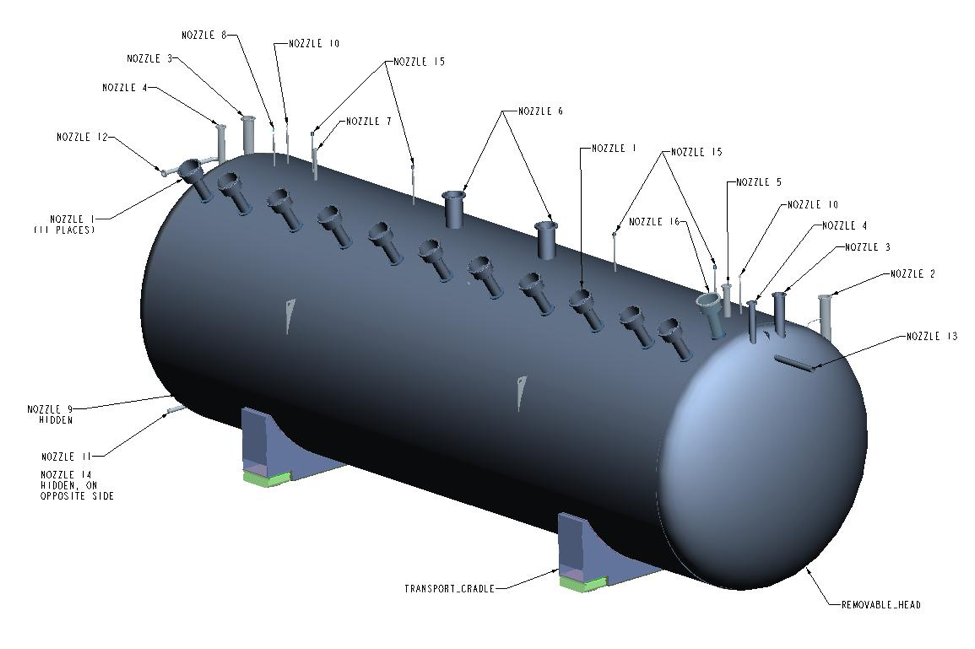}
\caption{MicroBooNE cryostat with nozzle penetrations labeled.}
\label{fig:cryostat-drawing}
\end{figure}

The cryostat is constructed to the American Society of Mechanical Engineers (ASME) boiler code requirements~\cite{pressure:1316452} and features a single-walled construction, cylindrical shape, and domed caps closing each end, as shown in figure~\ref{fig:cryostat-drawing}. The cryostat is 12.2 m in overall length, with an inner diameter of 3.81 m, and a wall thickness of 11.1 mm.  When empty the cryostat weighs $\sim$17,000~kg.  One end was removed for installation of the active detectors, welded back in place upon completion of that task, and then re-certified to the ASME code requirements.  

Ionization electrons must not be significantly attenuated, via attachment to electronegative contaminants in the liquid argon while they drift up to 2.5~m across the active volume. This dictates that the argon be kept free of electronegative contaminants to the level of 100 parts-per-trillion (ppt) oxygen-equivalent (O$_2$ equivalent), where the choice of O$_2$ equivalent units implies other polar molecules besides oxygen may be present, but the attachment rate constant for oxygen is used in calculations . The cryostat is designed to minimize outgassing (desorption) and to avoid leakage and diffusion of air into the system. This requirement imposes strict quality assurance demands on all welds for penetrations into the cryostat and on cleaning and handling procedures for the finished vessel.  Achieving the required level of purity is accomplished with a purification system, described in section~\ref{sec:purification}, that removes electronegative contaminants from the argon during the initial fill and those introduced over time by leaks and outgassing of system components.

The electron drift velocity ($v_{d} = 1600$~m/s at an electric field of 500 V/cm, with a liquid argon temperature dependence $\Delta v_{d}/v_{d} = -0.019\Delta T$) must remain constant in magnitude and direction throughout the active liquid argon volume to avoid distortion of the mapping of drift time into the position along the drift ($\hat{x}$) direction. This requirement limits the allowable temperature variations of the liquid argon to less than 0.1~K and the laminar and turbulent flow rate of liquid argon to less than 1~m/s. These requirements limit fractional errors in velocity, and therefore in the drift-coordinate determination, to be less than 0.1\%. The constraints on constancy of drift velocity affect the design by imposing limits on the acceptable heat flux through the insulation.

\begin{figure}[htb]
\centering	
\includegraphics[width=0.68\linewidth]{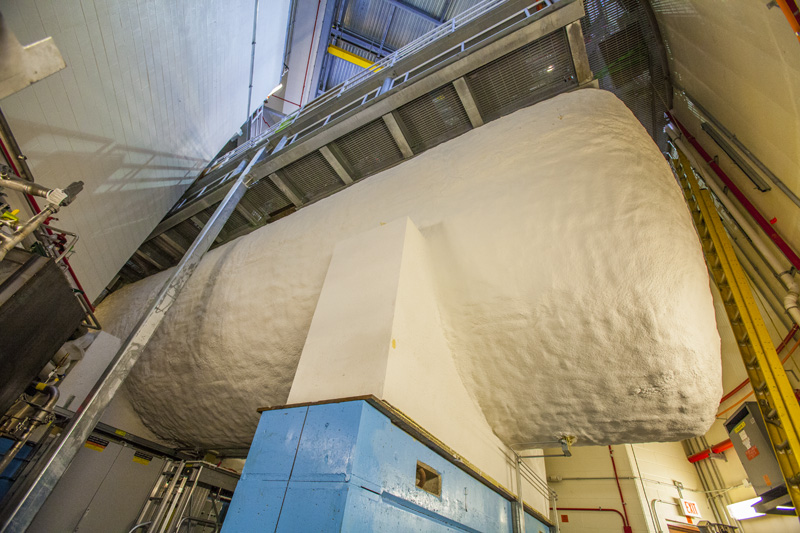}\\
\includegraphics[width=0.68\linewidth]{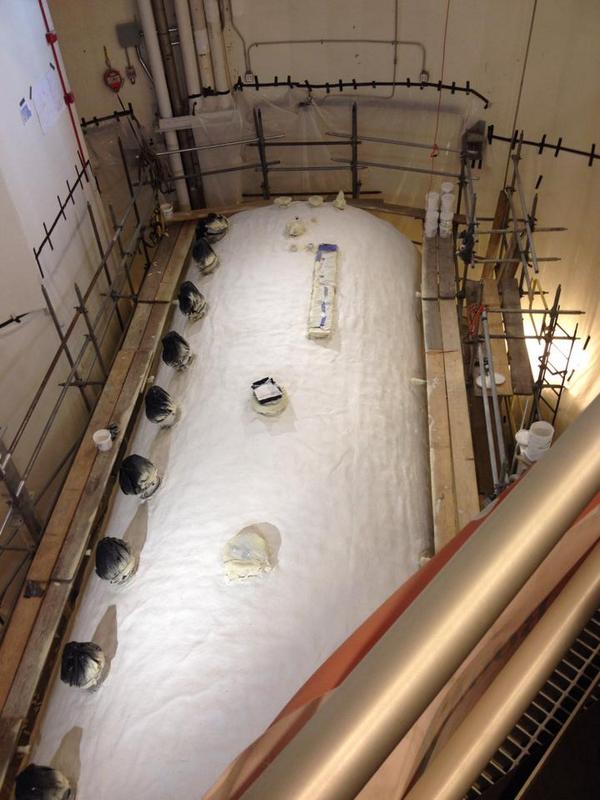}
\caption{Photographs of the cryostat after application of exterior foam insulation.}
\label{fig:cryostat-foam}
\end{figure}

Upon installation of the sealed cryostat in its final location at the LArTF, 41 cm of spray-on, closed cell, polyurethane insulation was applied to the exterior of the cryostat, as shown in figure~\ref{fig:cryostat-foam}.  At the LArTF, to avoid ground loops that could interfere with the \lartpc signals, the cryostat vessel is grounded in only one place, allowing it to act as a Faraday cage.  This grounding scheme is explained further in section~\ref{sec:lartf}.

The vessel surface has 34 nozzle penetrations for cryogenic and electrical services, detailed in table~\ref{tab:cryostat-feedthroughs}.  All nozzles are sealed with feedthroughs, flanges, or pipes that are suitable for operation at the nominal pressure and temperature of the cryostat.  

\begin{table}[!htb]
   \centering
    \caption{List of nozzle penetrations in the MicroBooNE cryostat, their function, and their flange/pipe type and outer-diameter dimension.  CF=ConFlat flanges, RFWN=raised face weld neck flanges, SS = stainless steel pipe.} 
    \begin{tabular}{lcr} 
    \hline
    Nozzle ID & Function & Flange\\
    \hline
     N1A-N1K & \lartpc Signal Feedthrough & 356 mm CF\\
     N2 & \lartpc HV Feedthrough & 203 mm CF\\
     N3A-N3B & Purity Monitor & 203 mm CF\\
     N4 & Temperature Signals & 152 mm CF\\
     N5 & Safety Vent & 102 mm RFWN\\
     N6A-N6B & Vacuum Pump-Out & 254 mm RFWN\\
     N7 & Condensor & 76 mm SS\\
     N8 & Top Instrument Port & 19 mm SS\\
     N9 & Bottom Instrument Port & 19 mm SS\\
     N10A-N10B & Liquid Level Probe & 19 mm SS\\
     N11 & Gas Circulation In & 51 mm SS\\
     N12 & Gas Circulation Out & 51 mm SS\\
     N13 & From LAr Filters & 76 mm SS\\
     N14 & To LAr Pumps & 51 mm SS\\
     N15A-N15B & Laser Calibration & 70 mm CF \\
     N16 & PMT Signal Feedthrough & 356 mm CF\\
     N17 & Spare & 152 mm CF\\
     N18 & Temperature Signals & 152 mm CF\\
     N19 & Spare & 152 mm CF\\
                  \hline
   \end{tabular}
   \label{tab:cryostat-feedthroughs}
\end{table}

\subsection{Liquid Argon Purification Subsystem}
\label{sec:purification}

The heart of the cryogenic system is the liquid argon purification subsystem.  The primary requirement of this subsystem is to keep the level of electronegative contamination to below 100 ppt of O$_2$ equivalent contaminants.  This requirement is determined by the physics needs of the experiment, namely the need to be able to reconstruct events at the longest drift distances in the LArTPC.  In addition to the requirement on the electronegative contamination, the system must maintain the level of nitrogen contamination in the argon, by minimizing the leak rate from the atmosphere, at less than 2 parts per million (ppm) \cite{Jones:2013bca} to keep the quenching and attenuation of the scintillation photons in the argon to a minimum.  

\begin{figure}
\centering 
\includegraphics[width=0.85\textwidth]{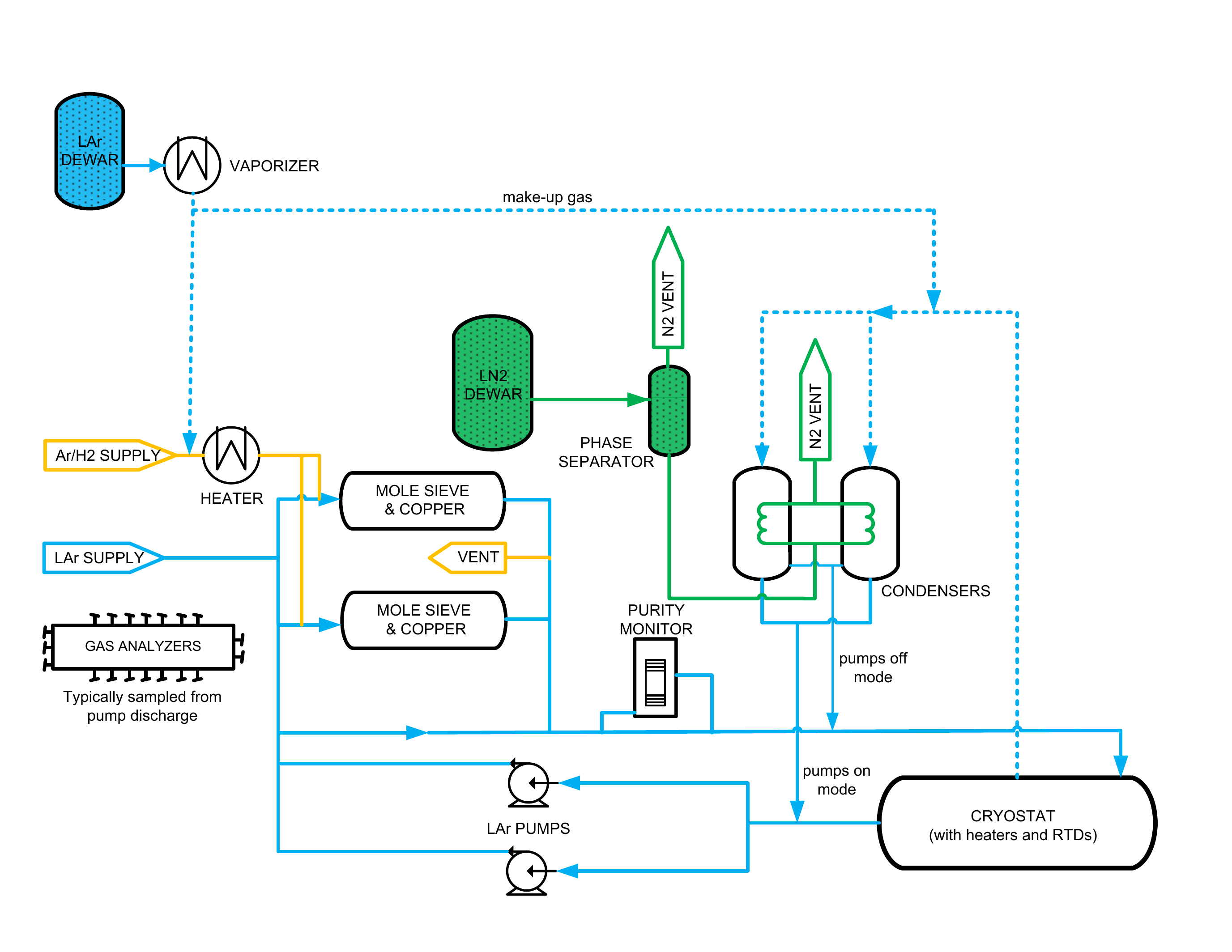}
\caption{Flow diagram of argon in MicroBooNE, showing direction of liquid and gaseous argon in the cryogenic system.  Dashed lines represent gas lines, solid lines represent liquid lines, and yellow lines are for the filter regeneration.  Gaseous argon from the cryostat is condensed and directed through the purification subsystem. Liquid argon drawn from the cryostat volume is directed into the purification subsystem.}
\label{flowdiagram}
\end{figure}

The MicroBooNE argon purification subsystem consists of liquid argon pumps and filters that serve to circulate the argon and remove impurities (e.g. O$_2$ and H$_2$O) that degrade the quality of the data collected by the active detectors.  It should be noted that the filters do not remove $N_2$ and so the ultimate $N_2$ contamination  is set by the quality of the delivered argon.  There are two pumps in the system arranged in parallel in order to allow for continuous recirculation while one pump is being serviced.  Similarly, there are two sets of filters arranged in parallel in the system.  Figure \ref{flowdiagram} schematically depicts the flow of liquid and gaseous argon in the MicroBooNE cryogenic system.

The recirculation pumps are Barber-Nichols~\cite{barber-nichols} BNCP-32B-000 magnetically-driven partial-emission centrifugal pumps.  Each pump isolates the liquid argon from the electric motor by a magnetic coupling of the impeller to the motor.  The impeller, inducer, and driving section of the magnetic coupling each have their own bearings that are lubricated by the liquid argon at the impeller.  The motor is controlled by a variable frequency drive (VFD) that allows adjustment of the pump speed to produce the desired head pressure and flow within the available power range of the motor.    

\begin{figure}
\centering 
\includegraphics[width=0.45\textwidth]{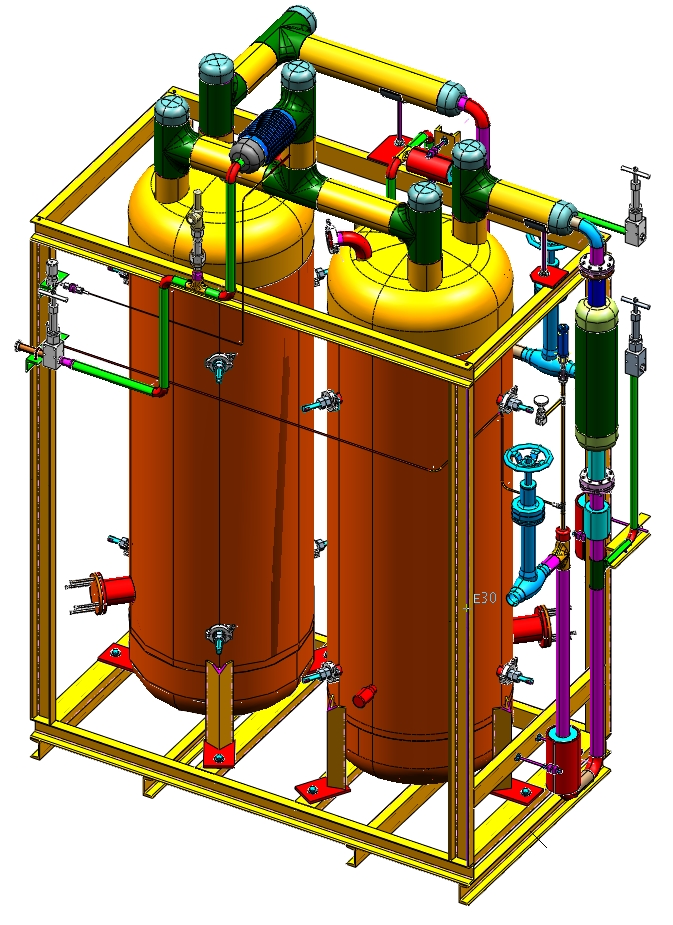}
\includegraphics[width=0.45\textwidth]{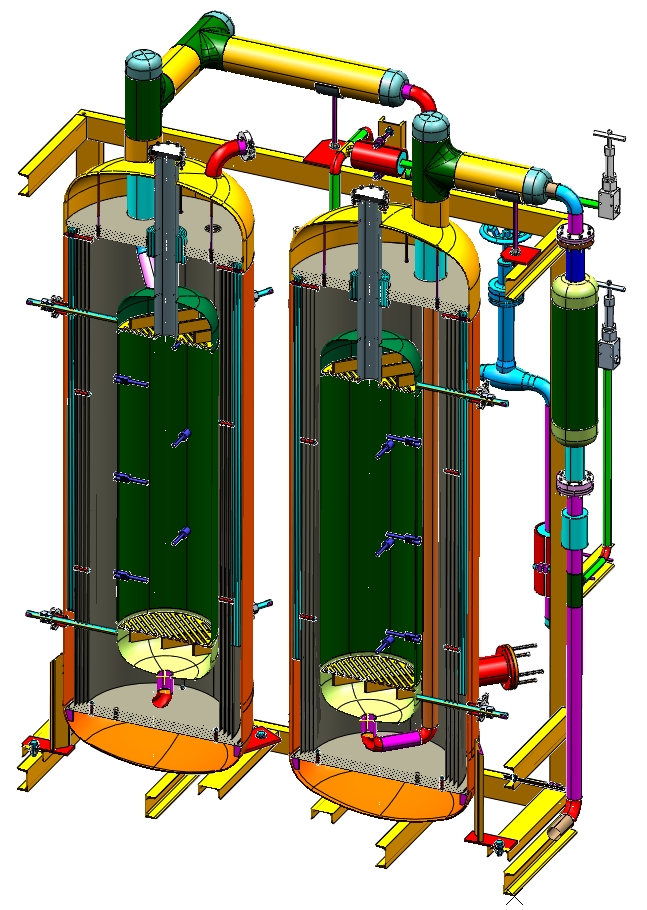}
\caption{Three-dimensional rendering of a MicroBooNE filter skid.  The left drawing shows the full skid, while the right drawing shows a cut-away of the vessels.}
\label{filters}
\end{figure}

Each filter skid contains two filters, as depicted in figure \ref{filters}, each having identically-sized filtration beds of 77~liters.  The first filter that the argon stream enters contains a 4A molecular sieve supplied by Sigma-Aldrich~\cite{sigma-aldrich} that primarily removes water contamination but can also remove small amounts of nitrogen and oxygen.  The second filter contains BASF~CU-0226~S, a pelletized material of copper impregnated on a high-surface-area alumina, which removes oxygen~\cite{basf} and to a lesser extent water.  Because the oxygen filter will absorb water and thereby reduce its capacity for removing oxygen, it is placed after the molecular sieve.  The oxygen-filtering material must be reduced to copper with the procedure described below before it can remove oxygen from the liquid argon.  The filters are insulated with vacuum jackets and aluminum radiation shields.  The metallic radiation shields were chosen because the filter regeneration temperatures, described below, would damage traditional aluminized mylar insulation.  Pipe supplying the filter regeneration gas is insulated both inside the filter vacuum-insulation space and outside the filter with Pyrogel XT, which is an aerogel-based insulation~\cite{aspen-aerogels} that can withstand temperatures up to 920 K. 

The filters are initially activated and then regenerated as needed {\it in situ} using heated gas, by a procedure developed for the LAPD.  The filters are regenerated using a flow of argon gas that is heated to 473 K, supplied by a commercial 500~liter liquid argon dewar.   Once the argon gas reaches 473 K, a small flow of hydrogen is mixed into the primary argon flow and exothermically combined with oxygen captured by the filter to create water.  Too much hydrogen mixed in with the primary argon flow would induce temperatures that are sufficiently high to damage the copper-based filter media.  The damage is induced by sintering of the copper, which reduces the available filter surface area.  To avoid such damage, precautions are taken to maintain a hydrogen fraction below 2.5\% of the heated gas mixture.  During the heated gas regeneration, five filter bed temperature sensors monitor the filter material temperature and the water content of the regeneration exhaust gas is measured.  To remove any remaining trace amounts of water, the filters are then evacuated using turbomolecular vacuum pumps while they cool. 

A particulate filter with an effective filtration of 10 microns, positioned between the cryostat and the filter skids, prevents any debris in the piping from being introduced into the cryostat.  The particulate filter consists of a commercial stainless steel sintered-metal cylinder mounted in a custom cryogenic housing and vacuum jacket.  Filtration is accomplished by flowing liquid argon to the interior, then outward through the walls, of the sintered-metal cylinder.  Flanges on the argon piping, along with flanges and edge-welded bellows on the vacuum jacket, allow removal of the particulate filter.


The argon purification piping is 2.54 cm diameter stainless steel that was pre-insulated by the manufacturer with 10.2~cm of polyurethane foam.  During the fabrication process, all piping was washed with distilled water and detergent to remove oil and grease, then cleaned with ethanol.  All valves associated with the argon-purification piping utilize a metal seal with respect to ambient air, either through a bellows or a diaphragm, to prevent the diffusion of oxygen and water contamination.  The exhaust side of each relief valve is continuously purged with argon gas to prevent diffusion of oxygen and water from ambient air across the O-ring seal.  Where possible, ConFlat flanges with copper seals are used on both cryogenic and room-temperature argon piping.  Pipe flanges in the system are sealed using spiral-wound graphite gaskets.  Smaller connections are made with VCR fittings with stainless steel gaskets.  

\subsection{Nitrogen Refrigeration}

The cryostat and purification systems that contain the liquid argon are subject to heat load from the environment, as well as from the active detectors that have electrical power enabled.  To keep these systems operating at a stable temperature and pressure, a liquid nitrogen refrigeration system is present to provide the necessary cooling power.  The liquid nitrogen system contains two condensers that are arranged in parallel.  One of these is utilized for normal operations and one serves as a backup on standby.  Each condenser contains two liquid nitrogen coils, an inner and an outer, with the gas argon on the shell side.  Typically only one coil is actively running and the second can be manually activated during situations where the system heat load is higher than usual.  Each condenser is sized to handle a heat load of approximately 9.5 kW.  With the vessel full of liquid argon and no pump or liquid argon circulation running, the condenser uses $\sim$2350 liters of liquid nitrogen per day, which equates to a 3.9 kW system heat load.  With the recirculation pumps and the electronics in operation, the usage rate is about 3400 liters/day corresponding to a total heat load from the cryostat system of about 6 kW.  
%

\subsection{Controls and Purity Monitoring}

MicroBooNE makes use of resistive thermal devices (RTDs) to measure temperatures throughout the experimental infrastructure. Twelve RTDs are located along the walls of the cryostat, and another ten RTDs are mounted inside screws attached to the structure of the LArTPC. Each of the filter vessels in the purification system contain nine RTDs.  An interlock based on the RTDs within the filter vessels prevents overheating that could potentially occur during filter regeneration with heated argon-hydrogen gas.

Liquid argon contaminations ranging between 300 and 50~ppt O$_2$ equivalent can be measured using double-gridded ion chambers, henceforth referred to as purity monitors, immersed in liquid argon. The design of the purity monitors is based on the design presented by Carugno et al.~\cite{Carugno:1990-purityMonitor}.  A description of the purity monitors, the data-acquisition hardware and software used in LAPD can be found in~\cite{Adamowski:2014-LAPD}. MicroBooNE uses the same type of purity monitors, and the same data-acquisition hardware and software. 

An estimate of the electron drift lifetime is made by measuring the fraction of electrons generated at the purity monitor cathode that subsequently arrive at the purity monitor anode $(Q_A/Q_C)$ after a drift time $t$. The ratio of $(Q_A/Q_C)$ is related to electron lifetime, $\tau$, such that

\begin{equation}
Q_A/Q_C = e^{-t/\tau}.
\end{equation}

Measurement of liquid argon purity in the MicroBooNE cryogenic system are provided by three purity monitors of various lengths. One purity monitor with a drift distance of 50~cm sits in a vessel just downstream of the filters and is used to monitor filter effectiveness. Two purity monitors, one with a drift distance of 19~cm and the other with a drift distance of 50~cm, sit within the primary MicroBooNE vessel at each end of the LArTPC.  They are installed at different heights to allow purity measurements at different depths of the argon.

 \subsection{Initial Purification}
 The MicroBooNE cryogenic system was designed to allow the cryostat to go from containing atmosphere (and the detector) to containing high purity liquid argon (and the detector) without ever evacuating the cryostat. Such a process is considered essential to the development of multi-kiloton experiments where the cost of an evacuable cryostat would be prohibitive.  While a successful test had been carried out in the LAPD \cite{Adamowski:2014-LAPD}, before MicroBooNE, this process had never been performed in a cryostat with a fully instrumented large detector.  A more complete description of the initial purification process with operational details is available in~\cite{zuckerbrot}.
 
The process for preparing the cryostat for filling with liquid argon involves three stages.  The first is the ``piston-purge'' stage where argon gas flows into the cryostat at its lowest point and the argon (being denser than the atmosphere) pushes the atmosphere out of the cryostat.   The second stage is a recirculation phase where the argon-gas loop is closed and the argon flows through a water-removal filter to reduce outgassing of water from detector materials.  The final stage is a cool down phase where the loop remains closed and the argon gas is cooled in a heat-exchanger and cools the detector to the point where the insertion of liquid argon will not damage the detector.

The so-called ``piston-purge'' was achieved using a single pipe for gas input at the bottom of the cryostat and an identical pipe for the output at the top; the pipes (referred to as ``sparger'' pipes) each have 4.76 mm diameter holes every 12.7 cm on both sides.  Calculations of the mutual diffusion of argon and nitrogen suggested that a rate of 770.4 Nm$^3$/hr. was adequate in avoiding turbulence and minimizing back-diffusion of the air ~\cite{voirin}. A total of about 13 volume exchanges took one week and resulted in contamination levels of $<$10 ppm $H_2 O$, $<$ 1 ppm $O_2$ and $<$ 1 ppm $N_2$. 

The recirculation phase used the same piping system as the piston-purge but at a significantly higher flow rate (3 cryostat changes/hr). As mentioned, the recirculation passed the gas through the water removal filter and over a period of three weeks, the water concentration was reduced to $<$ 1 ppm. To counteract the increase in $N_2$ and $O_2$ levels from outgassing observed during this phase, (typically a rise of a few ppb/hr), a small fraction of the gas was vented and replaced with fresh argon. 

The final stage involved cooling the cryostat and detector to prepare for filling with liquid argon. The TPC design imposed two requirements on the cooldown. One was that  the input gas be no more than 20~K colder than the frame of the TPC to avoid wire-breakage since the massive TPC frame takes a long time to cool while the wires immediately adopt the input gas temperature and shrink,  thereby increasing the stress on the wires.  The second requirement was that the temperature difference between the top and bottom of the TPC be less than 20~K at any time to avoid warpage of the frame.  A set of RTDs screwed into the TPC frame, RTDs in the inlet gas piping, and a set of RTDs attached to the anode side of the cryostat were used to measure and monitor the process. For the cooldown, the argon-gas was cooled using a three-pass counter-flow nitrogen heat-exchanger, and recirculated using the sparger pipes. The cooldown was declared complete after three weeks when the average temperature had reached 105 K and the temperature difference between the top and the bottom of the detector was $<$10 K. Upon reaching this state, the O$_2$ level had been reduced to 18 ppb and the H$_2$O concentration had fallen to 22 ppb. These levels are reduced by a factor of $>$800 in the liquid and presented an excellent environment for the start of filling.

\newpage
\section{Liquid Argon Time Projection Chamber}
\label{sec:tpc-all}

The MicroBooNE \lartpc drifts and collects charge to produce fine-grained images of the ionization that is liberated by charged particles traversing a volume of highly-purified liquid argon.  This section describes the design and implementation of the \lartpc in the experiment.

\begin{figure}
\centering	
\includegraphics[width=0.8\linewidth]{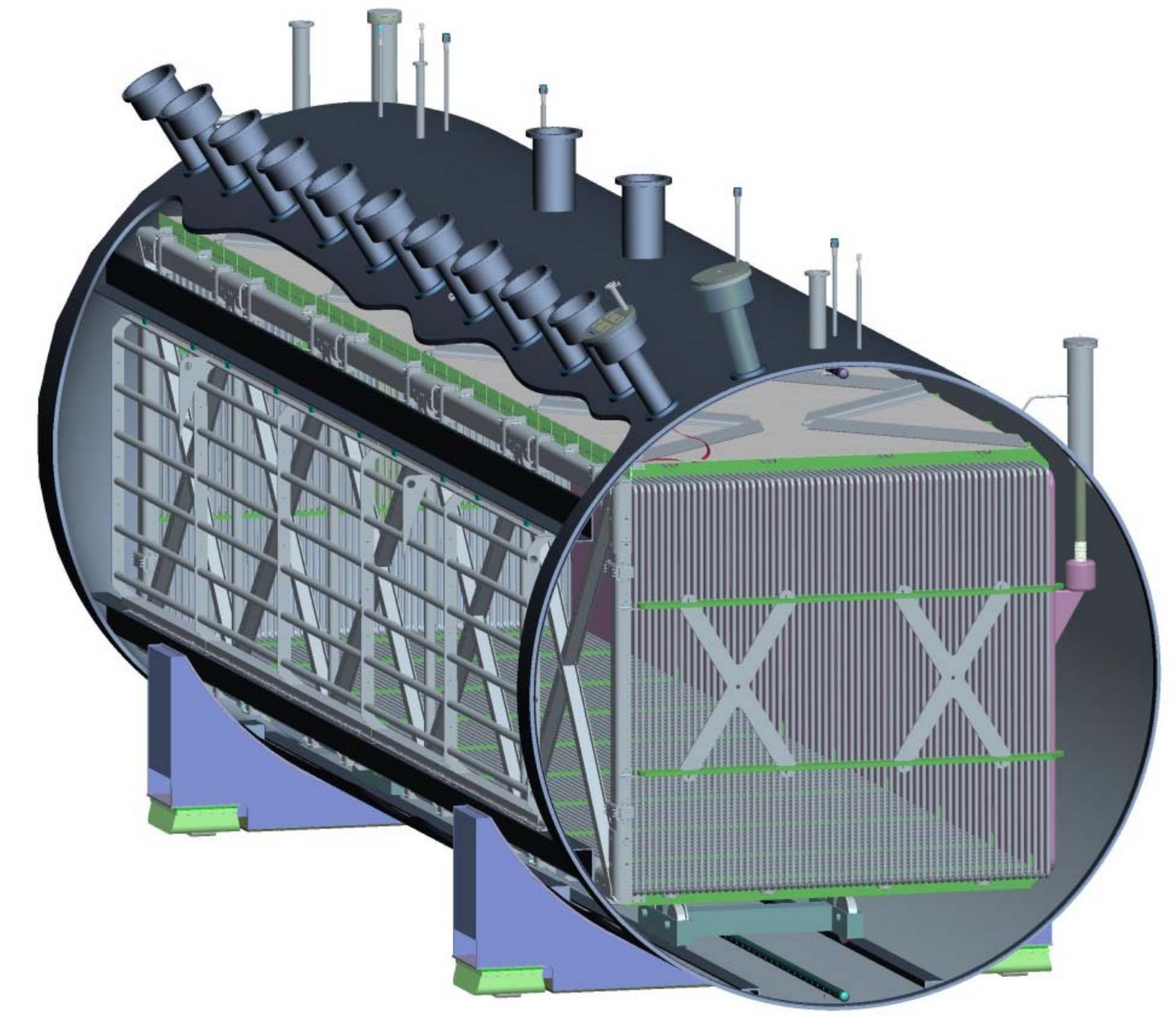}
\caption{Schematic diagram of the MicroBooNE \lartpc, depicted as it is arranged inside the cryostat.}
\label{fig:tpc-cryostat}
\end{figure}

The \lartpc is composed of three major structures: the cathode, the field cage, and the anode. A negative voltage is introduced via a feedthrough passing through nozzle N2 on the cryostat and applied at the cathode, which defines an equipotential surface.   A uniform electric field between the cathode and the anode planes is established by a series of field rings connected by a voltage divider chain starting at the cathode and ending at the anode plane.   Facing the cathode planes are the sense wire planes: two induction planes (referred to as the ``U'' and ``V'' planes) with wires oriented at $\pm60^{\circ}$ from vertical, followed by one collection plane (referred to as the ``Y'' plane) with vertically-oriented wires.   The wires of the anode planes are the sensing elements that detect the ionization created by charged particles traveling through the LArTPC.  Figure \ref{fig:tpc-cryostat} depicts the assembled MicroBooNE \lartpc after insertion into the cryostat, showing details of the cathode, field cage, and anode plane.  Table \ref{tab:tpcparam} lists the main parameters of the MicroBooNE LArTPC, which will be described in detail in this section.  


\begin{table}[!htb]
   \centering
     \caption{MicroBooNE \lartpc design parameters and nominal operating conditions.} 
    \begin{tabular}{lr} 
    \hline
    Parameter & Value \\
    \hline
    $\#$ Anode planes & 3\\
     Anode planes spacing& 3 mm \\
     Wire pitch & 3 mm  \\
     Wire type & SS, diam. 150 $\mu$m\\
     Wire coating & 2$\mu$m Cu, 0.1$\mu$m Ag\\
     Design Wire tension & 6.9N $\pm$ 1.0N\\
     $\#$ wires (total) & 8256 \\
     $\#$ Induction0 plane (U) wires & 2400 \\
     $\#$ Induction1 plane (V) wires & 2400 \\
     $\#$ Collection plane (Y) wires & 3456 \\
     Wire orientation (w.r.t. vertical) & +60$^{\circ}$,-60$^{\circ}$,0$^{\circ}$ (U,V,Y) \\
     \hline
     Cathode voltage (nominal) & -128 kV \\
     Bias voltages (U,V,Y) & -200 V, 0 V, +440 V \\
     Drift-field & 500 V/cm\\
     Max. Drift Time, Cathode to U (at 500 V/cm) & 1.6 ms\\
    \hline
    $\#$ Field-cage steps & 64\\
    Ring-to-ring voltage step & 2.0 kV\\
    \hline
   \end{tabular}
   \label{tab:tpcparam}
\end{table}

\subsection{Cathode}
\begin{figure}
\centering	
\includegraphics[width=0.62\linewidth]{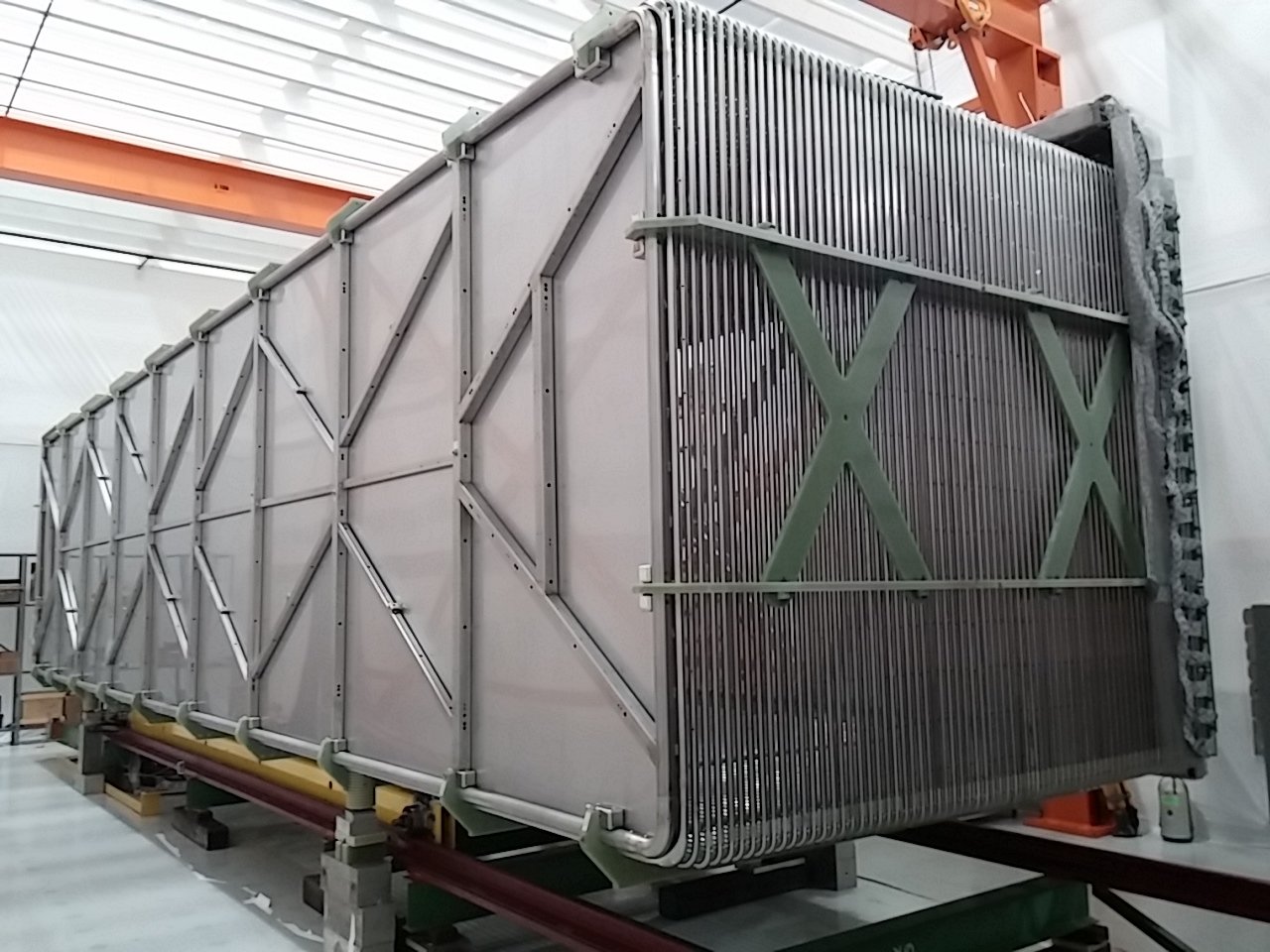}
\includegraphics[width=0.62\linewidth]{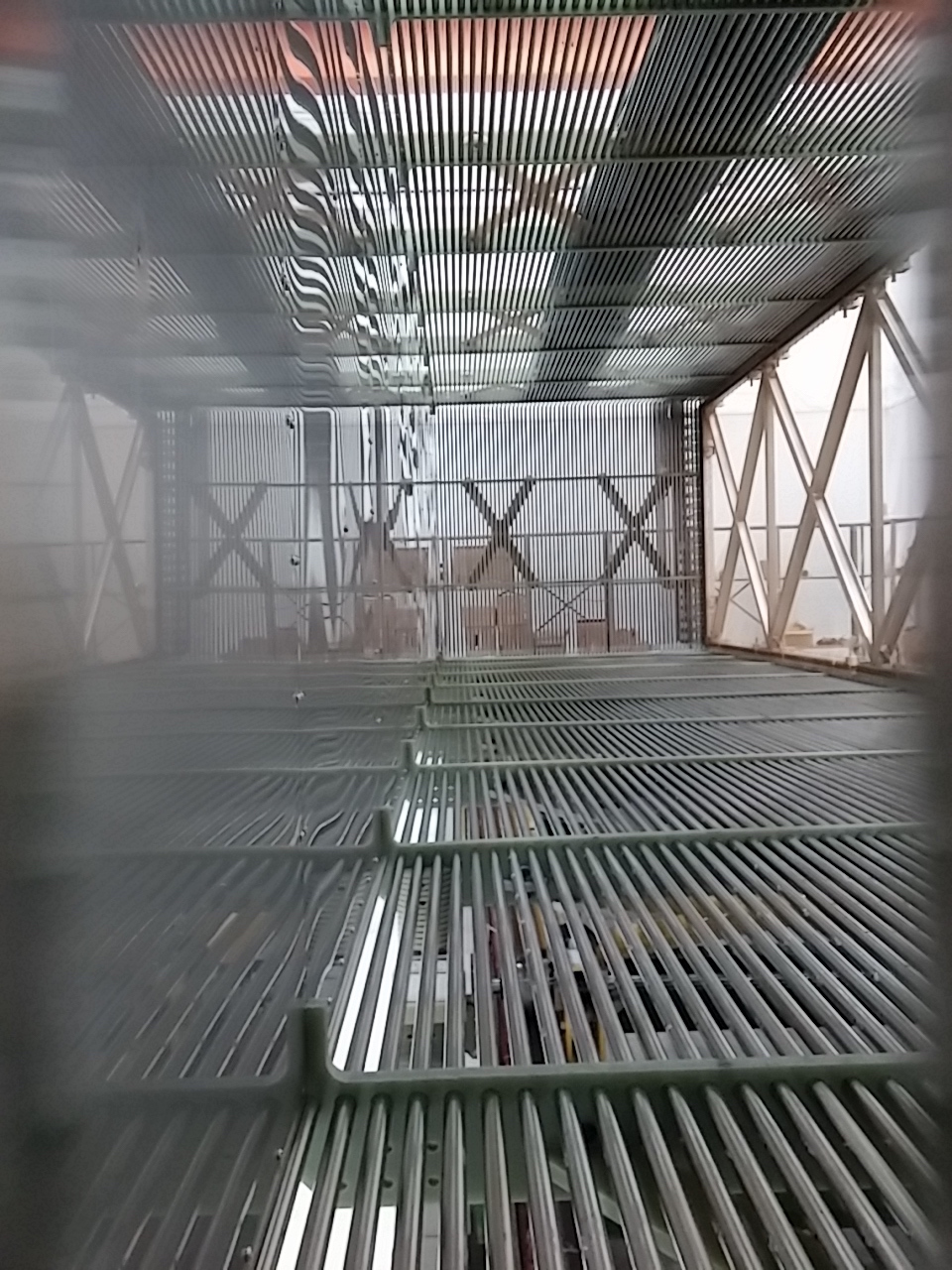}
\caption{Top: Exterior view showing the cathode frame and structural supports to which cathode sheets are fastened.  Bottom: Interior view of cathode plane as viewed from the upstream end of the \lartpc, showing cathode sheets.  Note that the cathode sheets are polished, so a reflection is is clearly present in this photograph.}
\label{fig:tpc-cathode}
\end{figure}

\begin{figure}
\centering
\includegraphics[width=0.75\linewidth]{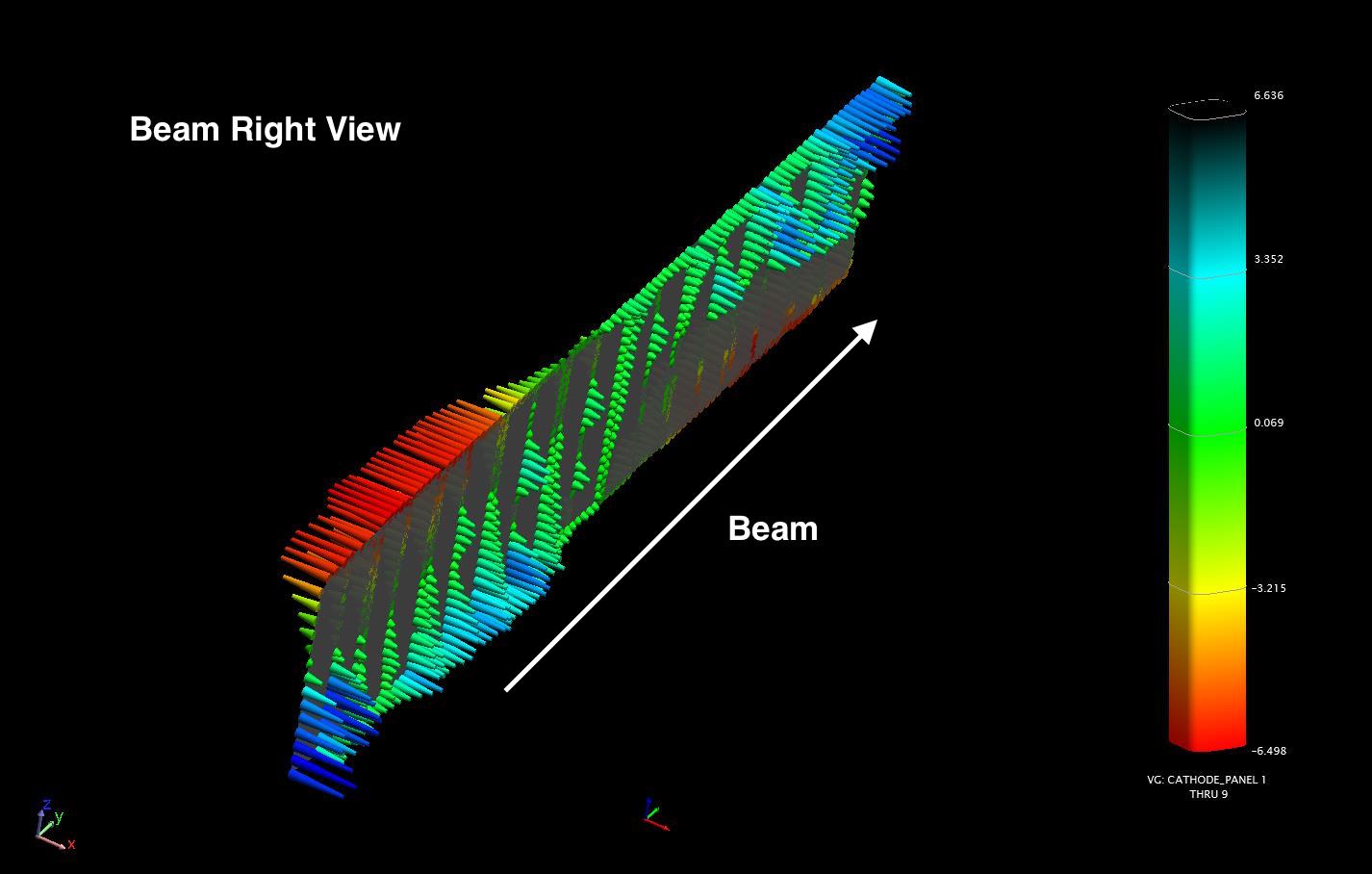}
\includegraphics[width=0.75\linewidth]{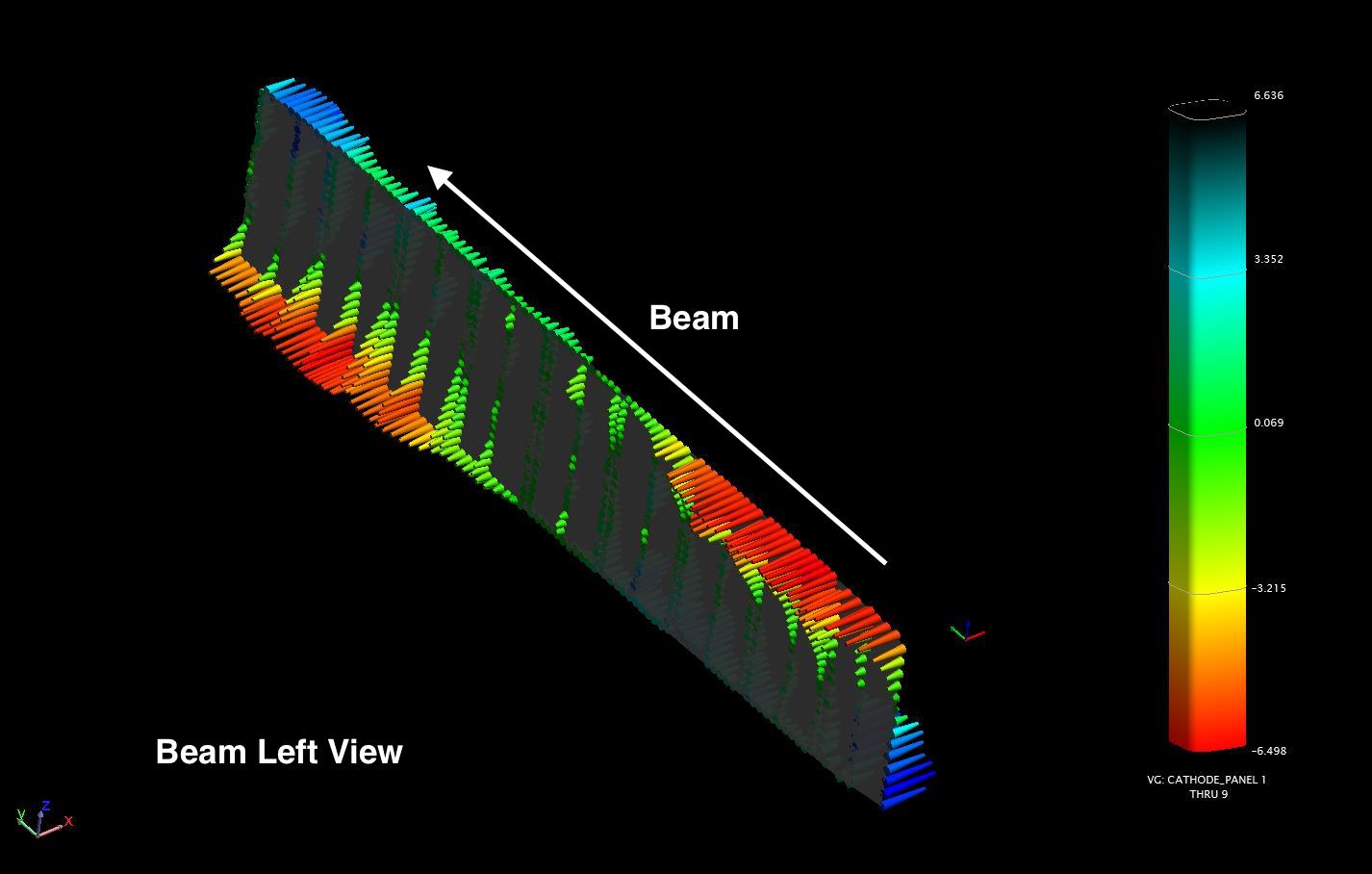}
\caption{Survey results showing the flatness of the cathode, as viewed from the interior (top) and exterior (bottom) sides, after shimming.  Color scale extends from -6.498 mm (red) to +6.636 mm (blue).}
\label{fig:tpc-cathode-survey}
\end{figure}

The cathode is assembled from 9 individual stainless steel sheets (Type 304, 2.3 mm thick) that are fastened to a supporting frame by hex-head stainless-steel button-screws. The outer edge of the cathode frame consists of round stainless steel tubes of 5.08 cm outer diameter and 3.18 mm wall thickness.  Within this outer edge, square tubes with 5.08 cm $\times$ 5.08 cm cross-sectional area, and 3.18 mm wall thickness, are fastened together with hex-head button-screws, forming a support structure upon which the cathode sheets are attached.  The individual components of the support structure are further welded together to eliminate sharp features from this high-potential surface.  The exterior frame and support structure of the cathode, and also an interior view, are shown in figure~\ref{fig:tpc-cathode}. The cathode plane sheets are shimmed according to survey data to make the cathode as flat and as parallel to the anode frame as possible, resulting in the two surfaces being parallel to within 0.0413$^{\circ}$.  Flatness of the cathode is evaluated relative to a best fit plane of survey data (more than 10000 survey points recorded with a laser tracker). The largest deviations of the cathode from the best fit plane are +6.6 mm and -6.5 mm. Approximately 55$\%$ of the measured survey points fall within +/-3 mm of the best fit plane, and more than 90$\%$ of the points fall within $\pm$5 mm.  Figure~\ref{fig:tpc-cathode-survey} shows the results of the survey, with deviations from flat represented as color-coded data extending away from the nominal plane of an ideal cathode. 

\subsection{Field Cage}
The field cage encloses the volume between the cathode plane and the anode wire planes, and creates a region with a uniform electric field.  The volume defined by the interior of the field cage, bounded by the anode and cathode planes, is referred to as the ``active'' volume.   The field cage structure consists of 64 individual sets of thin-walled stainless steel tubes (2.54 cm OD, 0.51 mm wall thickness), each shaped into a rectangular loop framing the perimeter of the active volume. These 64 loops are mounted parallel to the cathode and anode planes, as shown in figure~\ref{fig:tpc-cathode}, and are held in place by a G-10 rib support structure. Each field cage loop is electrically connected to its neighbors via a resistor divider chain (described in the following section), causing each loop to operate at a different electrical potential, which in turn maintains a uniform electric field between the cathode and anode planes. For a nominal -128~kV cathode voltage, the difference in potential between adjacent field cage loops is 2~kV, ramping down the total potential in equidistant steps from cathode to anode. The distance from center-to-center of adjacent field cage loops is 4.0~cm.

Each field cage loop is assembled from 2.07 m long vertical pipes on the upstream and downstream ends of the \lartpc, and on the top and bottom from two 5.18 m long horizontal pipes connected by a stainless steel coupling in the center.  Each tube has venting holes approximately every 15 cm to allow for effective purging from atmosphere and to avoid any trapped volumes.  

\begin{figure}
\centering	
\includegraphics[width=0.7\linewidth]{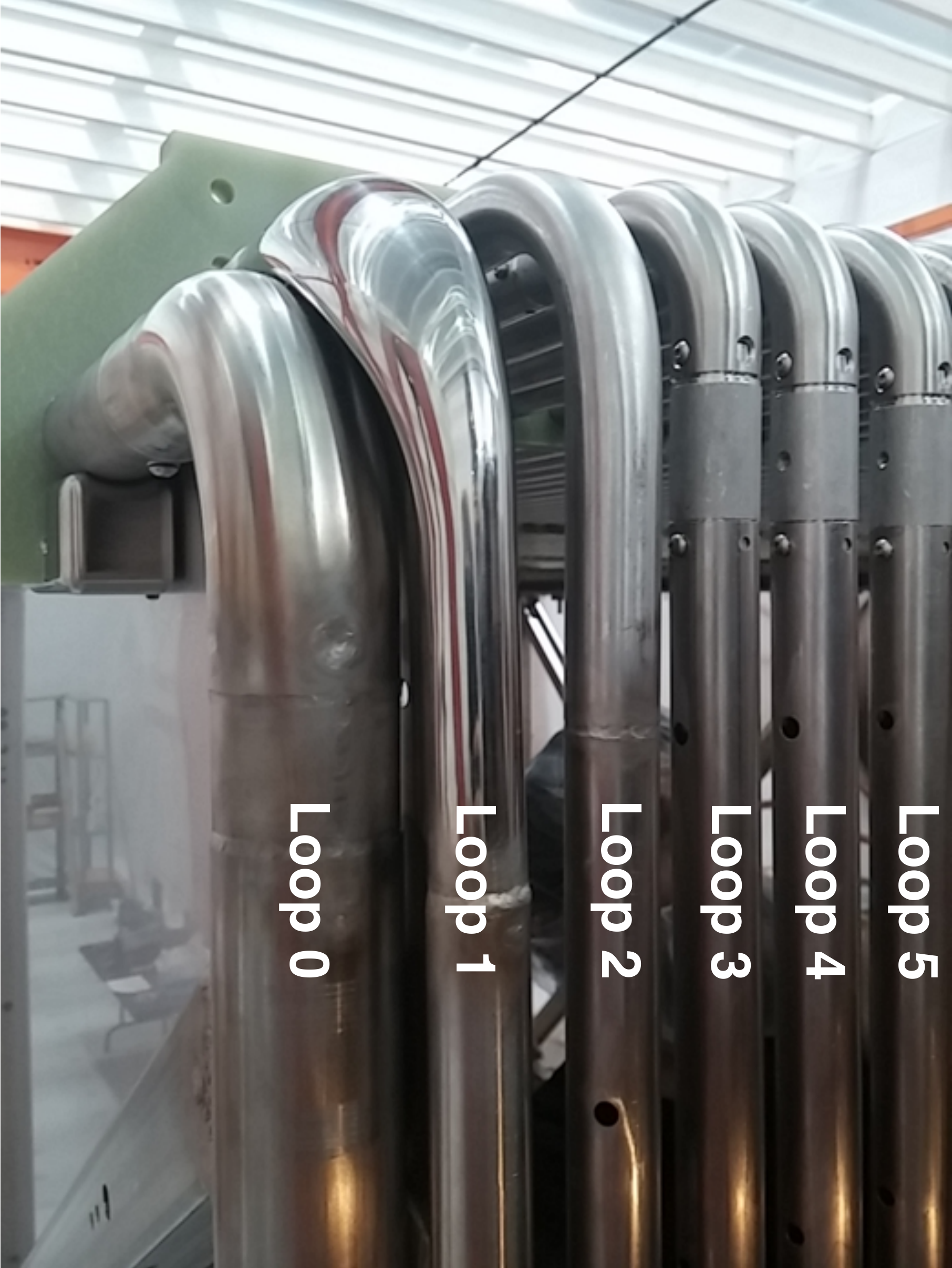}
\caption{Photograph of the field cage during construction, with loops 0 (cathode) through 5 labeled.  Field cage loops closest to (and including) the cathode are modified to reduce sharp edges that would result in higher electric fields.}
\label{fig:tpc-smooshed-elbow}
\end{figure}

The four corners of each field cage loop are curved with a radius of 5.24 cm. Each corner is formed by three parts: two couplings and an elbow, shown in figure~\ref{fig:tpc-smooshed-elbow}. The couplings make the connections between the pipes and the elbow. The thin-walled tubes and elbows slip-fit over the ends of the couplings with a 2.2 cm overlap. Each coupling has two 6-32~NC tapped holes and the connections to the adjoining pieces are made by hex-head button-screws and split-ring lock washers with no teeth.


In order to avoid electrical breakdown between the inner cryostat surface and field cage parts at high potential on or near the cathode, the electric field strength is minimized at the corners and edges of the field cage. Loops 0, 1, and 2 are each designed differently than the other field cage loops.  Loop 0 is a special case in that it is made from larger diameter piping of 5.08 cm OD, and frames the cathode and also acts as its mechanical support.  It operates at the same electrical potential as the cathode plane sheets attached to it.  Loop 0 has a slightly smaller area than the other field cage loops, as shown in figure~\ref{fig:tpc-smooshed-elbow}.  Loop 1 is the first of the 64 loops in the field cage with 2.54 cm pipe OD.  It surrounds the cathode plane and operates at cathode potential.  The elbow of loop 1 has a specially designed geometry in order to minimize the electric field potential.  The elbow of loop 2 has a larger radius of curvature than the standard elbows, also for the purpose of minimizing the electric field potential. For all three of these loops (loop 0, 1, and 2), connections at corners and joints are made by welding instead of screws to avoid sharp edges that would result in higher electric fields and greater chance of electrical breakdown.

Another precaution to minimize the electrical field between the loops and the cryostat surface is the positioning of the coupling screws: for the first 20 loops, the screws are positioned on the sides facing the screws of the neighboring loops instead of facing inward to the \lartpc active volume and outward toward the grounded cryostat surface. Hex-head button-screws and lock washers are also used here in order to minimize sharp metal edges.

\begin{figure}
\centering	
\includegraphics[width=0.7\linewidth]{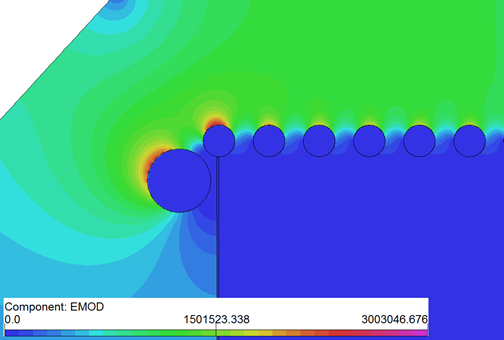}
\caption{Cross-section view showing the electric field simulation inside the field cage when the cathode is set to a voltage of -128 kV.  Loop 0 is represented by the larger diameter circle on the left of the image.  The legend shows the absolute values of electric field modulus in units of V/m.}
\label{fig:efield}
\end{figure}

Figure \ref{fig:efield} shows the simulated electric field values inside the MicroBooNE \lartpc, and some of the surrounding volume inside the cryostat, when the cathode is set to an operating voltage of -128 kV.  

\subsubsection{Resistor Divider Chain}

A resistor divider chain installed across the field cage loops steps the voltage down in magnitude from the cathode plane to the anode wire plane in equal steps.  For a nominal value of -128~kV on the cathode, this results in a potential difference of 2~kV between each pair of loops. The value of the equivalent resistance between loops within the divider chain was chosen to be low enough such that the current flow through the divider circuit is much greater than the signal current flowing through the \lartpc. The signal current in our case is dominated by the free ionization produced by the cosmic ray flux, and is estimated to be <50~nA.  An equivalent resistance of 250~M$\Omega$ between each pair of field cage loops, corresponding to a current flow of 8~$\mu$A, was chosen.  

The voltage divider chain is mounted on the inside of the field cage at the upstream end of the detector. The couplings at the top corner of each field cage loop have additional holes facing the inside of the field cage, where the resistors are mounted. On the first 16 field cage loops, pairs of Metallux HVR 969.23 499~M$\Omega$ resistors (rated to 23~W, 48~kV in air) are mounted electrically in parallel to establish the beginning of the voltage divider chain. On the remaining loops, four thick-film Ohmite Slim-Mox 104E metal-oxide epoxy-coated resistors with a lower power and voltage rating (1.5~W, 10~kV in air) are mounted in parallel, per loop.  Extensive testing was done on these two types of resistors \cite{Bagby:2014wva}. 

\begin{figure}
\centering	
\includegraphics[width=0.8\linewidth]{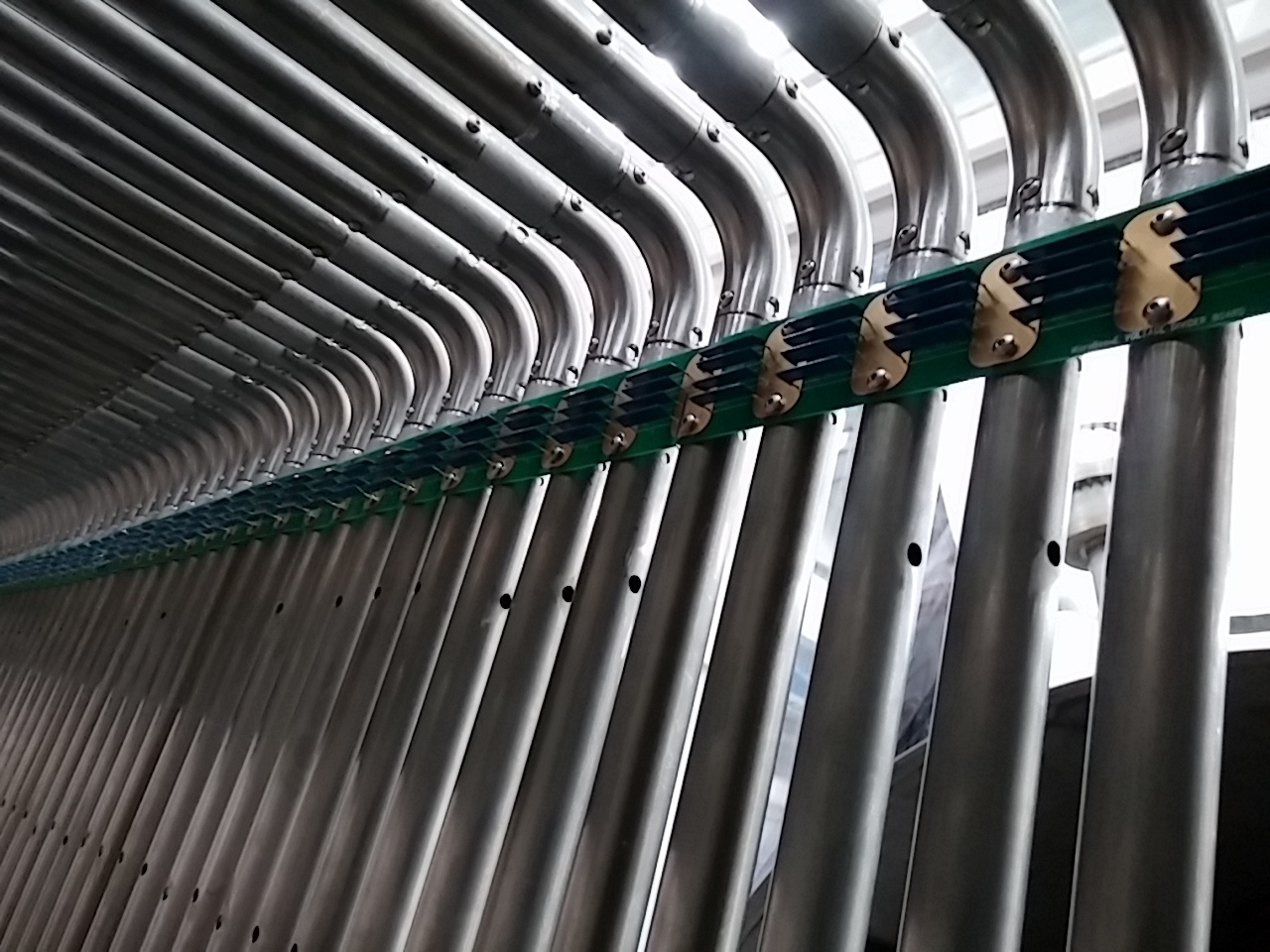}
\caption{The Ohmite Slim-Mox 104E resistors arranged in parallel sets of four on printed circuit boards that span eight field cage tubes.}
\label{fig:tpc-voltage-divider-slimmox}
\end{figure}

For the loops with the Slim-Mox resistors, printed circuit boards span across eight field cage gaps and therefore have eight 250~M$\Omega$ resistances in series, shown in figure~\ref{fig:tpc-voltage-divider-slimmox}. The electrical connection between the boards and each field cage loop is made by metal contact pads on the back side of the boards, held in electrical contact with the field cage tube by a hex-head button-screw and lock washer.

While the designed operating voltage difference across each resistor in the detector is 2~kV with a power flow of 4~mW, there is a slight possibility that the voltage drop and power could temporarily exceed the rating of the resistors in the case of discharge between the cathode plane or field cage loops and the cryostat wall, through the bulk liquid argon. Recent studies~\cite{Acciarri:2014ica} have shown that the value of the minimum breakdown electrical field decreases with the increasing argon purity; for purities as high as that required in the MicroBooNE detector, breakdown has been observed at electric fields as low as 40~kV/cm.

The field cage behaves like a capacitance network at high frequencies. Based on measurements and simulations, the total energy stored inside the field cage when fully charged is estimated to be approximately 24~J. In the case of a discharge between the cryostat and the cathode or one of the field cage loops close to the cathode, simulations show that voltages of up to 80~kV peak, with a discharge time constant of a few seconds, can develop across the resistors. The observed peak voltages in such discharge scenarios decrease the further the breakdown occurs from the cathode, such that discharges occurring between the cryostat and field cage loops 32 through 63 do not exceed the 10~kV rating of the resistors.

Two strategies have been implemented to protect the resistors nearest the cathode from damage due to discharge.   The first is the use of the Mettalux resistors on the first 16 loops given their higher voltage rating of 48~kV as compared to the Slim-Mox resistors on the remaining loops.

\begin{figure}
\centering	
\includegraphics[width=0.8\linewidth]{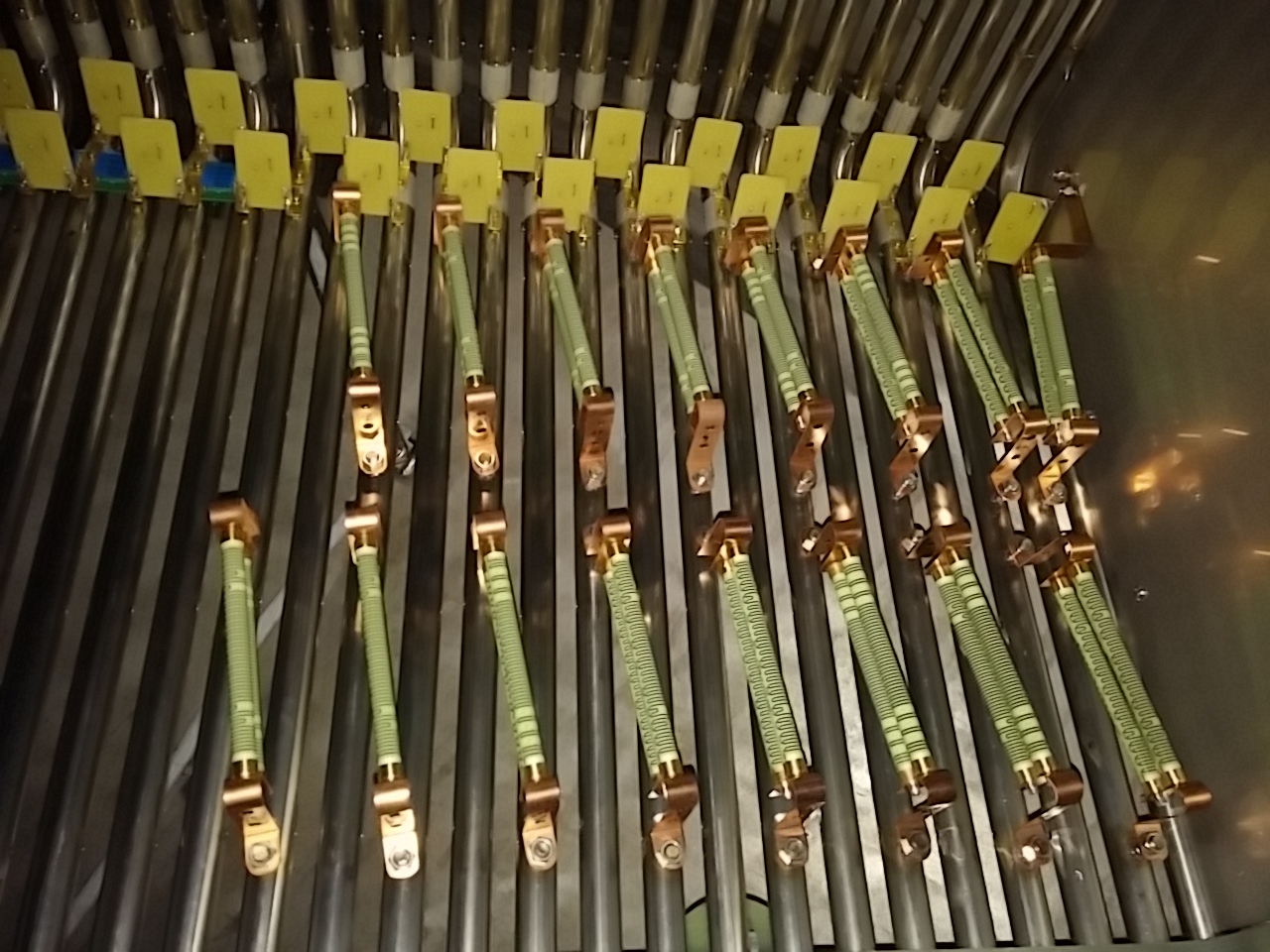}
\caption{The Metallux HVR 969.23 resistors mounted on the 16 field cage loops closest to the cathode.}
\label{fig:tpc-voltage-divider-metallux}
\end{figure}

Since the Metallux HVR 969.23 resistors are significantly larger physically than the loop-to-loop distance, they are mounted diagonally between each pair of field cage loops. They are held by copper brackets, which are attached to studs welded onto the field cage tubes, shown in figure~\ref{fig:tpc-voltage-divider-metallux}.

\begin{figure}
\centering	
\includegraphics[width=0.8\linewidth]{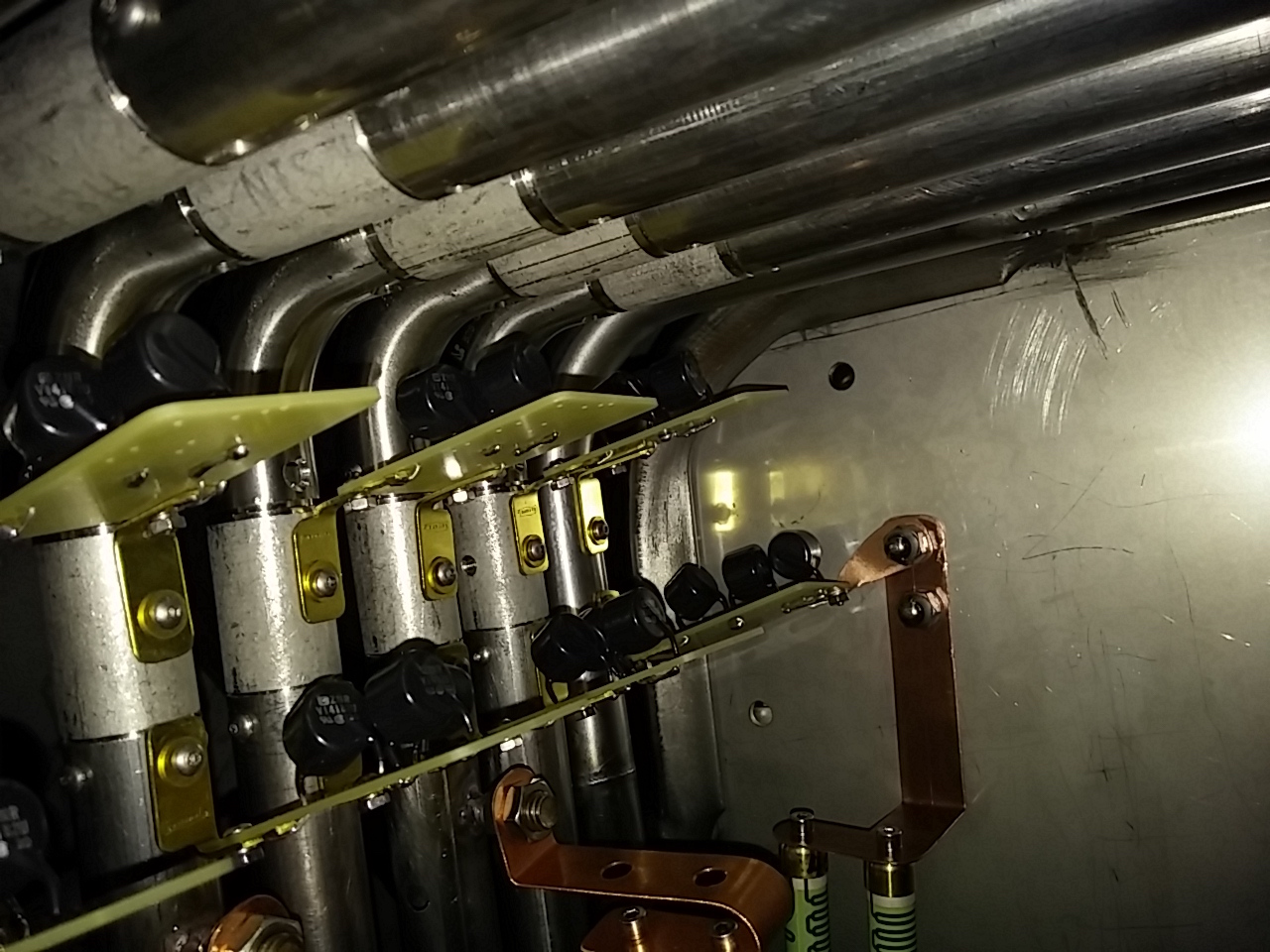}
\caption{Surge-protecting varistors (small black disks in the photograph) are installed in parallel with the voltage divider resistors for the first 32 field cage loops. Here, they are shown mounted on small boards in sets of 3, and attached to the field cage by means of 6-32 hex-head button-screws.}
\label{fig:tpc-voltage-divider-varistors}
\end{figure}

The second protective measure is installation of surge protection circuits on field cage loops 1 through 32.  The chosen surge protection devices are designed to short the circuit in the case of a voltage spike, which protects any other electrical components installed in parallel.  Below their clamping voltage, they exhibit a very high resistance and do not influence the circuit. The behavior of Gas Discharge Tubes (GDTs) and varistors in liquid argon has been studied extensively for application in the MicroBooNE field cage \cite{Asaadi:2014iva}. The surge protection device chosen is a Panasonic ERZ-V14D182 varistor with a clamping voltage of 1700 V. In order to obtain a very high resistance in normal operation and a clamping voltage above the 2~kV in normal operation mode, three of these devices are mounted in series across a block of four Slim-Mox 104E or two Metallux 969.23 resistors. These additional varistor boards make electrical contact with the field cage via brass mounting brackets that are fastened to the field cage with button-head screws, as shown in figure~\ref{fig:tpc-voltage-divider-varistors}.

\subsection{Anode Planes}

The anode frame holds the induction and collection plane sense wires at tension and provides overall structural support for the beam-right side of the LArTPC.  Individual sense wires for all anode planes are held in place by wire carrier boards, which are printed circuit board assemblies that position the wires as well as provide the electrical connection to the electronic readout system of the experiment.

\subsubsection{Mechanical structure}

\begin{figure}
\centering	
\includegraphics[width=0.8\linewidth]{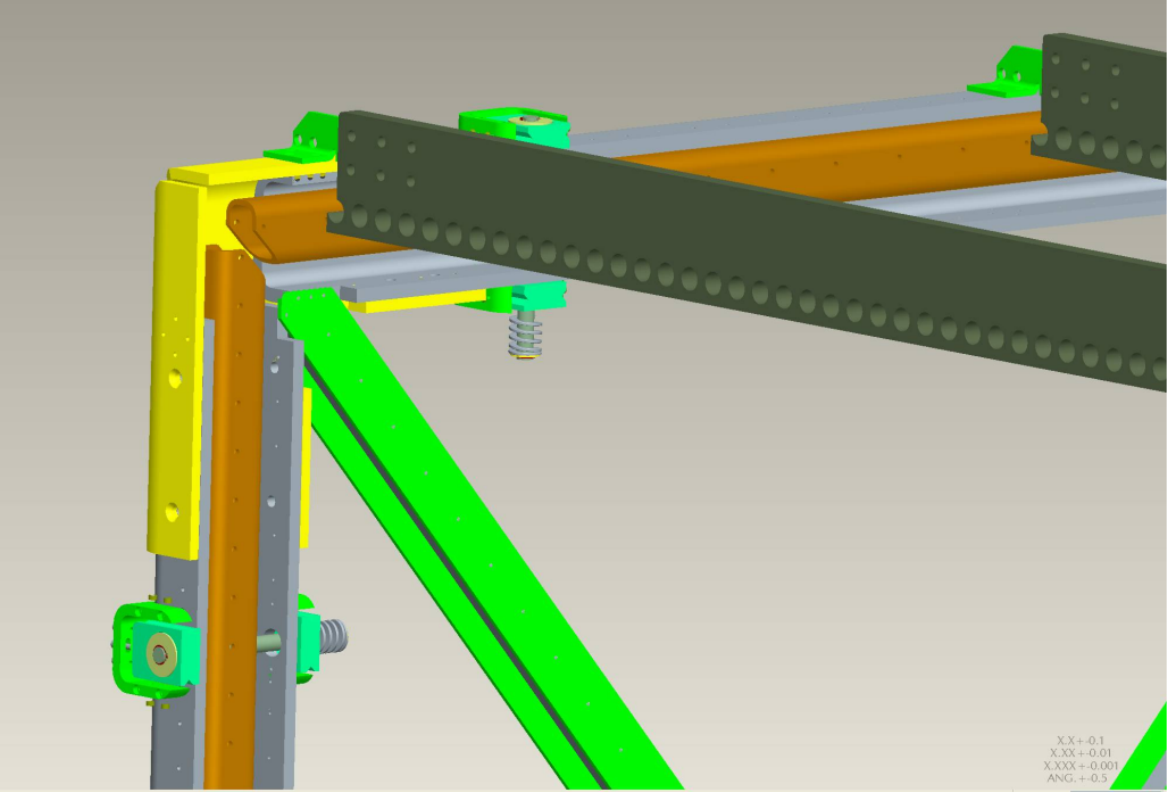}
\caption{Rendering of the anode frame assembly.  The C-channel is depicted in gray, and the adjustable tensioning bar assembly is shown in orange.}
\label{fig:anode-frame-3dmodel}
\end{figure}

The anode frame is comprised of a stainless steel C-channel hosting adjustable tensioning bars to which the wire carrier boards are attached. The C-channel and tensioning bar assembly is depicted for one corner of the anode frame in figure~\ref{fig:anode-frame-3dmodel}.  Wire carrier boards attach to precision alignment pins distributed along the length of the tensioning bars. 

\subsubsection{Wire winding and quality assurance}

\begin{figure}
\centering
\includegraphics[width=0.5\textwidth]{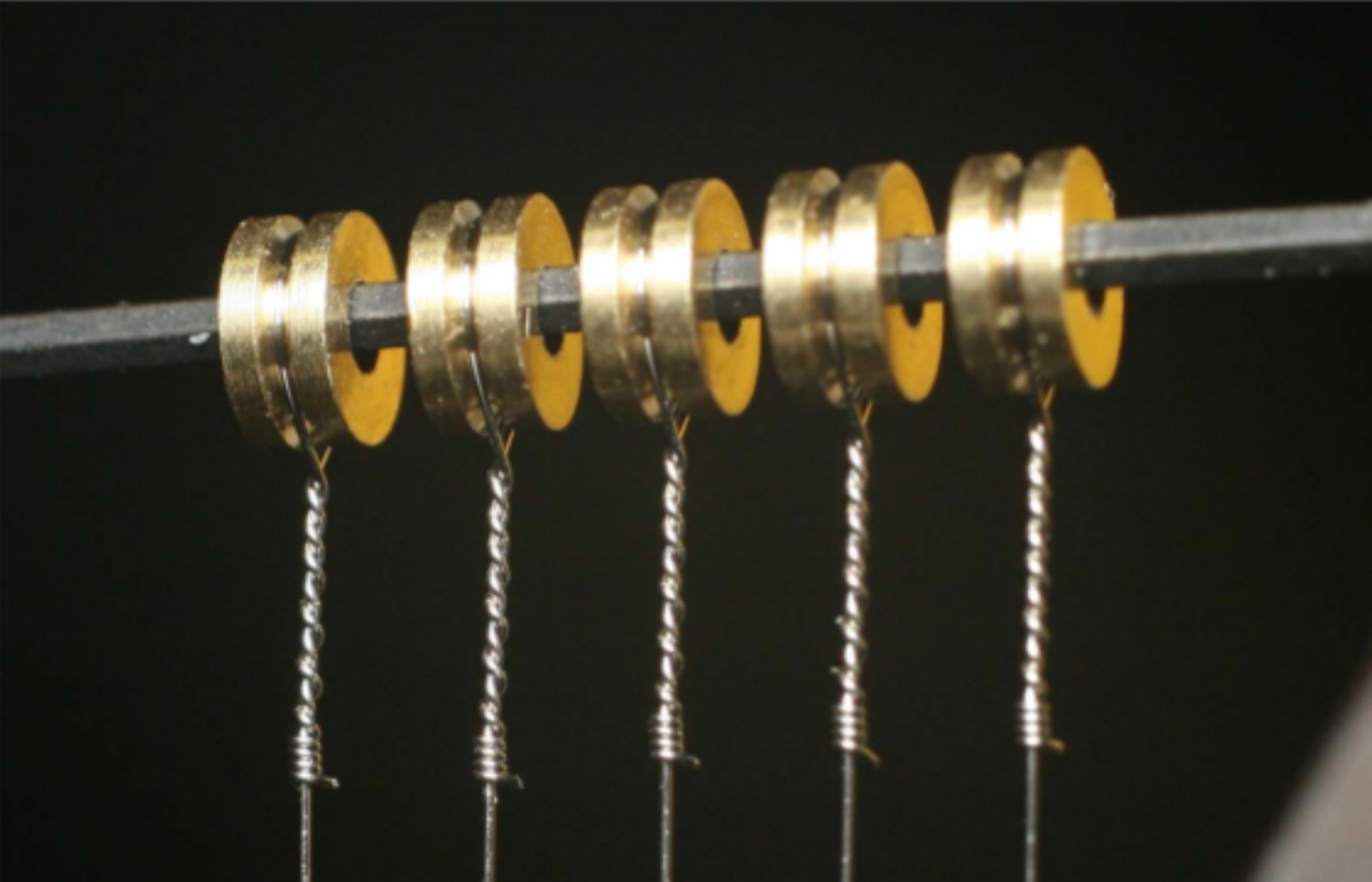}
\caption{Photograph of the wire termination on the brass ferrules.  Each ferrule is 3~mm in diameter, and 1.5~mm thick.}
\label{fig:ferrules}
\end{figure}

The three anode planes are constructed from wire carrier boards that have individually-prepared wires attached to them in groups of 16 (for the U- and V- angled planes) or 32 (for the vertical Y-plane).  Consistent quality in wire preparation was achieved by a semi-automated winding machine, which terminated the ends of each wire via wrapping around 3~mm diameter brass ferrules as shown in figure~\ref{fig:ferrules}.   The wire termination method via wrapping around a brass ferrule similar to that used in the ICARUS T600 detector.  

\begin{figure}
\centering
\includegraphics[angle =0,width=0.7\textwidth]{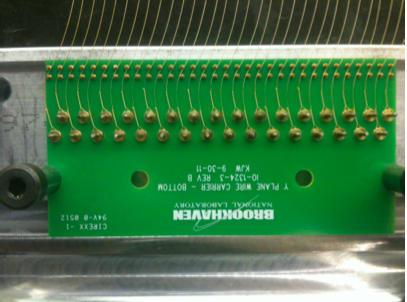}
\caption{Photograph of a collection plane wire carrier board that has been filled with wires, but has not yet had the cover plate installed.}
\label{fig:carrier-boards}
\end{figure}

Each wire was tested for strength on a tensioning stand where a load of 2.5~kg (more than 3 times the nominal load of 0.7 kg) was applied for 10 minutes, ensuring that the wire preparation did not leave any weaknesses that could result in a breakage.  Upon successful completion of the quality assurance testing, each wire was placed onto a wire carrier board, shown in figure~\ref{fig:carrier-boards}.

\begin{figure}
\centering
\includegraphics[angle=0,width=0.5\textwidth]{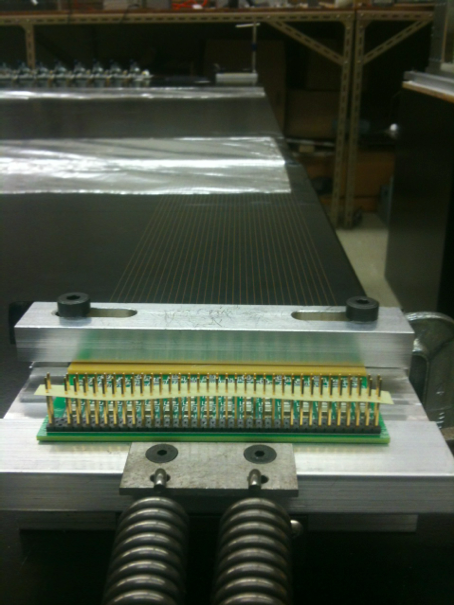}
\caption{Photograph of a collection plane wire carrier board on the tension stand.}
\label{fig:stress-stand}
\end{figure}

When installed on the wire carrier boards, the wires make contact with gold pins which are connected to a trace that routes to the cold electronics.  Once the wire-carrier board was filled with wires, a cover plate was installed and press-fit rivets were installed to hold the assembly together. The assembled wire carrier board was then placed onto a tension stand, to reapply a 2.5 kg tension/wire to the whole board for 10 minutes. This is to ensure that the wires were not weakened during the board assembly process. The tension stand is depicted in figure \ref{fig:stress-stand}.   A comprehensive description of the MicroBooNE wire preparation and associated quality assurance studies can be found in \cite{Acciarri:2016ugk}.

\subsection{Parts Preparation}

The majority of the parts that make up the \lartpc are either stainless steel or G-10. These two material types, as well as any others used in the LArTPC, were tested in the Fermilab Materials Test Stand (MTS)~\cite{Rebel:2011-MTS}, whose purpose was to investigate the suitability of materials for use in LArTPCs.  The MTS confirmed that none of the materials used in the \lartpc assembly would contaminate the liquid argon.  Before assembly, all \lartpc parts were cleaned according to the procedures described in the following sections.

\subsubsection{Cleaning stainless steel}

The delivered stainless steel parts were often greasy due to machining, and those with holes or interior cavities generally had a significant amount of trapped metal shavings due to the machining processes. Many of the pieces also had markings from permanent ink pens, dirt smears, rust spots, and/or dried oil from machining. These pieces were scrubbed with ScotchBrite 7447 general-purpose hand pads before cleaning.

Parts that were small enough to fit in an ultrasonic bath were prepared according to the following prescription. A pre-rinse with tap water was performed to remove particulate matter, followed by deburring of sharp edges. The first ultrasonic wash was 15 minutes in heated distilled water with a 3\% solution of Citranox acid detergent~\cite{citranox}. After a first rinse in distilled water, a second ultrasonic wash was performed, again for 15 minutes, but using heated distilled water with a weak solution of Simple Green detergent~\cite{simplegreen}. A second rinse in distilled water was performed, followed by a final rinse in a fresh bath of distilled water. The parts were then wiped dry with lint-free cloths, air dried completely, and wrapped in plastic film for storage.

Stainless steel parts that were too large to fit in the ultrasonic bath were prepared by a simpler prescription out of necessity. A pre-rinse with tap water was followed by deburring to remove sharp edges. Both the first and second washes were done with tap water and a weak solution of Simple Green detergent, scrubbing with brushes and lint-free sponges. Two tap water rinses were done, and the parts were then wiped dry with lint-free cloths. As a final additional step, each part was wiped with 200-proof ethyl alcohol, and then air dried completely. These parts were also wrapped in plastic film for storage.

\subsubsection{Cleaning G-10}

G-10 is known to absorb large quantities of water, which would outgas in the argon and could inhibit reaching the required argon purity in the detector. For this reason all G-10 parts were cleaned and then baked to remove moisture. The largest G-10 parts on the detector are beams that span the distance between the cathode and anode. These were washed in 1900-liter ultrasonic baths that are overseen by Fermilab Accelerator Division, typically used for cleaning large sections of accelerator beam pipes. An initial pre-wash was done with tap water to remove as much particulate matter as possible, since the machining process left a large amount of dust on the machined edges.  Pieces were then placed in the ultrasonic bath with heated deionized water and a 2\% solution of Elma Clean 65 (EC 65) neutral cleanser~\cite{elmaclean}. Two ultrasonic bath rinses were performed, and the pieces were then sealed in plastic bags with clean dry nitrogen gas. In order to remove the absorbed water, the large G-10 parts were then transported to the Fermilab Technical Division where they underwent an outgassing procedure to remove any remaining absorbed moisture.  They were baked in a large oven under vacuum until a plateau in the outgassing rate was reached, as reported by a monitor inside the oven. Upon completion of the outgassing procedure, the parts were resealed in plastic bags for storage.

\subsection{Assembly}
\label{sec:assembly}

\begin{figure}
\centering	
\includegraphics[width=0.48\linewidth]{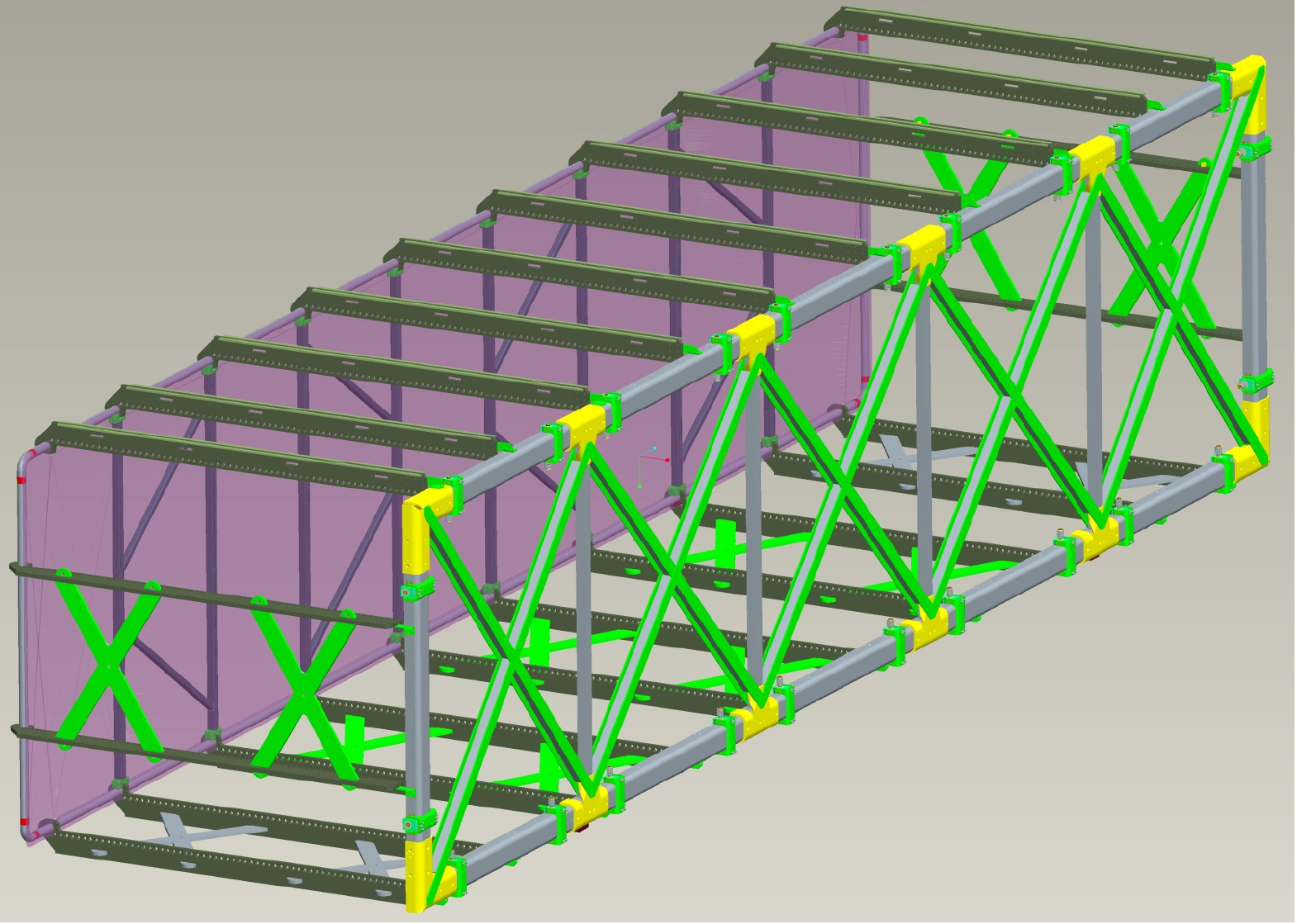}
\includegraphics[width=0.48\linewidth]{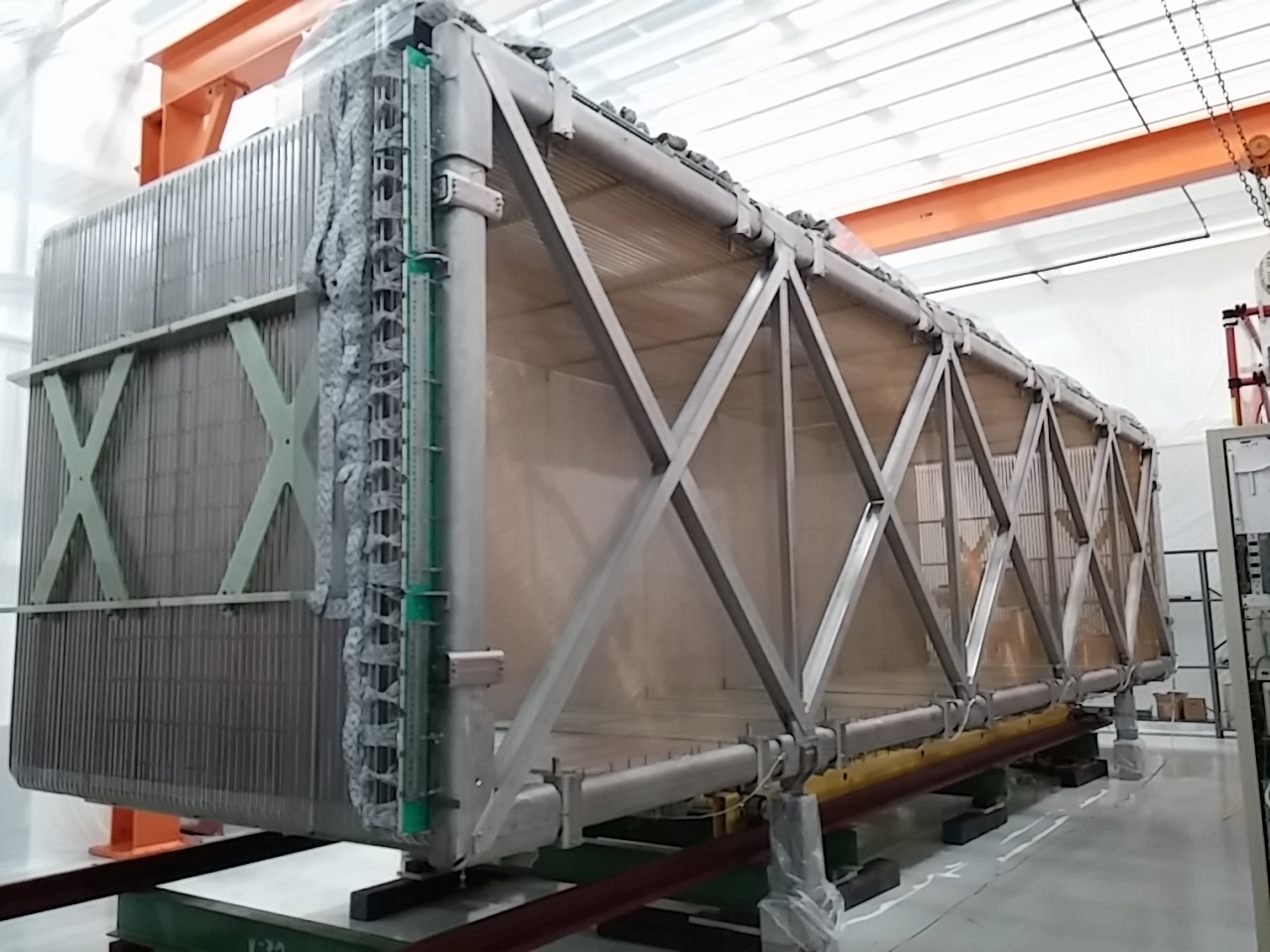}
\caption{Left: Rendering of the full \lartpc frame assembly.  Right: Assembled \lartpc after wire and electronics installation.}
\label{fig:tpc-full}
\end{figure}

The full mechanical structure of the \lartpc is shown in figure~\ref{fig:tpc-full}, with the left image depicting the cathode frame as semi-transparent to show the support structures on which the cathode sheets are attached. The anode frame is on the right of this image, with I-beams configured in a crossed pattern to maintain the shape and rigidity of the outer C-channel structure. Ribs of G-10 connect the anode and cathode, electrically isolating them from each other while also providing mounting holes to hold in place each of 64 field cage loops that define the active volume of the LArTPC.  The field cage loops are visible in the photograph on the right of figure~\ref{fig:tpc-full}.

\begin{figure}
\centering	
\includegraphics[width=0.48\linewidth]{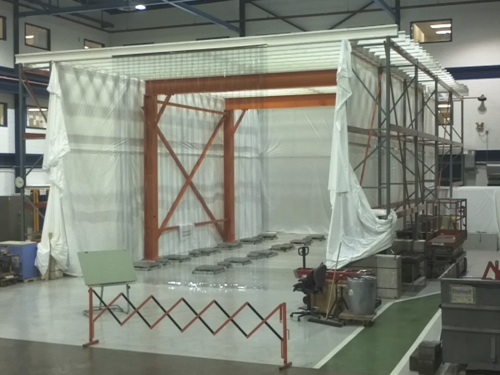}
\includegraphics[width=0.48\linewidth]{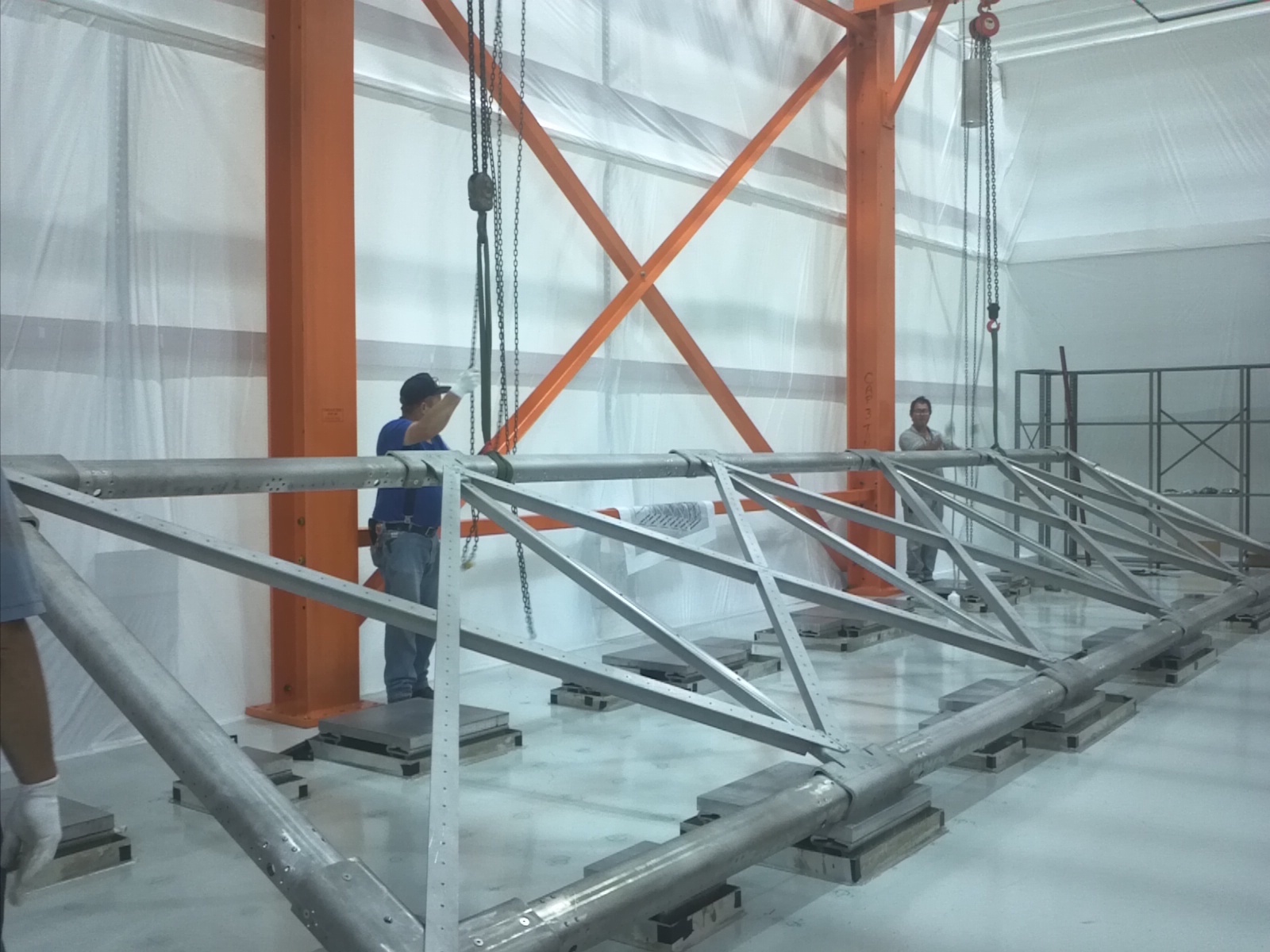}
\includegraphics[width=0.48\linewidth]{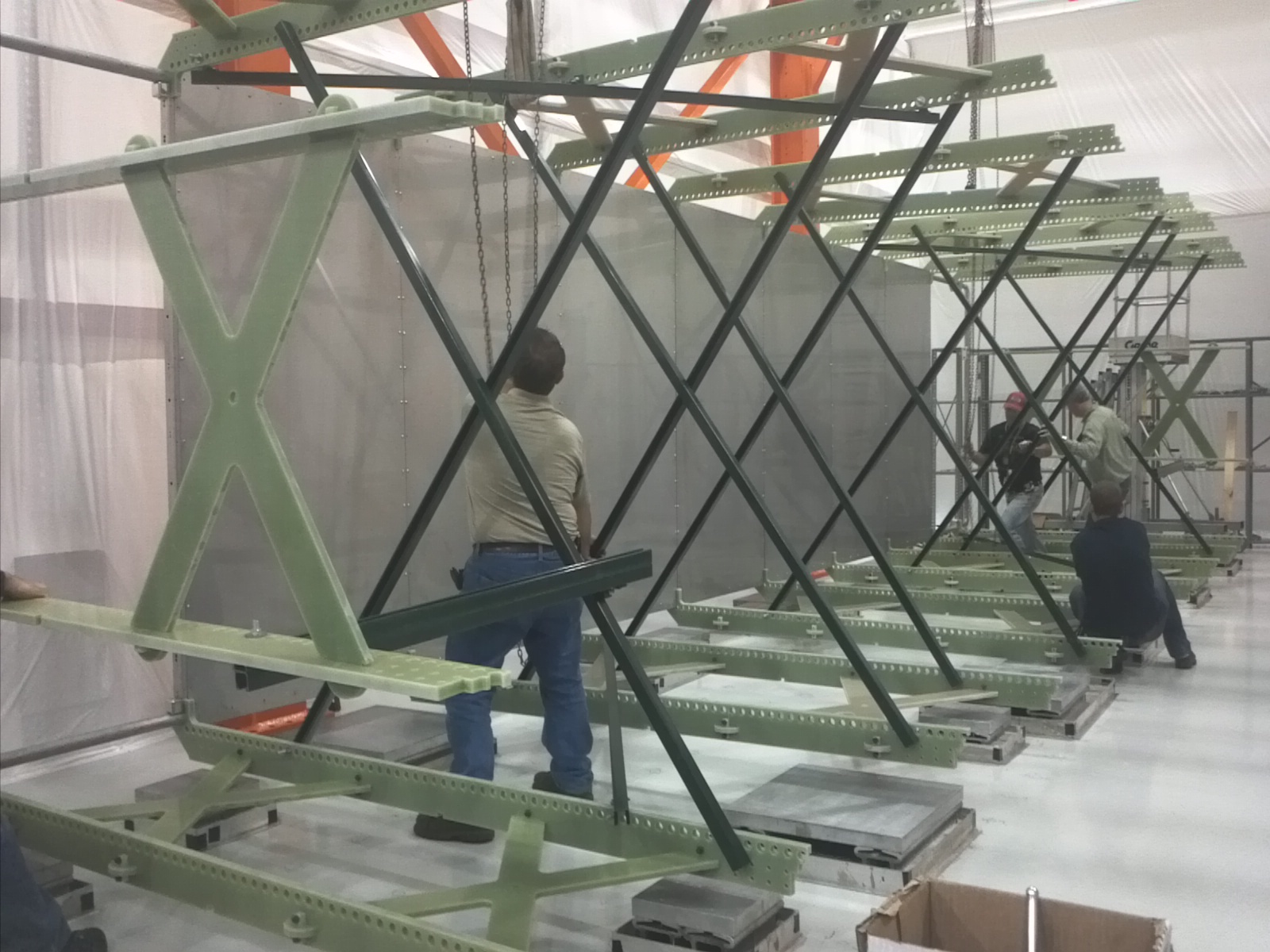}
\caption{Top Left: Clean tent where MicroBooNE \lartpc assembly was conducted. Top Right: Anode frame in the process of being moved from the metal assembly platforms. Bottom: Cathode frame and G-10 ribs on metal assembly platforms.}
\label{fig:tent}
\end{figure}

Assembly was done inside of a clean tent, shown on the top left in figure \ref{fig:tent}, on a flat surface made up of adjustable-height metal platforms that were installed on the assembly room floor.  These platforms were leveled to better than 0.5~mm before beginning assembly.  The anode frame was the first part of the detector to be assembled on this surface, shown on the top right of figure~\ref{fig:tent}. It was temporarily placed aside, and the cathode frame was assembled on the same set of platforms along with the G-10 ribs, which stood vertically with the help of temporary unistrut support pieces.  The combined cathode and G-10 frame was then lifted and rotated to the proper orientation, with G-10 ribs extending horizontally from the cathode to the anode, as shown on the bottom in figure~\ref{fig:tent}. Finally, the anode frame was brought back over and attached to the G-10 ribs, and the stainless steel tubes that make up the field cage loops were fed through the holes in the G-10 ribs to complete the mechanical structure of the LArTPC. 

\subsubsection{Wire installation and tension measurements}
\begin{figure}
\centering
\includegraphics[width=0.95\textwidth]{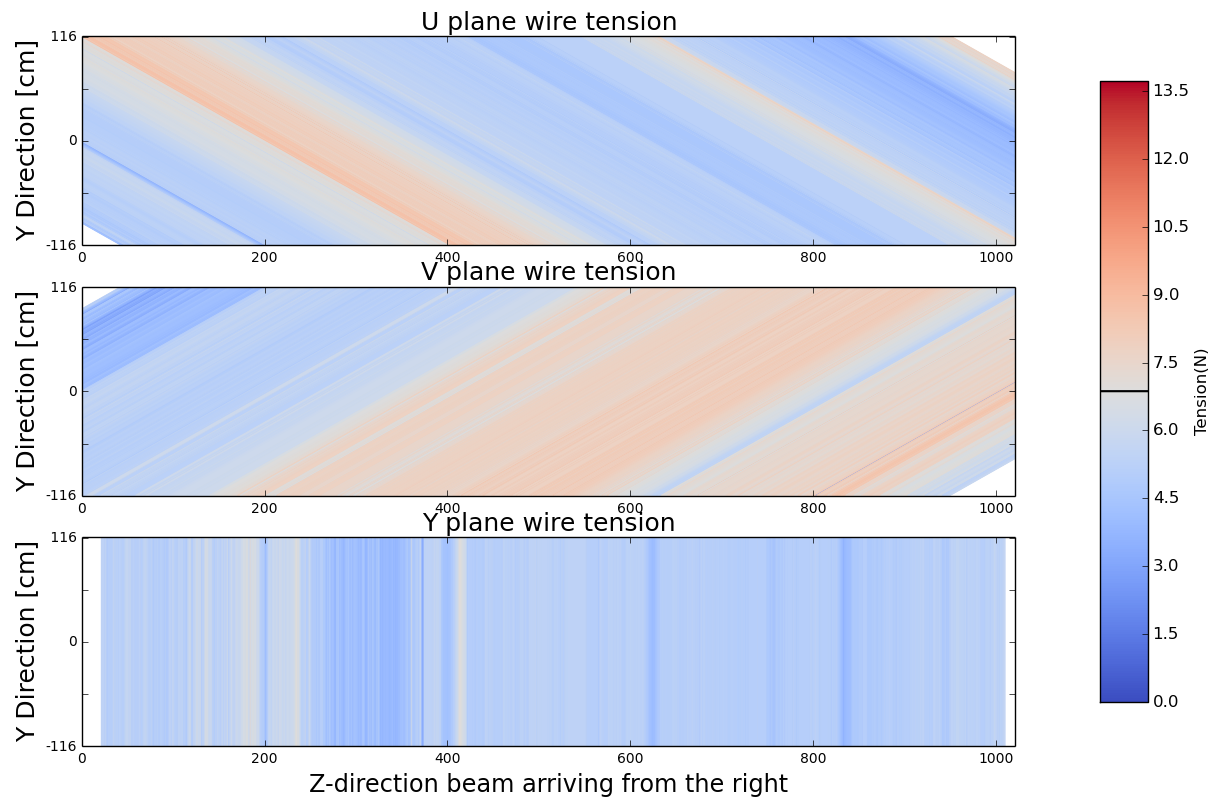}
\caption{Final survey results for wire tension of the MicroBooNE LArTPC.}
\label{fig:heatmap}
\end{figure}

During detector assembly, the completed wire carrier assemblies (consisting of wires and supporting carrier boards on either end) were manually installed onto the adjustable tensioning bars residing in the C-channel of the supporting anode frame.  A team of two people installed each assembly onto the anode frame.  The collection plane was the first installed, followed by the middle induction plane, and then finally the inner induction plane.  Once all three anode planes were completely installed, the tensioning bars were adjusted and a survey was taken of the tension of all anode wires.  Tension was set according to the design criteria that it be small enough to prevent wire breakage during cool down and large enough to limit the maximum wire sag due to gravity to under 0.5~cm for any 5~m long U or V wire.   Tension was recorded through measurement of the resonant frequency of a laser beam reflected from a plucked wire and incident on a photodiode connected to a spectrum analyzer program \cite{SpectrumLaboratory}.  The tension measuring equipment was developed and produced by the University of Wisconsin Physical Sciences Laboratory.  The tensioning bars were adjusted iteratively until the surveyed tension of all wires was within a range, approximately $\pm$1.0 N of the nominal value of 6.9 N, where no single wire was too taught or loose to create detector performance issues.  Figure \ref{fig:heatmap} shows the final surveyed tension of the wires for each plane.

\subsection{High Voltage System}
\label{sec:hv}

To create the drift field, negative voltage (referred to as the ``high-voltage'' or ``HV'') is supplied to the \lartpc cathode.  This potential is generated outside of the cryostat by a Glassman LX150N12 power supply.  Before entering the cryostat, the output of the power supply is passed through a current-limiting resistor chain that serves as both a low-pass filter for the power supply ripple, and a partition for the stored energy in the system.  The resistor chain is a set of eight 10 M$\Omega$ resistors connected in series and submerged in a transformer oil in an aluminum container.  This assembly was successfully tested to -200~kV in a dedicated test dewar.

\begin{figure}
\centering{
\includegraphics[width=\textwidth]{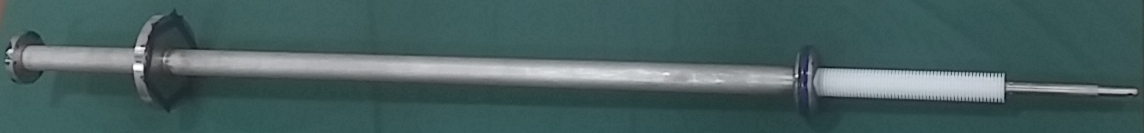}\\
\includegraphics[width=\textwidth]{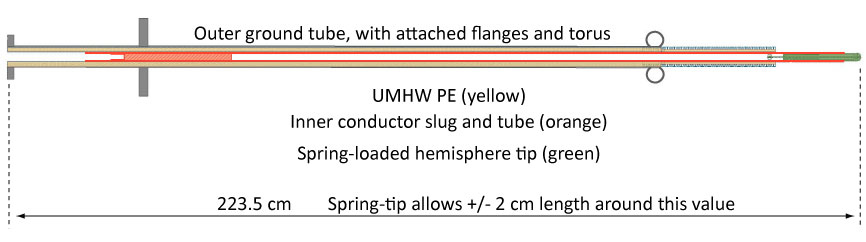}
}
\caption{Photograph and drawing of the production HV feedthrough. The spherical probe tip is attached to the end of the inner conductor, on the right side of these figures.}
\label{fig:hv_ftpic}
\end{figure}

The potential is introduced into the cryostat by a custom-designed HV feedthrough.  The feedthrough is based on an ICARUS design~\cite{Amerio:2004-T600}; a 2.54 cm diameter stainless steel inner conductor is surrounded by a 5.08 cm outer diameter ultra-high molecular weight polyethylene (UHMW PE) tube that is encased in an outer ground tube.  A photograph and drawing of the production feedthrough are shown in figure~\ref{fig:hv_ftpic}.

\begin{figure}
\centering{
\includegraphics[width=0.55\textwidth]{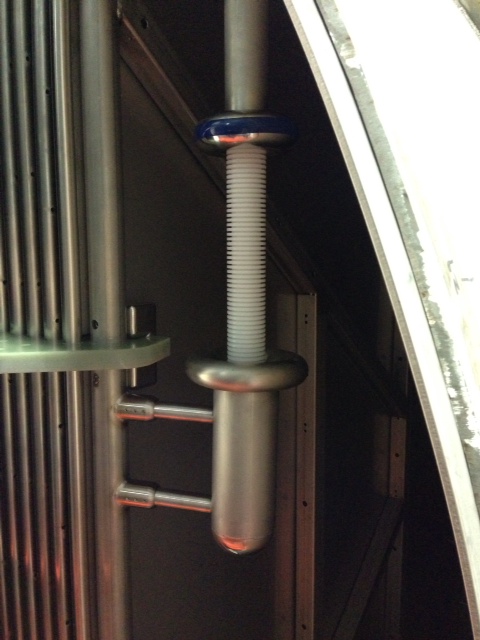}
}
\caption{The production HV feedthrough inserted into the cathode receptacle cup inside the cryostat.}
\label{fig:hv-FT-cryostat}
\end{figure}

Figure~\ref{fig:hv-FT-cryostat} shows the HV feedthrough extending into its receptacle cup attached to the MicroBooNE LArTPC cathode.  The lower termination of the outer ground tube is a torus chosen to reduce the electric field between both the feedthrough and the cathode plane, and along the feedthrough itself.  The electrical connection to the cathode is made with a hemispherical spring-loaded tip attached to the inner conductor of the feedthrough.  The UHMW PE tube is machined with grooves and extends into the receptacle cup attached to the cathode.

\newpage


\section{Light Collection System}
\label{sec:light-collection}

Liquid argon is a bright scintillator, and sampling the light from interactions in the argon can bring powerful new capabilities and information to complement the charge information from the \lartpc.  The light collection system in MicroBooNE is designed to meet MicroBooNE's physics goals for light collection, which are twofold.  First, for accelerator-induced events, the light collection system is designed to enable the experiment to trigger on $\geq$40~MeV kinetic energy protons produced in beam neutrino interactions.  Second, for non-beam events,  the system is designed for efficient observation of 5 to 10~MeV electrons from supernova neutrino interactions.   

The light produced by neutrino interactions in MicroBooNE is an important input for both event selection and reconstruction.  
One of the critical capabilities the light collection system provides is the ability to form a beam-event trigger when a pulse of light is observed in coincidence with the beam spill.  Because typically only one beam spill in 600 will produce a neutrino interaction in the detector, such a trigger will substantially reduce the overall data output rate.  
For non-beam physics studies, the light system provides triggering and an event $t_0$ for the \lartpc system.   For accelerator-induced events, the time of the beam spill (1.6 microsecond duration) provides an adequate $t_0$ for tracks from the event.  However, because of the long window over which the ionization electrons of an event drift (about 1.6 ms maximum drift time at 500 V/cm field), almost every ($>$99\%) accelerator-induced event will include one or more cosmic ray muons crossing the detector during the TPC livetime.  Utilizing the distribution of hits across the photodetectors, one can better reject cosmic ray muon tracks.  
The light can also be used to select and trigger on specific types of cosmic-ray calibration events (Michel electrons, straight-through muons, etc.) and non-beam events (supernova neutrinos, cosmic background events to proton decay studies, etc.).

The light collection system consists of primary and secondary sub-systems.  The primary light collection system is made up of ``optical units,'' each one consisting of a PMT located behind a wavelength-shifting plate.  In total, 32 optical units were installed, yielding 0.9\% photocathode coverage.  The secondary system consists of four light guide paddles.   These paddles were introduced for R\&D studies for future LArTPCs, and are placed near the primary optical units to allow a comparison of their performances.  A flasher system, used for calibration, consists of optical fibers bringing visible light from an LED to each PMT face.

\begin{figure}
	\centering
              \includegraphics[width=0.95\textwidth]{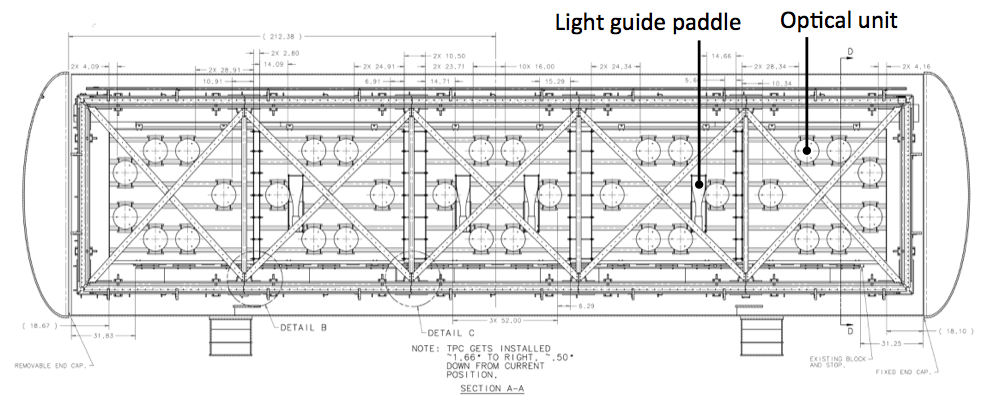} 
        \caption{The MicroBooNE light collection system consists of a primary system of 32 optical units and a secondary optical system of four lightguide paddles~\cite{Katori:2013wqa}. These are mounted behind the anode wire planes such that the view is not obscured by structural cross bars of the LArTPC. }\label{fig:lightlayout}
\end{figure}

The light collection detectors are located in the $y$-$z$ plane behind the anode planes of the LArTPC, as shown in figure~\ref{fig:lightlayout}.  The combined transparency of the three anode planes is 86\% for light at normal incidence.  This transparency value assumes 100$\%$ of VUV photons impinging on the wires are absorbed.  The detectors were placed so as not to be obscured by the \lartpc structural cross-bars, shown in figure~\ref{fig:lightlayout}.  Locating the light detectors behind the anode plane places them in a very weak electric field due to the +440 V bias of the collection plane.  To test for an effect from weak electric fields, the response of a PMT placed between a +700 V mesh and ground, separated by 50 cm, was studied.  The PMT zenith was 25 cm from the voltage source.   No effects on the signal were observed.  This was expected, as the photocathode is held at ground, effectively acting as a Faraday cage.  

Throughout the design and construction of the light collection system,  substantial R\&D was performed.   The reader should refer to~\cite{Jones:2013bca,Bugel:2011xg,Katori:2011uq,Chiu:2012,Baptista:2012,Briese:2013wua,Jones:2013mfa,Katori:2013wqa,Moss:2014ota,Conrad:2015xta,Moss:2015hha} for detailed results of these studies.    A useful overall review is available in~\cite{Jones:2015bya}. 

\begin{figure}
	\centering
    \includegraphics[width=0.95\textwidth]{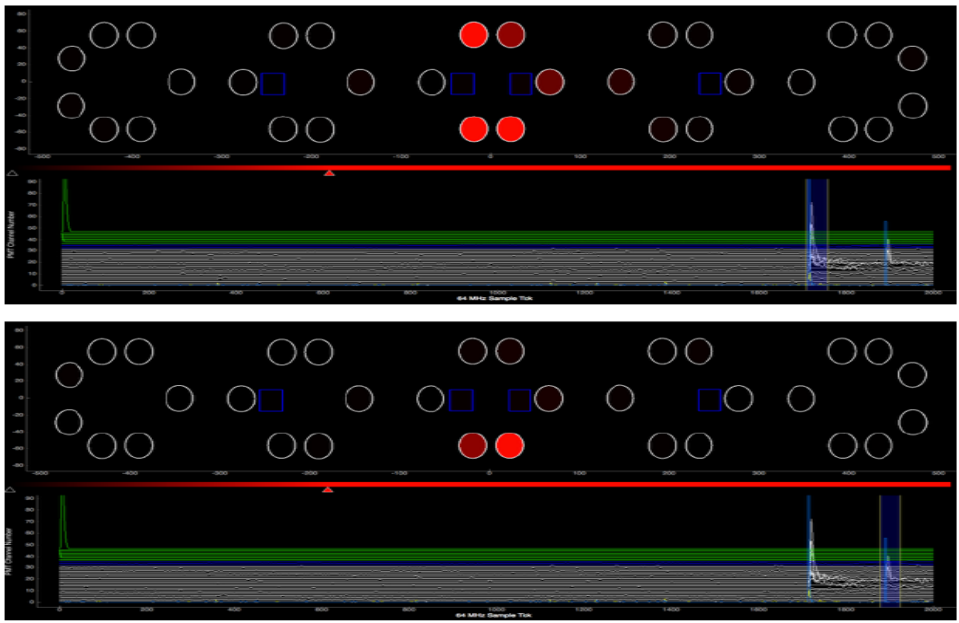} 
    \caption{Displays showing two sequential events in the argon as seen by the light collection system.  The sequence is consistent with a muon that stops (top) and decays (bottom).  The circles correspond to the optical units. Red circles indicate those units with hits within the time range indicated by the vertical bars drawn on the PMT waveforms, which are shown at the bottom of each display.}\label{fig:muondecaylight}
\end{figure}

Figure~\ref{fig:muondecaylight} shows the light observed in two sequential events in the argon, consistent with a muon entering the detector followed by a Michel electron from the decay.   One can see that the light is relatively well localized.   This allows the light to be correlated with specific tracks in the detector.  This ``flash-track matching'' is used to identify and reconstruct the tracks that are in time with the beam spill--an important goal of the light collection system.

\subsection{Light Production in Argon \label{sint}}

Light produced in liquid argon arises from two processes: scintillation and Cherenkov radiation.  
Scintillation light is produced by the formation and eventual radiative decay of excited argon dimers (or eximers) and is emitted in an isotropic distribution.  Liquid argon is an excellent scintillator: it produces a large amount of light per unit energy deposited (about 24,000 photons per MeV at 500 V/cm drift field) and is transparent to its own scintillation. 
The scintillation light has a prompt and slow component with decay times of about 6 ns and 1.6 $\mu$s, respectively. The two lifetimes correspond to the two lowest-lying eximer states with the prompt component coming from the decay of a singlet state and the slow from the decay of a triplet state.  The prompt to slow ratio is about 1:3 for minimum ionizing particles and varies with ionization density and particle type. Both components consist of photons with a wavelength of 128 nm.

The effective lifetime of the triplet component may be modified by quenching (non-radiative dissociation of excimers by impurities)  \cite{Acciarri:2008kv}.  Other factors that can affect the arrival of the light include Rayleigh scattering, absorption by impurities, and obstructions.  For detailed discussion of the physics of scintillation light production and propagation in MicroBooNE, see~\cite{Jones:2015bya}. Table~\ref{tab:lightparam} summarizes information about the scintillation light.  

\begin{table}
   \centering
    \caption{Important properties of scintillation light in liquid argon that affect detection in MicroBooNE.} 
    \begin{tabular}{llr} 
    \hline
    Property & Value & Reference\\
    \hline
    Wavelength & 128 nm & \cite{SUZUKI1979197}\\
    Singlet, Triplet state time constants & 6 ns, 1.6 $\mu$s &\cite{PhysRevB.27.5279} \\
    Photons/MeV for $E=500$ V/cm & 24000 & \cite{PhysRevB.17.2762}\\
     Triplet lifetime quench due to N$_2$ at 1 ppm  & 20$\%$ & \cite{Acciarri:2008kv} \\
     Attenuation length due to N$_2$ at 1 ppm & 66 m &\cite{Jones:2013bca} \\
     Rayleigh Scattering Length & 66 cm &\cite{1997NIMPA.384..380I} \\
     Transparency of the wireplanes at normal incidence & 86$\%$ & -\\
     Photocathode coverage of the 32, 8-inch PMT array & 0.9\% & -\\
    \hline
   \end{tabular}
   \label{tab:lightparam}
\end{table}

\begin{figure}
	\centering 
\includegraphics[width=\textwidth]{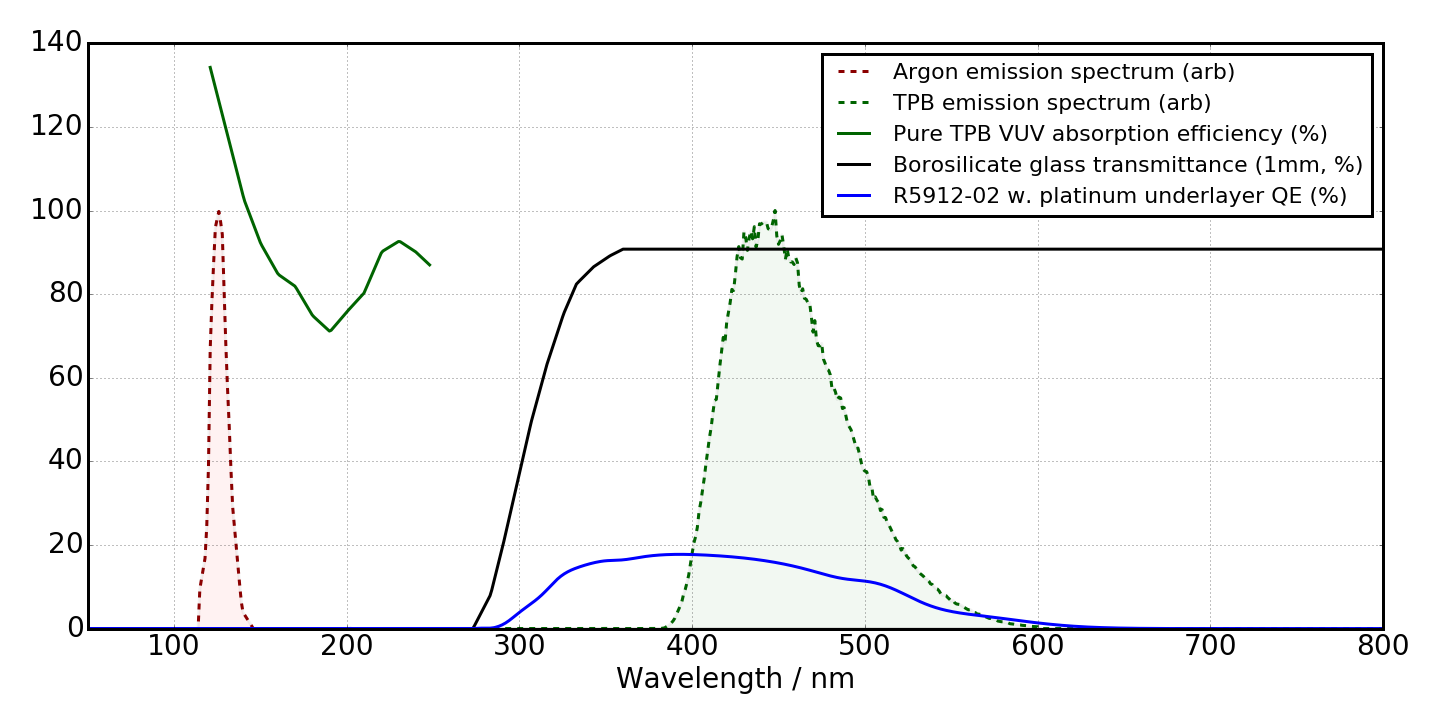} 
        \caption{Scintillation light emission spectrum (red) and TPB re-emission spectrum (green), in arbitary units.  Superimposed are relevant efficiencies (plotted in $\%$ on the $y$ axis):  Dark green line -- absorption of VUV light by TPB;  Black line -- transmission of borosilicate glass;  Blue line -- efficiency of a R5912-02mod cryogenic PMT \cite{Jones:2015bya}.}\label{fig:lightchallenge}
\end{figure}

Because scintillation photons have a wavelength of 128 nm, they are are very difficult to detect using conventional photodetectors.  Figure~\ref{fig:lightchallenge} summarizes the challenges involved in detection of the 128 nm scintillation light.  
In order to detect the scintillation light (red distribution shown in figure~\ref{fig:lightchallenge}), the VUV photons must be shifted into the visible region.   MicroBooNE employs tetraphenyl-butadiene (TPB).
This organic fluor absorbs in the UV (green line) and emits in the visible with a peak at $425\pm20$~nm (green hatched region), the peak wavelength having a slight dependence on the micro-environment of the fluors.  This is a favorable wavelength for detection by the PMTs employed by MicroBooNE.  The efficiency for transmission through borosilicate PMT glass (black) and the quantum efficiency of the cryogenic tubes used in MicroBooNE (blue) are overlaid on the TPB spectrum.

\subsection{The Primary Light Collection System}

\begin{figure}
	\centering
           \includegraphics[width=0.45\textwidth]{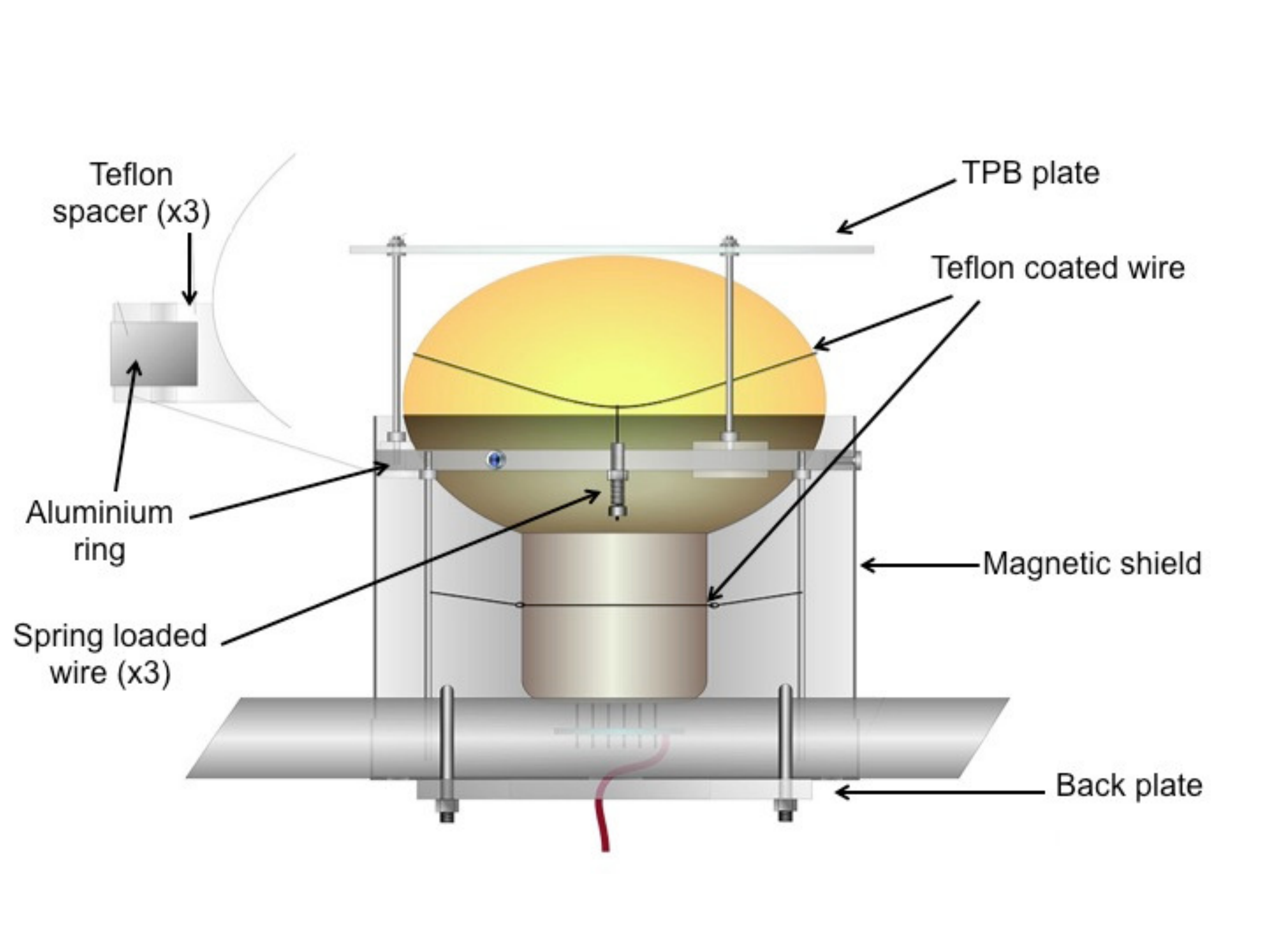} 
           \includegraphics[width=0.45\textwidth]{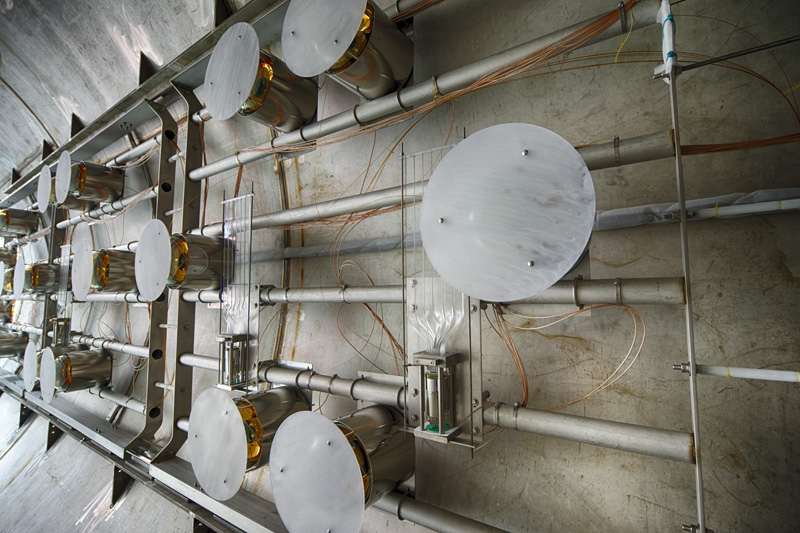} 
        \caption{Left: diagram of the optical unit; Right: units mounted in MicroBooNE, immediately prior to \lartpc installation.}\label{fig:unitmounted}
 
\end{figure}

Each of the 32 optical units of the primary light collection system consist of a cryogenic Hamamatsu 5912-02MOD PMT seated behind an acrylic plate coated with a TPB-rich layer and surrounded by a mu-metal shield. 
Figure~\ref{fig:unitmounted} shows a diagram of one unit (left) and a photograph of the installed units (right).  Past experiments have directly coated PMTs with wavelength shifter \cite{Amerio:2004-T600}.  However, 
the MicroBooNE design separates the PMT from the wavelength-shifting plate for simplicity of quality control and installation.  This proved important, as R\&D indicated that TPB is particularly vulnerable to environmental degradation (see section~\ref{environ}).  In this section, a description is provided for each component of the optical unit, as well as for the overall assembly.

\subsubsection{Photomultiplier Tubes, Bases, and Initial Tests}
\label{sec:pmt-bases}

Reference \cite{Briese:2013wua} provides detailed information on the selection 
and testing of the 200 mm (8.0 in) diameter Hamamatsu R5912-02mod cryogenic PMTs employed in MicroBooNE.   
In this section a brief summary of the findings from this testing is presented.

\begin{figure}
\centering 
\includegraphics[width=0.65\textwidth]{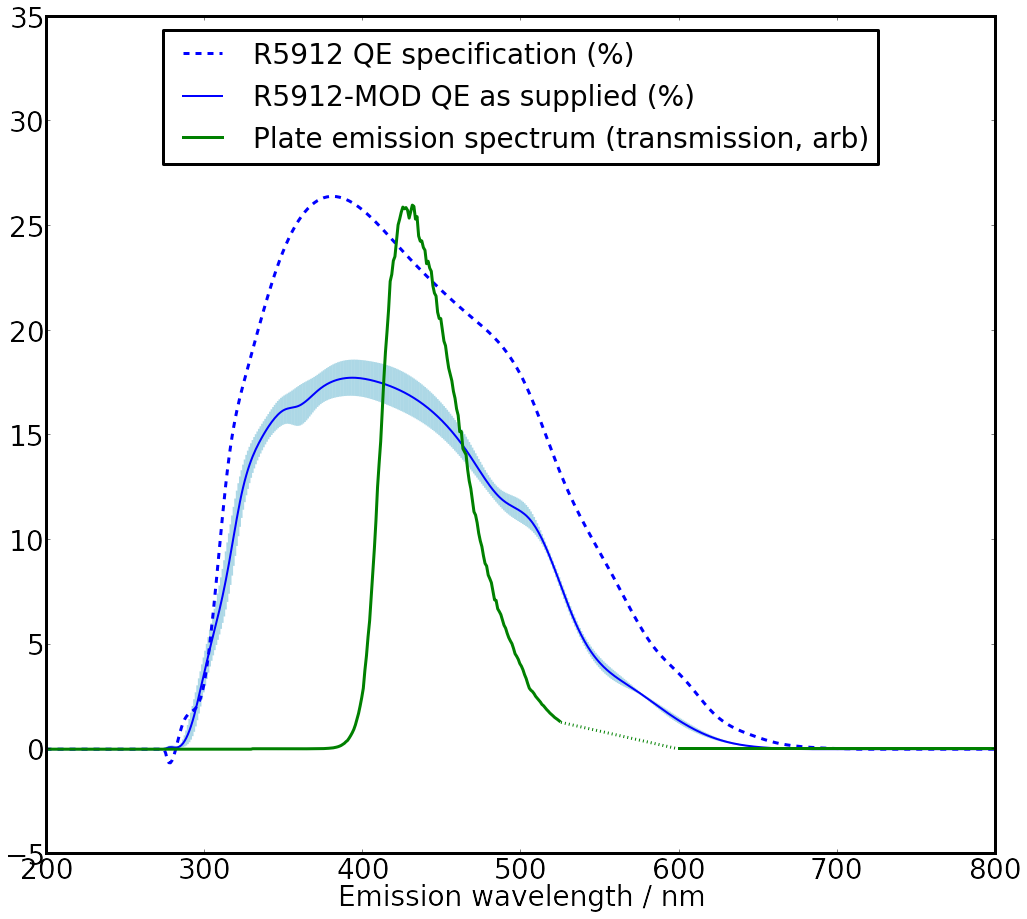}
\caption{The specification for the non-platinum undercoated PMT from \cite{Hamamatsu-Datasheet8inch}.  The blue band shows the mean and standard deviation of the four quantum efficiency curves provided by Hamamatsu for the installed PMTs. Also shown is the measured emission spectra of the MicroBooNE wavelength-shifting coatings, discussed in section \ref{sec:wavelengthshift}.  \cite{Jones:2015bya}
 \label{fig:TPBSpectraPlate}  }
\end{figure}

The R5912-02mod employs a bi-alkali photocathode.  Because the PMT is designed for cryogenic use, the R5912-02mod also features a thin platinum layer between the photocathode and the borosilicate glass envelope to preserve the conductance at low temperatures.   While this allows the PMT to function below 150 K, absorption in the platinum reduces the efficiency of the PMT by 20\%.
Figure \ref{fig:TPBSpectraPlate} provides quantum efficiency curves for these PMTs. The manufacturer's specifications do not include the effects of the platinum photocathode coating, but a wavelength dependent quantum efficiency was provided by Hamamatsu for 4 of the 32 installed PMTs.  The mean and standard deviation of these curves are shown in figure \ref{fig:TPBSpectraPlate}. 

The  R5912-02mod is a 14-stage PMT.
The high gain at room temperature ($10^9$ at $\sim$1700 V), compensates for known reduced gain at 87 K in the liquid argon.  
The high gain also has the additional advantage of allowing operation at lower than nominal voltage which reduces heat-loss in the PMT bases and the potential for high voltage breakdown at the feedthroughs.

The PMT base is designed such that the photocathode is 
grounded and the anode is held at large, positive voltage. 
Thus the PMT bulb, which is closest to the \lartpc anode plane, 
is at ground and does not disturb the electric field on the wires.  
The result is that the high voltage (HV) can be provided and the signal can be extracted from the PMT using a single cable, 
reducing the cable volume in the vapor region and removing a possible source of out-gassing water impurity.

The flat PC-board base was made of Rogers RO4000-series woven glass-reinforced laminate, 
which is the same material as the \lartpc cryogenic front end boards.  Use of Rogers 4000 series as the base material avoids the potential risk of contamination of the liquid argon. The Rogers material has a similar temperature expansion coefficient as the surface mounted components which enhances the reliability of the electronics assembly when operated at cryogenic temperatures.  The base is attached $\sim$1 cm from the bottom of the PMT and is the closest possible distance of safe approach to the PMT vacuum seal tip.   
A schematic and photos of the PMT base are provided in reference~\cite{Briese:2013wua}. 
The passive components include only metal film resistors and C0G/NP0 capacitors, 
which have the minimum temperature coefficients, and the performance of all components at cryogenic temperatures was tested.  
Because the supplied HV and return signal share a single cable, the signal must be split from the HV through an AC-coupling capacitor, as is discussed in section~\ref{LCUnitimplement}.

All installed PMTs were tested in a PMT test stand, both
at room temperature and in liquid nitrogen at 77 K.  
The details are described in reference~\cite{Briese:2013wua}.  
In brief, the test stand consisted of a light-tight 346~liter, 
liquid nitrogen filled dewar into which up to four PMTs could be installed. 
The PMTs were immersed and maintained in the dark environment 
for up to three days before most measurements of the dark rate and gain were performed.  
A fiber brought in light from a pulsed blue LED, which was tested for linearity with bias voltage.  


Among the important results from cryogenic testing  were the following \cite{Briese:2013wua}:
\begin{itemize}
\item After the required period in the dark and cold, the PMTs could be ramped to voltage quickly in the cold environment, 
and after 30 minutes, the gains were found to be stable.
\item If the room temperature PMTs were immersed in liquid nitrogen, 
dramatic changes in the gain were observed after initial turn-on. 
The PMT gain remained high in the first $\sim$5 hours after immersion, 
and then suddenly dropped by more than a factor of two, 
afterwards reaching a stable value with a small drift.
\item The PMT response showed good linearity up to 100 photo-electrons (PE), 
which was the maximum attainable by the PMT test stand LED.  
\item The HV for each PMT was selected to produce a gain of $3\times 10^7$ in liquid nitrogen,
and was typically chosen to be $\sim$1300 V.
\item The dark current plateaus extended up to 1800 V in liquid nitrogen, and the dark current is higher in liquid nitrogen than at room temperature.
\item The PMT performance depended on rate of the pulsed LED (see later in this section).  
\end{itemize}
No PMTs were rejected on the basis of the testing.  
However, there were three unexpected results to note here.


The first unexpected behavior was that, at room temperature, 
most of the PMTs showed gains that were 10 to 30\% higher than manufacturer's specifications.
As expected at cryogenic temperatures, the gain is reduced by $\sim$10\% to 50\%.
To measure the PMT gain, the LED was set to produce one to two PEs.  
The gain was found from the separation of the single PE peak from the pedestal, where the single PE response was fit using the procedure described in~\cite{Bellamy:1994bv}.   

Second, it was found that the PMTs are noisier in the cryogenic environment. 
It would typically be expected that thermal emission is suppressed at cryogenic temperatures and one would expect a lower dark current 
for PMTs operating in this regime.  
However, the dark current measured in the liquid nitrogen was higher than at room temperature. 
Although the cause is unknown, this phenomenon has previously been observed~\cite{Meyer:2008qb}. A proposed explanation is provided in~\cite{MeyerDark}.

Third, an LED pulse-rate-dependent gain shift was found during testing, as shown in figure~9 of reference~\cite{Briese:2013wua}.  This behavior is described qualitatively in \cite{HamamatsuBook}.  With 10 kHz LED pulsing, the gains of cold and dark-adapted PMTs were shown to steadily increase, requiring nearly 24 hours from turn-on to stabilize. The effect was not observable at 10 Hz.   This is relevant to MicroBooNE because the
cosmic muon rate in the MicroBooNE detector is $\sim 5$ kHz.  Therefore a similar effect is expected in the MicroBooNE detector.  Preliminary results from measurements of the PMTs installed in the LAr-filled MicroBooNE cryostat shows the expected effect. 


\subsubsection{Wavelength-Shifting Plates}
\label{sec:wavelengthshift}

In order to be sensitive to 128~nm VUV liquid argon scintillation photons, the optical assemblies use a wavelength-shifting coating to convert this VUV light to visible wavelengths that are detectable by PMTs.  In MicroBooNE, the active ingredient of this coating is TPB, an organic fluor which absorbs efficiently in the vacuum ultraviolet with an emission spectrum peaked around 425~nm \cite{Burton:1973} as shown in figure~\ref{fig:lightchallenge}. 

\begin{figure}
\centering 
\includegraphics[width=\textwidth]{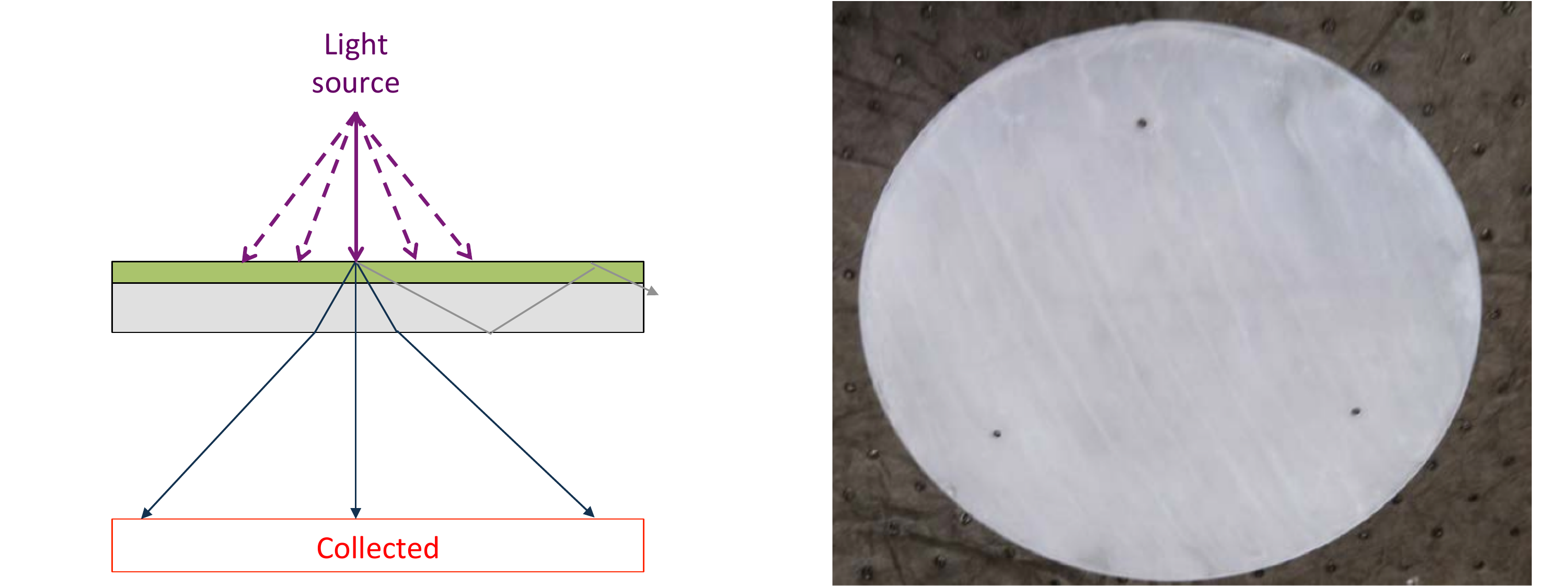}
\caption{Left : Illustration of transmission mode, used by the optical units. Right: photograph of a coated plate. \label{fig:PlateCoating}}
\end{figure}

The optical unit PMTs observe light transmitted through TPB-coated, 305 mm diameter acrylic plates (see figure~\ref{fig:PlateCoating}).
The coating consists of a 1:1 TPB-to-polystyrene ratio, with 1 g of each dissolved in 50 ml of toluene.  A small amount of ethyl alcohol is added as a surfactant.
The coating is applied to the acrylic plate in three layers by brush-coating.  The solution dries in air at room temperature.
This leads to a final layer which is oversaturated with TPB, and white crystals form on the surface as the coating dries.  The presence of surface crystallization gives the MicroBooNE plates a white, opaque finish. Details of the process are described in reference~\cite{Ignarra:2014yqa}.

The emission spectrum in figure \ref{fig:TPBSpectraPlate} was measured using a Hitachi F-4500 fluorescence spectrophotometer with an incident beam of wavelength 270 nm, selected by a diffraction grating from a xenon lamp.  A standard rhodamine dye calibration sample was used to correct for drift in the lamp and spectrometer.  The measurement was made in transmission mode, and an artifact peak at a harmonic of twice the incident wavelength was observed between 525 and 600 nm.  This region is omitted from the reported spectrum of figure \ref{fig:TPBSpectraPlate}.  

As shown in figure \ref{fig:TPBSpectraPlate}, although the absolute quantum efficiency for the platinum-undercoated cryogenic PMT is lower than the non-cryogenic version, the wavelength dependence is similar, and overlap between the TPB emission spectrum and the sensitive wavelength range of the PMT remains high.  Using the measured TPB emission spectrum for plate coatings and the PMT quantum efficiency curve provided, the spectrum-averaged PMT quantum efficiency is 15.3\% $\pm$ 0.8\% per visible photon incident on the photocathode. 

\begin{figure}
\centering 
\includegraphics[width=0.6\textwidth]{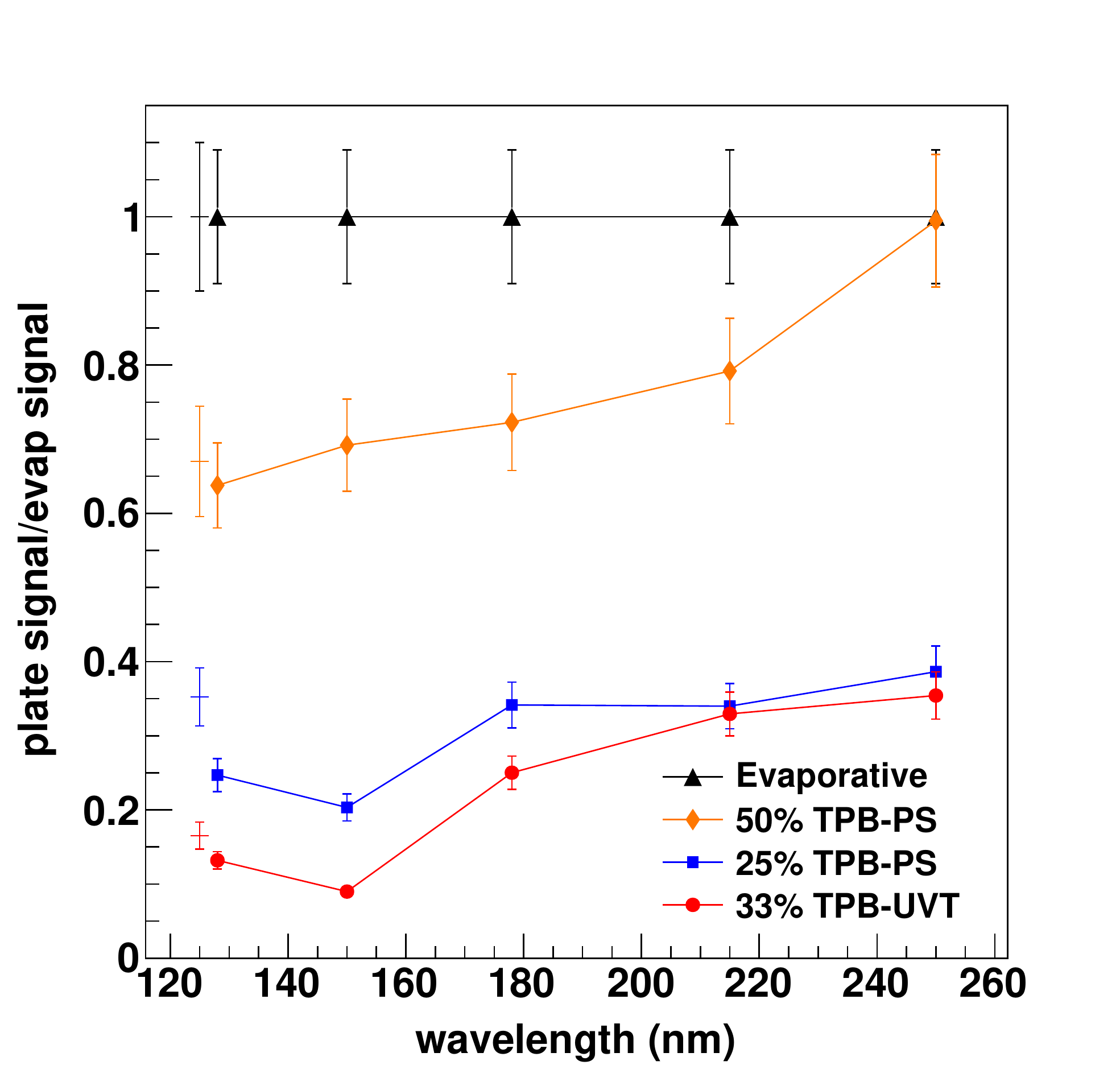}
\caption{Measured efficiencies of various wavelength-shifting coatings, from \cite{Ignarra:2014yqa}. In this plot, 50\% TPB-PS is the MicroBooNE plate coating used in the optical units.  The 33\% TPB-UVT is the light guide coating for the secondary light collection system described below.  Connected points were measured in a vacuum monochromator at room temperature, and non-connected points were measured in liquid argon with 128 nm scintillation light. All points are normalized to the performance of an evaporatively coated plate. \label{fig:CoatingEfficiency}  }
\end{figure}

The wavelength-shifting performance of the coating was measured at 128~nm relative to evaporative coatings of the type studied in \cite{Gehman:2011}.  Coating efficiencies were measured as a function of wavelength between 128 and 250~nm using a vacuum monochromator at room temperature.  They were also measured in liquid argon using 128 nm scintillation light, relative to the same evaporatively coated plate.  These data are shown in figure~\ref{fig:CoatingEfficiency}, and more information about these measurements can be found in \cite{Ignarra:2014yqa}.  
  
The absolute efficiency of the MicroBooNE coatings can be obtained by multiplying the relative efficiencies of figure \ref{fig:CoatingEfficiency} by the measured absolute efficiencies from \cite{Gehman:2011}, and accounting for the temperature dependence in the wavelength-shifting efficiency of pure TPB reported in \cite{Francini:2013-jinst}.  The expected number of visible photons per incident 128~nm photon is $0.98 \pm 0.17$ for the MicroBooNE plate coating.

\subsubsection{UV Light Protection for the Wavelength-Shifting Plates \label{environ}}

TPB coatings have been shown to degrade under exposure to ultraviolet light \cite{Chiu:2012} through a radical-mediated photo-oxidation to the UV-blocker and photo-initiator benzophenone \cite{Jones:2013}.  Several measures were taken to ensure that degradation was minimized during the construction of the experiment.  The TPB powder and coated elements were stored in the dark at all times, with coated plates and light guides kept wrapped in foil and stored in a dark container before installation.  The detector construction area was covered with a UV blocking plastic \cite{LexanThermoclear}, and test plates were placed at various positions in the clean tent to check for degradation from stray light.  After several weeks of exposure, one test plate with a clear line of sight to the tent entrance demonstrated a few percent degradation, and all others showed no observable loss of efficiency. The open end of the MicroBooNE cryostat was shielded from light by a black curtain after installation, and the feedthroughs of the cryostat were blocked when not in use to prevent stray light from entering.  The coated plates were the final component of the optical system to be installed into the detector to give the minimum possible light exposure during the detector construction process.

\subsubsection{Cryogenic Mu Metal Shields}


The trajectories of electrons within the PMT, particularly those between the photocathode and the first dynode, can be deflected by the Earth's magnetic field thereby reducing the PMT response.  This effect can be reduced or removed by surrounding the PMT with mu metal, a metal of high magnetic permeability.  Commonly used mu-metal fails to provide shielding at cryogenic temperatures. Two types of cryogenic mu metal, Cryoperm 10 and A4K, both products of Amuneal, were identified that did provide shielding at cryogenic temperature. 

\begin{figure}
\centering 
\includegraphics[width=0.8\textwidth]{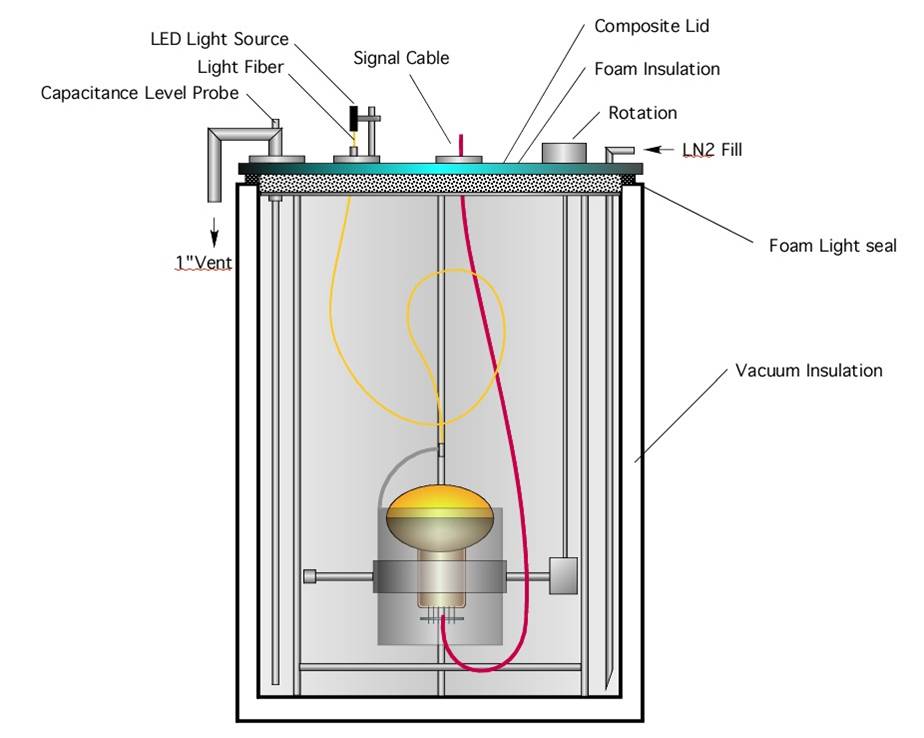}
\caption{Schematic of the system used to study the mu-metal shields.  The design allows rotation along all three axes. \label{fig:shieldcartoon}  }
\end{figure}

The mu metal shields were tested in the apparatus shown in figure~\ref{fig:shieldcartoon}.  The system allowed for the PMT to be positioned at an angle relative to the vertical axis, with the rotator set to 30 positions from 0$^{\circ}$ to 348$^\circ$.    The set-up was on a dolly that allowed for rotation about the vertical axis.  PMT tests were performed in air and in liquid nitrogen.   PMTs were dark adapted for 5 hours before testing.  
A blue LED provided 1 to 2 single PEs of light through an optical fiber.   The fiber was fixed to the PMT mount such that the endpoint was fixed with respect to the PMT as the system was rotated.   

\begin{figure}
\centering 
\includegraphics[width=0.6\textwidth]{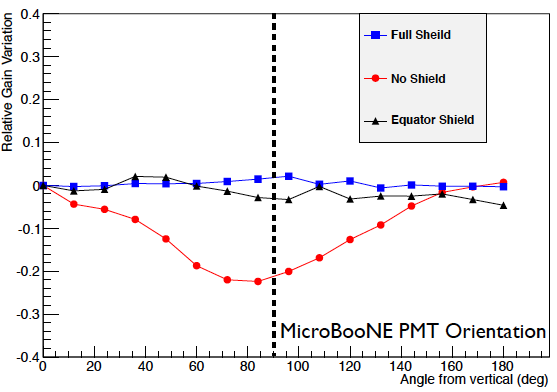}
\caption{Red:  Angular dependence of a PMT response with no shield;  Black:  for a shield that reaches the tube equator;  Blue: for a shield that fully covers the tube to the zenith. \label{fig:shield}  }
\end{figure}

The mean charge from the PMT as a function of angle was recorded.  The error was primarily systematic.   The 10$^6$ LED pulses per data point give $<1\%$ statistical error on the mean.  However, 24-hour studies of a single point showed $\sim$5\% variations in collected charge.  Results showed that A4K and Cryoperm function essentially identically, to within the measurement error, and so MicroBooNE chose A4K, based on significant cost savings.    A plot of the relative PMT gain variation versus angle from vertical, figure~\ref{fig:shield}, shows that adding the shield significantly improves PMT performance.  This plot shows the effect of the A4K shield with the top aligned with the equator of of the PMT (black) and aligned with the zenith of the bulb (blue), compared to no shield (red) as a function of PMT azimuthal angle.  Given the 5\% systematic error, the two shield positions are indistinguishable.  

As a result of these tests, the A4K shields were designed to extend just past the equator of the PMT.  The shield has small holes in the backplate that allow the PMT cables to exit the shield.   MicroBooNE is the first \lartpc to use cryogenic mu metal shields in its light collection system.

\subsubsection{Implementation of the Primary System \label{LCUnitimplement}}

\begin{figure}
\centering	
\includegraphics[width=0.65\textwidth]{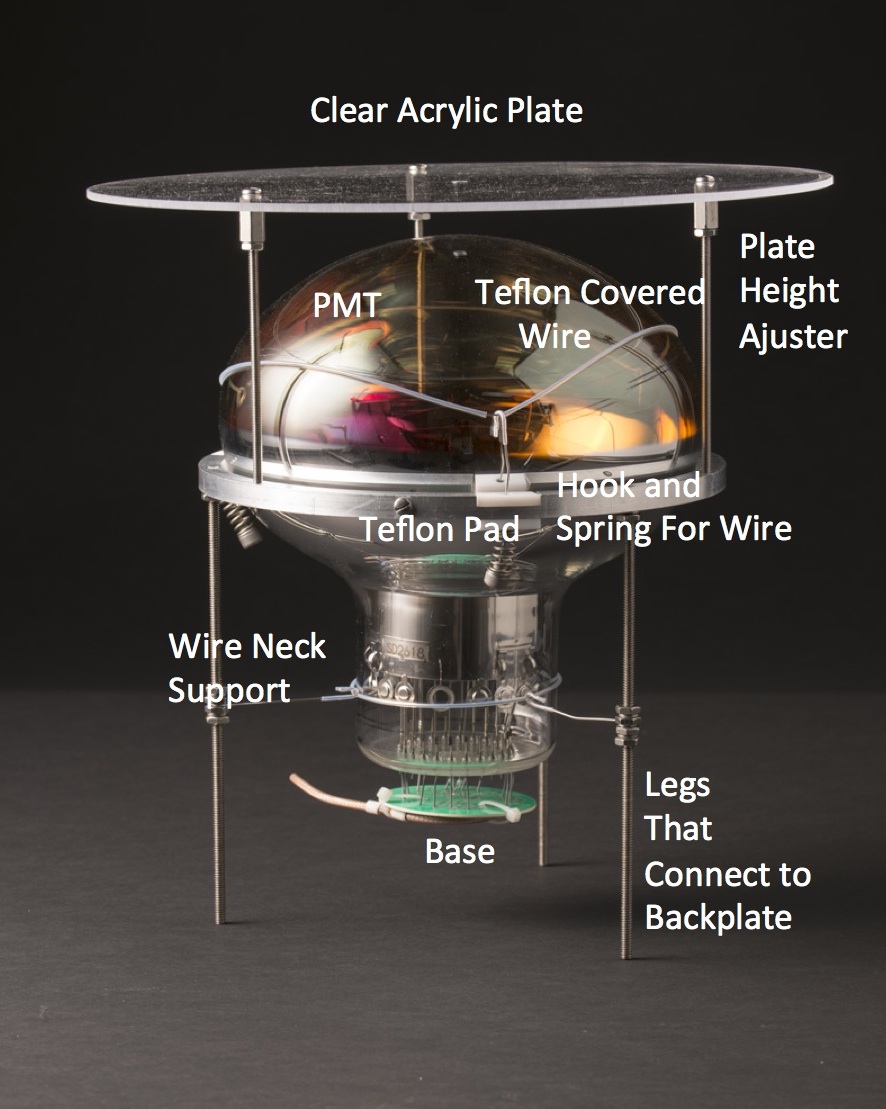}
\includegraphics[width=0.65\textwidth]{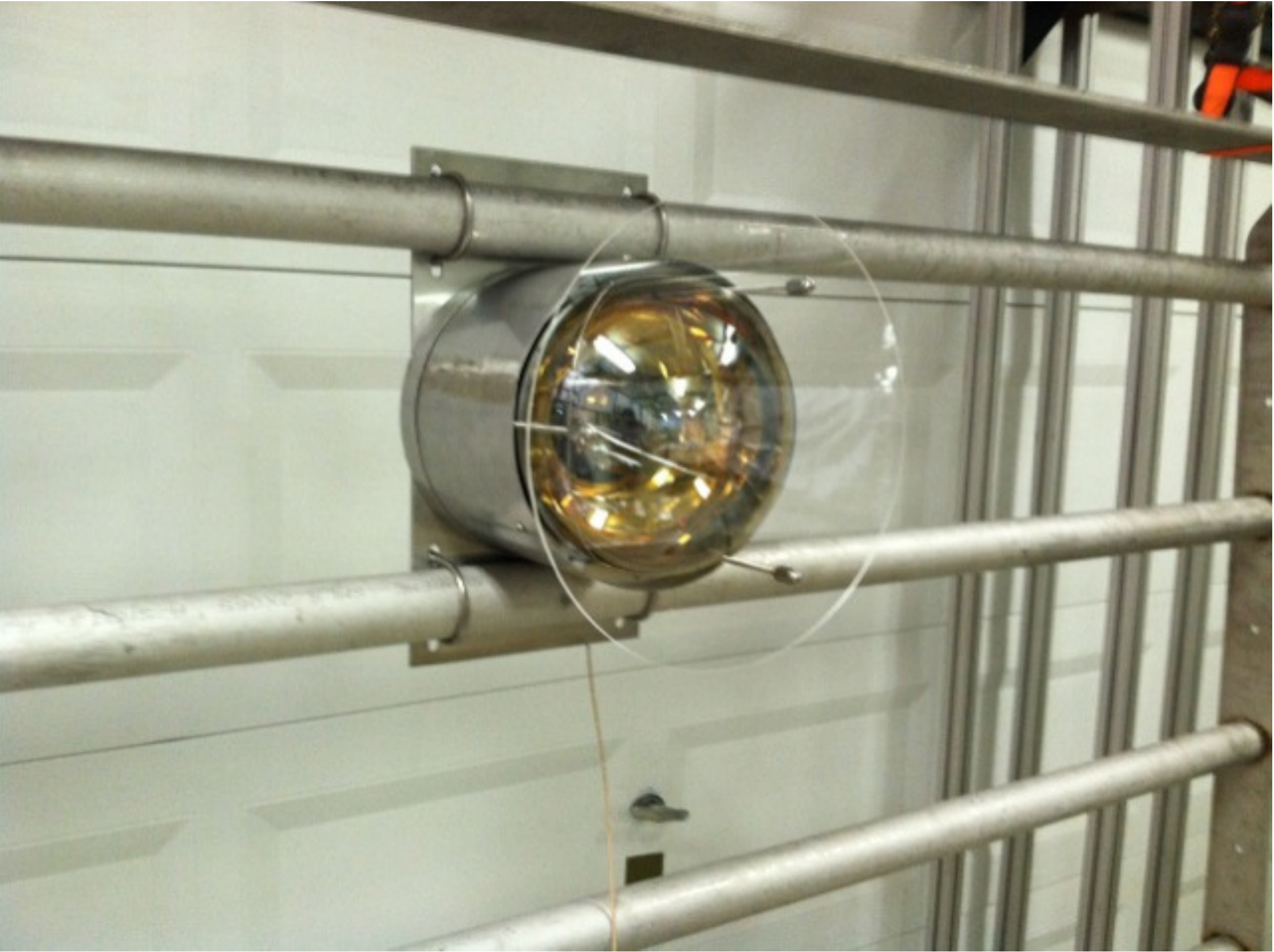}
\caption{Top: The optical unit mount internal to the shield, with components labeled; Bottom: Unit mounted on rails.  The clear plates were replaced with TPB-coated plates immediately before \lartpc installation, as discussed in the text~\cite{Briese:2013wua,Katori:2013wqa}.}
\label{fig:unitmodel}
\end{figure}

The light collection system is composed of optical units assembled as shown in the top picture of figure~\ref{fig:unitmodel}.  The PMT is seated within the mu metal shield on three teflon pads attached to an equatorial support ring.  The neck of the tube slides inside a loose wire guide-loop that prevents the PMT from tipping.  The PMT is held within this assembly using teflon-encased wires that extend across the bulb and connect to wire hooks attached to the equatorial ring with stainless steel springs. Legs extending from the support at the equator are screwed into a backplate for final mounting.  Concern about differences in contraction of the materials led to this design which holds the PMT in place, but with only moderate rigidity.  The units were tested to ensure the PMTs would not be displaced during installation and filling.  Three posts extend upward from the equatorial ring to hold the plate $\sim$3 mm above the apex of the bulb.   The optical units slide into a cylindrical mu-metal shield, which screws into the equatorial ring.  The unit is then mounted on stainless steel back-plates affixed to a support rack, as shown in the bottom picture of figure~\ref{fig:unitmodel}.

The support rack consists of five stainless steel components, or modules, for ease of installation.  Each module has vertical height 1.83 m and horizontal length 2.07 m, resulting in a total horizontal length of 10.36 m.  Unlubricated Thomson bearings fitted to the lower edge allow each module to slide into the cryostat on rails mounted in the vessel.   The system was designed to allow the light detection system to slide into the vessel after the \lartpc was installed.  However, in the end, scheduling permitted installation of the system before the \lartpc installation.  This had the advantage of making installation and surveying easier, but the drawback that the system would be exposed to UV light for a longer period.  Therefore, the units were installed with dummy clear acrylic plates, and the TPB coated plates were installed only just before the \lartpc was moved in and the detector could be easily protected from light.  During optical unit installation, each rack module was supported by a temporary mounting rail.  The optical units were then mounted in positions chosen to avoid obstruction by the \lartpc cross-bars, as shown in figure~\ref{fig:lightlayout}.  As the units were mounted and slid into the cryostat, the cables were loosely tied to the bars of the rack for support and constraint.  

\begin{figure}
\centering 
\includegraphics[width=0.9\textwidth]{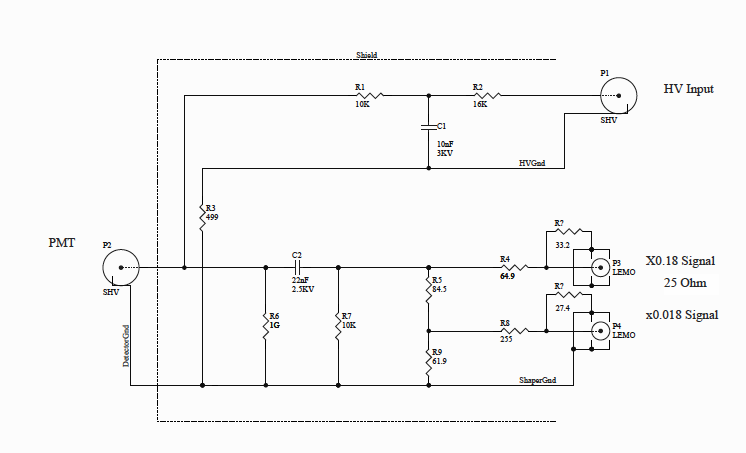}
\caption{The ``splitter'' circuit.  The circuit connects the HV source to the PMT. It also provides a pathway for signal pulses from the PMT to reach the readout electronics via an AC-coupling capaciter (C2).  The signal is split into two copies, one provided with an attenuation factor of 0.18 and another at 0.018. Both signal sources are recorded by the readout electronics in order to provide two dynamic ranges.}
\label{fig:splitter}
\end{figure}

The ``splitter'' circuit, located outside of the cryostat, is shown in figure~\ref{fig:splitter}. The splitter separates the HV of the PMT from its output signal which is subsequently split into a high-gain (HG) and a low-gain (LG) channel. The HG and LG channels respectively carry 18$\%$ and 1.8$\%$ of the output signal.  This allows a wide dynamic range for ADC readout of the PMT pulses.  The capacitance was chosen to minimize reflections, since the bases are not back-terminated.  The HV is supplied to the splitter using BiRa Corporation, Model 4877PS modules.

The PMT cable system delivers HV and returns signals between the external splitter and the optical unit in the cryostat.   A single cable runs from an external connector, through a feedthrough that is filled with epoxy, into the cryostat and to the PMT base. The RG316/U coaxial cable has 50 $\Omega$ impedance.  Cables were terminated with Pasternack PE4498 SHV to accommodate that the cable carries HV to the tube as well as signals from the tube.
The cable carries an AC voltage rating of 1100 V; however tests showed the DC rating to be at least three times higher and so suitable for this use.  The cables were routed through feedthroughs consisting of a pipe filled with solidified epoxy mounted on a conflat disk.  On the warm side of the feedthrough, the cables were terminated at a patch panel with SHV connectors. The SHV connector impedance has a negligible effect on the 20-30 ns PMT signals.  SHV cables connect the patch panels to the splitters.  The impedance of every channel was tested at the feedthrough patch panel for a stable and correct value for the base resistance, which was 4.04$\pm$0.02 M$\Omega$. 

\subsection{PMT Testing and Quality Assurance}

A vertical slice test (VST) of the MicroBooNE optical system was performed in the Bo Cryostat at Fermilab.  The Bo cryostat was a 250-liter vacuum-insulated vessel with an inner diameter 56 cm and a depth of 102 cm used for R\&D studies, and, relevant to this paper, a vertical slice test of the MicroBooNE optical units.  The system is described in detail in reference~\cite{Jones:2015bya}.  The cryostat can be filled with purified LAr with oxygen and water levels below 1 ppb and a typical nitrogen contamination below 1ppm.  The light collection system was tested with light from visible (420 nm) and UV (250 nm) LEDs piped in via fiber, as well as scintillation light from $^{210}$Po alpha sources and cosmic rays.

The slice consisted of two PMTs with base electronics, mu-metal shield, TPB plates,  cable feed throughs, splitters, the HV power supply and the interlock system.   Tests were performed without and with the mu-metal shield.   The test made use of the data acquisition components described in section~\ref{sec:electronics}, including the shaper, FEM, trigger card, control card, and server.  The MicroBooNE trace impurity monitors were also used.

\begin{figure}
\centering 
\includegraphics[width=0.9\textwidth]{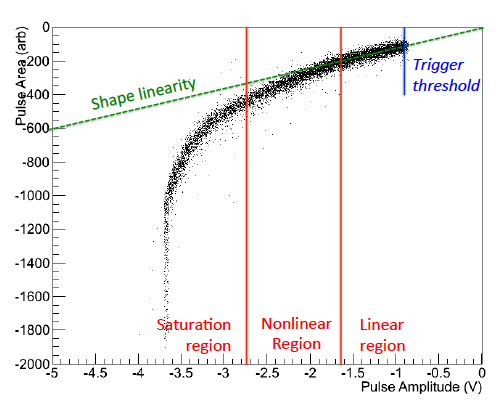}
\caption{\footnotesize As shown by the VST, linearity of cosmic-ray induced PMT pulses is maintained up to amplitudes of around -1.7 V (300 PE), and amplitude saturation occurs at -3.7 V (670 PE) \cite{Jones:2015bya}
 \label{fig:nonlinear}  }
\end{figure}

The VST informed the final design of many components, as well as producing results relevant to understanding the running conditions and performance expectations.   For example, during studies of the response of the slice to 128 nm scintillation light from the alpha source, 
valuable information was gathered on the single photo-electron dark rate and cosmic ray rate that could be applied to the MicroBooNE detector expectations.    As a second example, these runs allowed characterization of the pulse shape nonlinearities of the optical units, as seen in figure~\ref{fig:nonlinear}.  These were shown to be significant at $\sim$300 PE in pulse amplitude.  Full amplitude saturation occurred at $\sim$670 PE. Thus, it was concluded that for pulses of more than 300 PE, pulse shape cannot be described by a linear superposition of single photo-electron pulses. 

\subsection{Secondary System: Acrylic Light Guides for R\&D}

A secondary light collection system consisting of four lightguide paddles was also installed.   Their design has several advantages for future large detectors such as DUNE. First, the collection area per channel is larger than the optical units, providing more coverage for the same number of electronics channels, cables, and feedthroughs.  Second, the detectors have a narrow profile so they can be slid between chambers in a multi-\lartpc detector, minimizing space requirements of the light collection system.  In the case of MicroBooNE, the design gradually guides light in bent acrylic bars to a PMT.  This design was an early alternative to a perfectly flat design that guides the light to SiPMs.  Running this system will provide long-term information on performance of lightguide based systems.  It also enhances the MicroBooNE dynamic range, since the lightguide detectors saturate at a much higher light level than the optical units.

\begin{figure}
\centering 
\includegraphics[width=\textwidth]{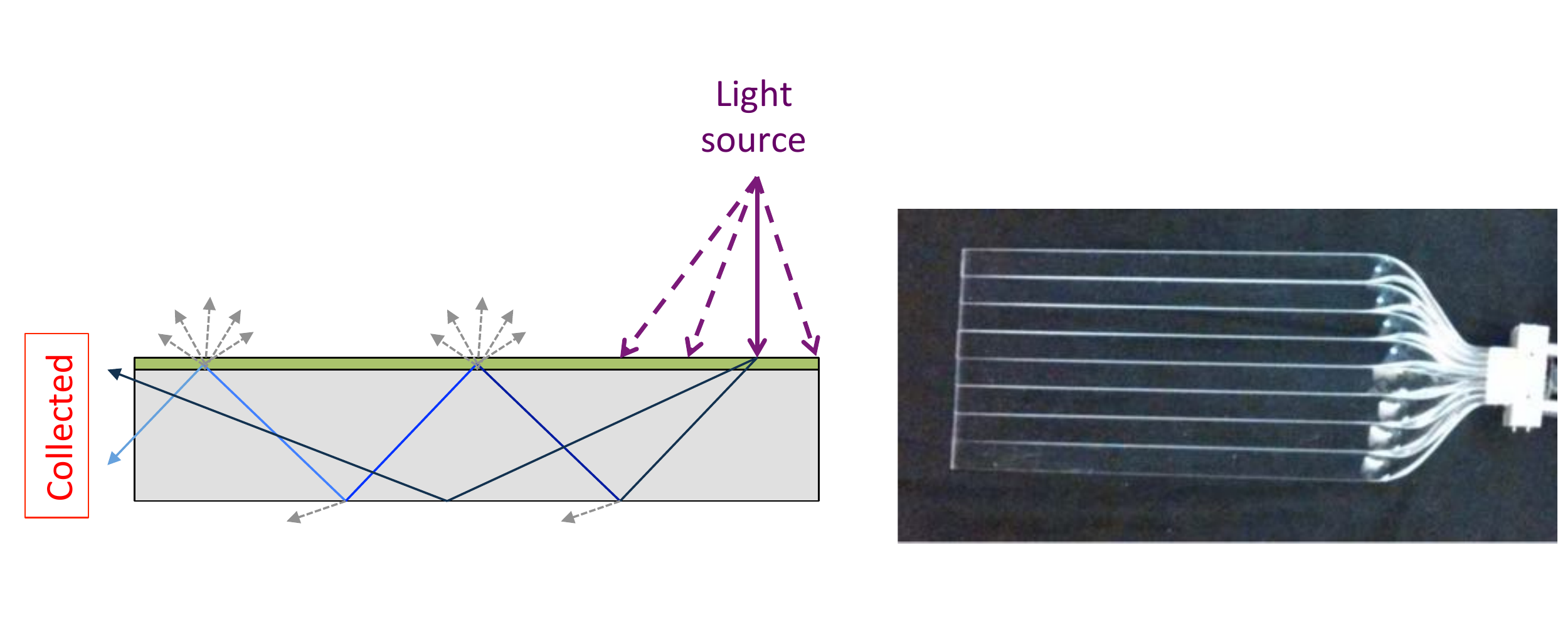}
\caption{Left : Illustration of guiding mode, used by the paddles. Right: photograph of a coated paddle. \label{fig:LGCoating}}
\end{figure}

In the case of the lightguides, the 128 nm light is absorbed and shifted by a clear wavelength-shifting coating, and the re-emitted light is guided to a 5.08 cm (2.0 in) Hamamatsu R7725-MOD PMT, as illustrated in figure~\ref{fig:LGCoating}, left. The installed paddles consist of six bars.  A photograph of one coated paddle with eight bars is shown in figure~\ref{fig:LGCoating}, right.  The active length of each bar is 50.8 cm.  This system was added for R\&D purposes and made use of 4 of the 8 spare channels available of HV, cables, feedthroughs, and electronics.  As shown in figure~\ref{fig:lightlayout}, each paddle is installed next to an optical unit for direct comparison of performance.

The coating requirements for plate assemblies and light guides are different, and so the composition and coating methods for each were separately optimized.    In the case of the light guide coatings, the figure of merit is the light emitted in guided mode.  Guided mode light is the light that is detected at one of the ends of a test sample, which is orthogonal to the illuminated face of the sample.  In addition to the wavelength-shifting efficiency of the active layer, the detected light yield is affected by the reabsorption and scattering losses in the coating as visible light propagates along the bar. The light guide assemblies have a TPB coating of 33\% TPB to 67\% UVT acrylic by mass, also with ethanol surfactant. The coating is applied as a single layer and the TPB remains suspended in the acrylic matrix as the coating dries, leading to a smooth, visibly transparent surface.  The performance and attenuation behavior of similar light guides to those installed in MicroBooNE were studied experimentally in \cite{Baptista:2012}.  The reported non-exponential attenuation suggests that surface losses dominate over bulk losses as the attenuation mechanism, and that the fractional loss per reflection within the light guide is of order 2-3\% \cite{Jones:2013}. 

\begin{figure}
\centering 
\includegraphics[width=0.65\textwidth]{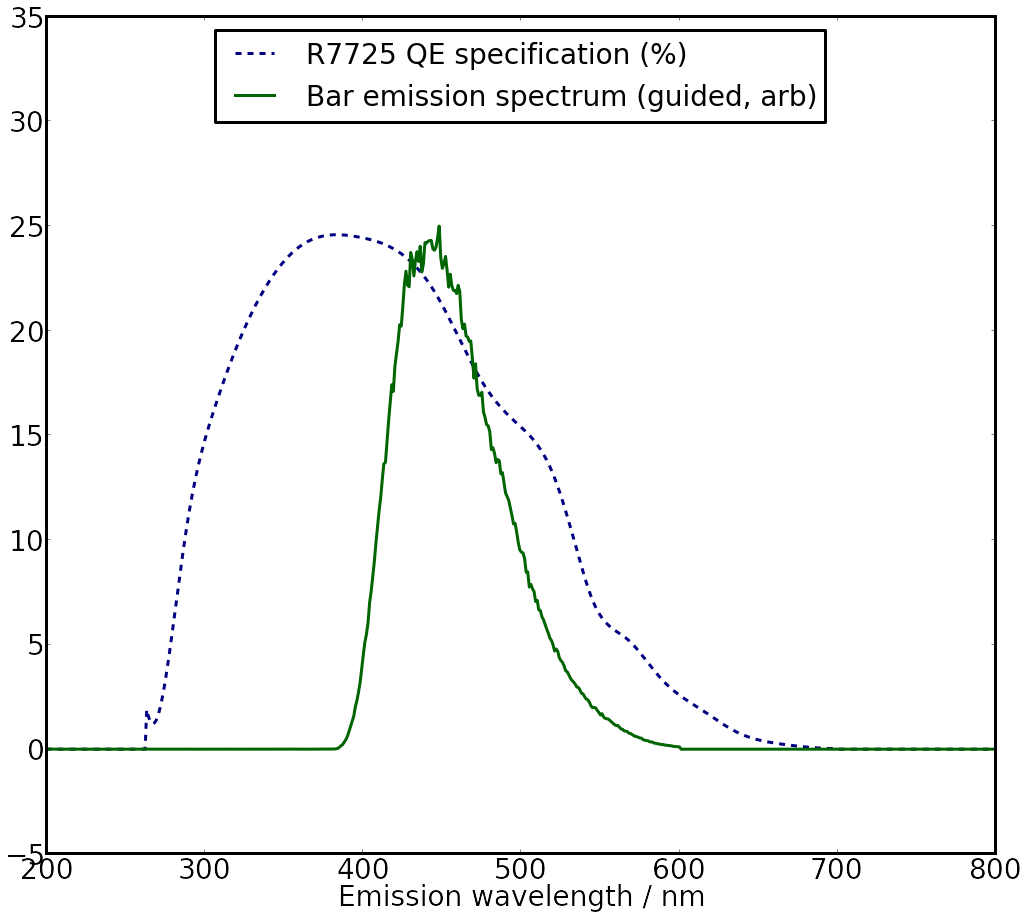}
\caption{Measured emission spectra of the light guide coating in guided mode, and the R7725 quantum efficiency.  Only the quantum efficiency of the non-undercoated PMT model is shown, from \cite{Hamamatsu-Datasheet2inch}.
 \label{fig:TPBSpectra}  }
\end{figure}

In the light guide coatings, the TPB is suspended in an acrylic matrix which leads to a slight broadening of the emission spectrum compared to the spectrum from the plates. This is an expected effect--TPB fluorescence has been shown to have dependence upon its microenvironment \cite{Birks:1959,Birks:1961,Francini:2013-jinst,Hanagodimath:2008, Liu:1997}, and reference~\cite{Francini:2013-jinst} demonstrated spectral broadening in the presence of a polystyrene substrate.  Using the monochromator described for the plate spectrum studies, the light guide coating spectrum was measured in guided mode. A 10 cm section of light guide, with the incident beam perpendicular to the TPB coated surface, was used.  Figure \ref{fig:TPBSpectra} shows the results of this measurement.  Based on reference \cite{Francini:2013-jinst} it is expected that the emission spectrum for the light guide coating, with TPB embedded in the substrate, will not change significantly as it cools to 87 K.  The expected efficiency for the light guide coating is found to be $0.25 \pm 0.05$ emitted visible photons per incident 128~nm photon.

\subsection{Calibration}

The flasher system for the optical units and the light guides is described in reference~\cite{Conrad:2015xta}. This system was developed to check the timing of the installed optical units, exercise the optical units during construction and commissioning, and to calibrate them during detector operations.  The reference provides engineering drawings and details.  The system is briefly summarized here.

A control board pulses an array of 400 nm LEDs, each of which is
coupled through an optical feedthrough to 10 m optical fibers within the cryostat.  The custom feedthrough/patch panel design encases the fibers in Arathane CW 5620 blue with HY 5610 hardener.  The internal fibers are Molex FVP polymide fibers with diameter of 600 $\mu$m, cladding of 30 $\mu$m, and an additional buffer layer of 25 $\mu$m.  Each PMT has an individual calibration fiber.  Each fiber is routed along the rack  and attached to the PMTs by a fiber holder constructed of an aluminum standoff with a nylon-tipped set screw at a distance of about 5 cm from the PMT glass.

\begin{figure}
\centering 
\includegraphics[width=\textwidth]{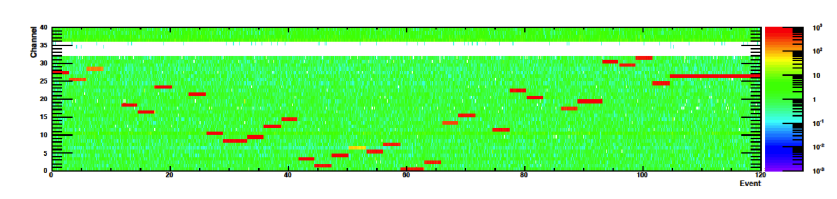}
\caption{Charge detected in each PMT due to flasher tests, shown as a function of time.  The y axis is the PMT channel number, the x axis follows the sequencing of signals to the individual PMTs, an the response is given by the color scale.   Each PMT is flashed for a short period. The white band indicates unused channels. \label{fig:flasherresult}  }
\end{figure}

Figure~\ref{fig:flasherresult} shows the results of flasher tests on all 32 optical units and four PMTs on the paddles.  Time is along the $x$ axis and PMT channel number is along the $y$ axis.  The white region separating the optical unit PMTs from the paddle PMTs represents unused channels.  The colored bars indicate charge detected in the PMTs during flashing.  One can see that all tubes respond properly.

\subsection{Coupling of PMT Signals to the Anode Wires}

ICARUS observed an unexpected cross-talk between their PMTs and collection wire plane \cite{Ankowski:2008aa}. The Argontube test stand at the University of Bern also observed this effect \cite{BernPrivate}.   Therefore, while the \lartpc was under final testing, before being rolled into the cryostat, an experiment was devised in order to determine the implication of this cross-talk for MicroBooNE.  One of the production 200 mm (8.0 in) Hamamatsu R5912-02mod PMTs was placed with its face 12.7 cm from the \lartpc collection wire plane. This PMT was encased in a dark box with an optical fiber delivering light from a LED flasher. The collection wire plane was read out using a test-stand that reflects the final data acquisition design.

\begin{figure}
\centering 
\includegraphics[width=0.75\textwidth]{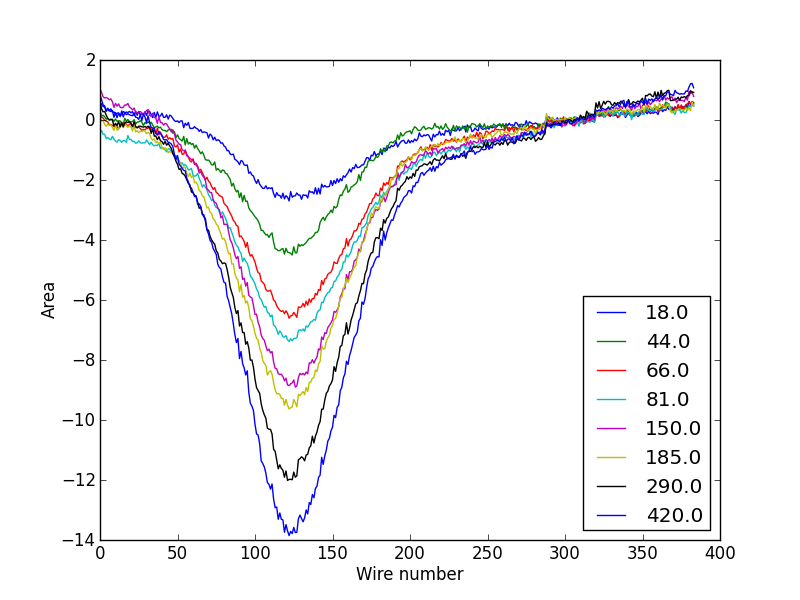}
\caption{Area of signal pulses recorded on collection plane (in arbitrary units) as a function of wire number (arb. offset). The legend indicates the calibrated photo-electron count. The signal pulses are averaged over 50,000 repetitions. The plots are pinned together at wire 300, as the distribution baseline fluctuated over the course of the experiment due to intermittent noise.
 \label{fig:PMTxtalk}  }
\end{figure}

A clear signal was observed on the collection plane when the PMT produced a signal. The magnitude of the signal was characterized as a function of photo-electron count (figure~\ref{fig:PMTxtalk}) and separation between the PMT and wire-plane. The signal induced on the collection plane was estimated to be no more than $\sim$10~ADC counts for a typical cosmic ray under normal operating conditions. This was reduced to $\sim$2~ADC counts when the PMT was encased in the $\mu$-metal shields. This was deemed an acceptably small amount that further shielding was not required.

The effect appears to saturate with photo-electron count, and is reduced when shorting the resistors between the anode and last dynode. A later test with Argontube showed that the signal was drastically amplified when using capacitors not able to withstand cryogenic temperatures in the PMT base \cite{BernPrivate}. This suggests that the effect is electrostatic in nature, and probably due to capacitive coupling between the PMT and wire-plane.  This is also suggested by the similarity between the signals observed and those produced by anode-coupled readout of a light collection system \cite{Moss:2015hha}.

\subsection{Initial Performance of the MicroBooNE Light Collection System}

\begin{figure}
\centering 
\includegraphics[height=0.7\textheight]{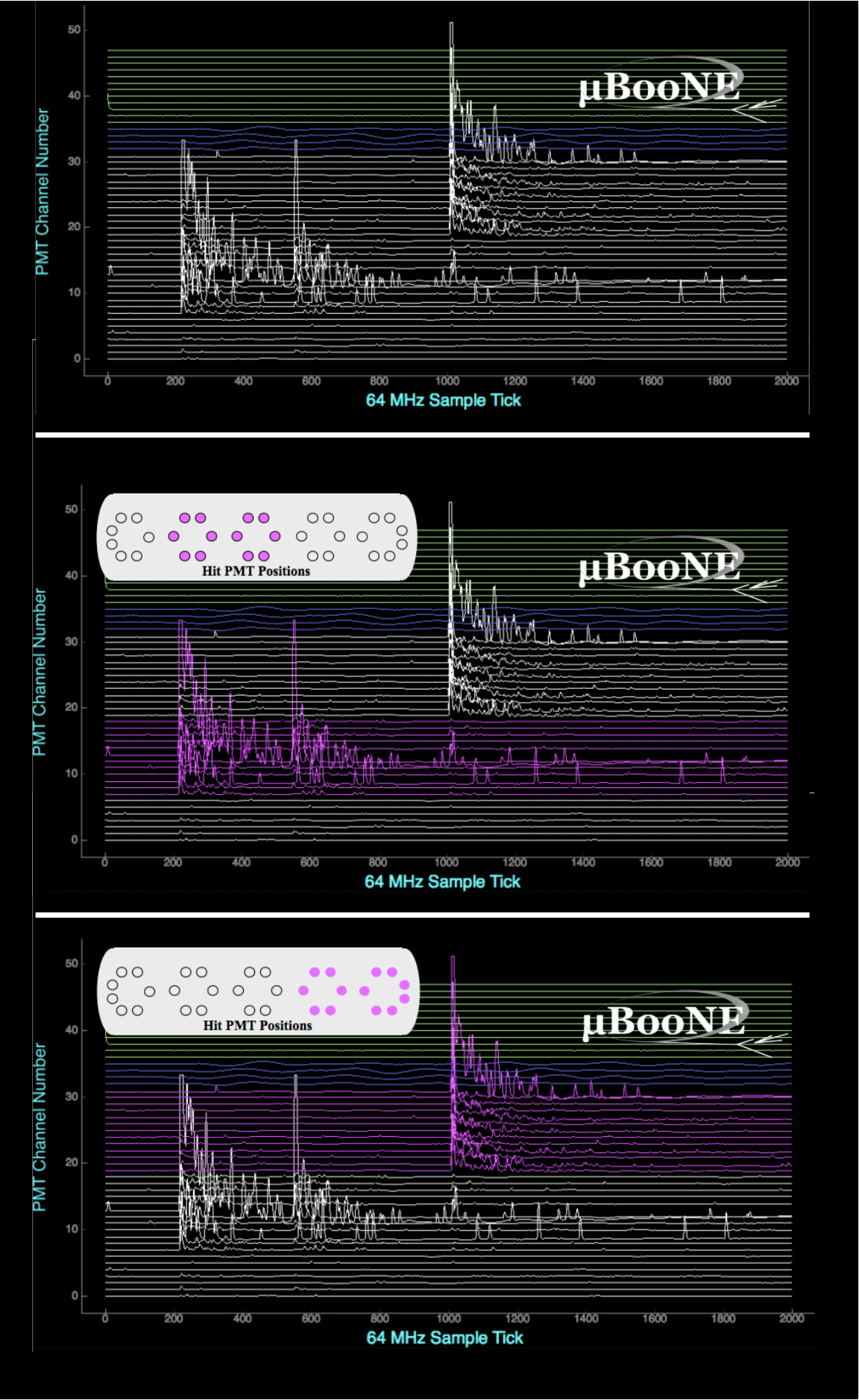}
\caption{Top:  Example waveforms from all 32 PMTs of the primary light collection system over a 31.25 microsecond readout window.  The waveforms from the PMTs are in white.  The blue waveforms are from the secondary lightguide PMTs, which are off in this picture.  Waveforms from channels reserved for logic inputs are shown in green.  In this image, the PMTs see two successive flashes of light at different parts of the detector.  Middle and Bottom:   Magenta highlights the early and late pulse, with the insets showing the PMTs which fired.   Each pulse is likely from two cosmic ray muons traveling through the detector.}
 \label{fig:example_readout}
 \end{figure}

\begin{figure}
\centering 
\includegraphics[width=0.7\textwidth]{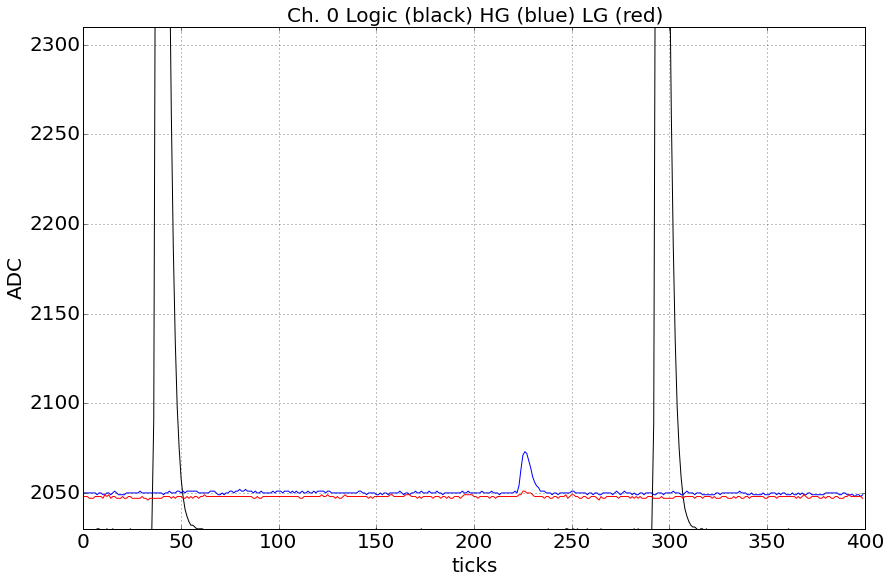}
\caption{Example waveform captured by the PMT readout electronics during single photo-electron calibrations.  The black waveforms are logic pulses that mark the time at which an LED in the flasher system is driven.  After some delay coming from PMT cable lengths and the flasher system, a candidate single photo-electron pulse is seen (at $\sim$230 ticks).}
 \label{fig:example_spe_wfm}
 \end{figure}

\begin{figure}
\centering 
\includegraphics[width=0.7\textwidth]{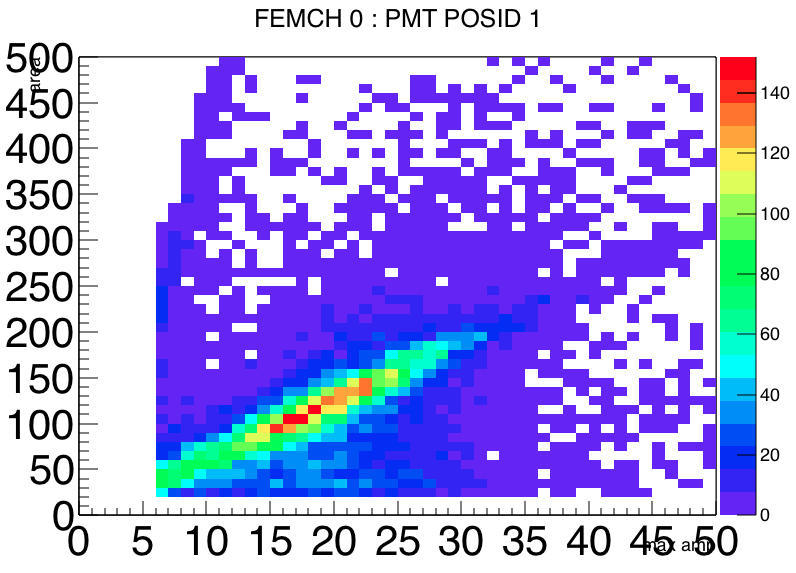}
\caption{Distribution of the maximum amplitude and charge of pulses collected during an LED flasher calibration run. The central distribution of events is due to single photo-electron pulses.}
 \label{fig:example_spe_pulsedist} 
\end{figure}

The system was first powered on after the cryostat had been filled with liquid argon.  All the PMTs in both the primary and secondary system were found to be operational.  Figure~\ref{fig:example_readout} shows the waveforms for all the PMT channels around the time that a large pulse, potentially from a cosmic ray muon, is observed.  After initial checks of the system's health, the one photo-electron response of each PMT was set such that a one photo-electron pulse has an amplitude of 20 ADC counts as seen by the PMT readout system.  The flasher LED system, described previously, is used to set this response. Figure~\ref{fig:example_spe_wfm} shows a candidate single photo-electron pulse following arrival of a TTL logic pulse driving the LED flasher system.  The LED light level is set so that the majority (about 80\%) of waveforms see no response in the region where pulses from the LED are expected to occur.  This ensures that for windows with pulses, the pulses are of single photo-electrons.  Figure~\ref{fig:example_spe_pulsedist} shows an example of the area vs. maximum amplitude of such pulses seen by a PMT during the flasher runs.


\newpage
\section{Electronics and Readout Systems}
\label{sec:electronics}

\begin{figure}
\centering
\includegraphics[width=0.95\linewidth]{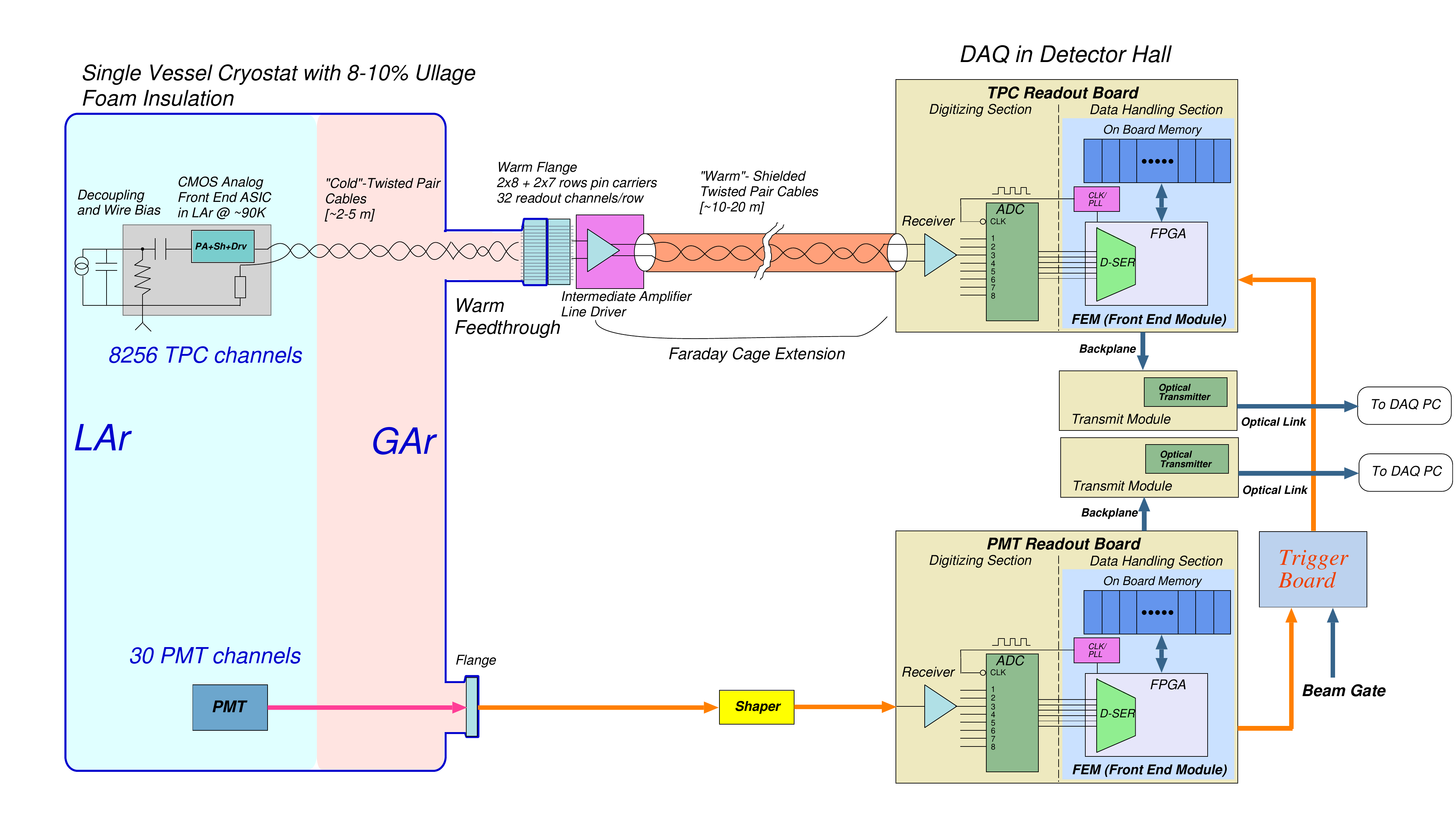}%
\caption{\label{readout_1}MicroBooNE \lartpc and PMT signal processing and readout stages.}
\end{figure}

The analog signals that develop on a \lartpc during its operation must be amplified, digitized, and written to disk for use in analysis.  Custom low-noise electronics that are capable of operating in the liquid argon environment have been developed for this purpose in MicroBooNE.  The data from these \lartpc electronic channels, as well as from the PMTs, is sent to a readout system that digitizes and organizes the information before passing it along to a data acquisition (DAQ) system that stores it on disk.  The stages of signal processing are illustrated in figure~\ref{readout_1}.  The following subsections describe the \lartpc cold electronics, the \lartpc and PMT readout electronics systems, and the DAQ system in more detail.  Details of the trigger capabilities available in MicroBooNE are also provided in this section.

\subsection{Cryogenic Low-Noise Electronics}
\label{sec:coldelectronics}
To obtain optimum detector performance, MicroBooNE uses cryogenic low-noise front-end electronics for readout of the LArTPC. To reduce electronic noise, the interconnection length between the \lartpc wires and preamplifier should be as short as possible thus minimizing the total capacitance seen at the preamplifier input. To accomplish this, the analog front-end ASICs, which include a preamplifier, shaper, and signal driver are located inside the cryostat in addition to the wire bias voltage distribution system, decoupling capacitors, and calibration networks. The front-end ASIC and associated circuits are implemented on a cold mother board which is directly attached to wire carrier boards on the \lartpc itself. Cold cables are used to transmit output signals from cold motherboards to warm interface electronics installed on the top of the signal feed-through flanges.

\subsubsection{CMOS ASIC}

The analog front end ASIC is designed in 180~nm CMOS technology, which integrates both the preamplifier and shaper on a single chip. Each chip has 16 channels to read out signals from 16 wires. Each channel also has a charge injection capacitor for precision calibration. In MicroBooNE, the shaper has four programmable gain settings (4.7, 7.8, 14 and 25 mV/fC) and four programmable peaking time settings (0.5, 1.0, 2.0 and 3.0 $\mu$s) that provide increased flexibility to the readout system. The ASIC also has programmable baseline settings (200 or 900 mV) to accommodate different detection wire configurations: either collection or induction plane. It has a selectable AC/DC coupling mode with a 100 $\mu$s time constant for the AC coupling mode, which can be used to reduce low frequency noise. The ASIC also has built-in band-gap reference and temperature sensors to facilitate biasing and monitoring.

\begin{figure}
\centering
\includegraphics[width=0.75\linewidth]{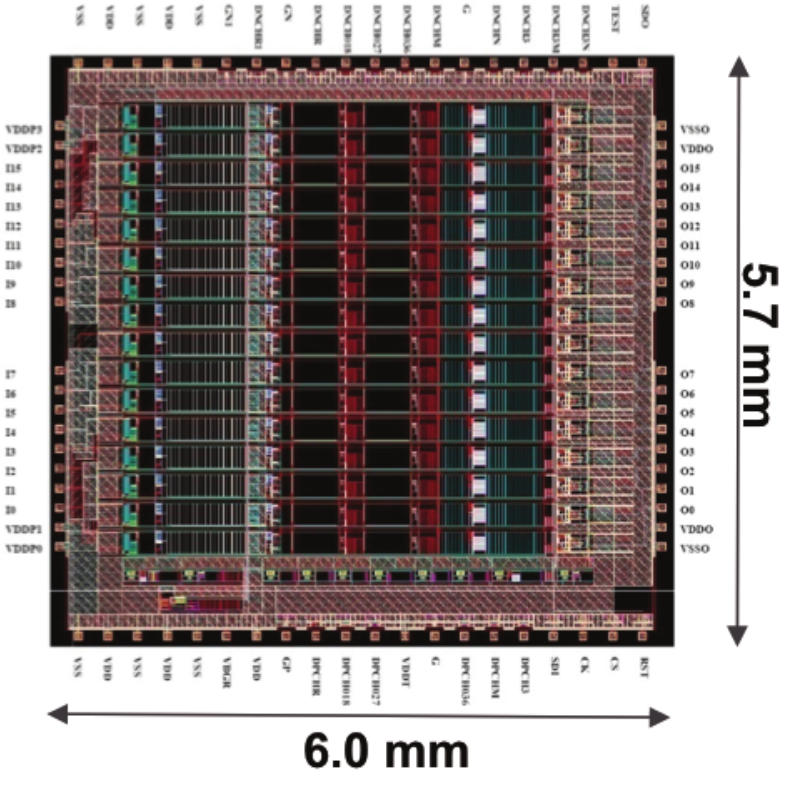}
\caption{\label{fig:figasic}Layout of the CMOS analog front end ASIC}
\end{figure}

The CMOS ASICs consume only 6~mW/channel in their default configuration. The front end ASICs of the entire detector generate 50 W of heat load that is easily handled by the cryogenics system. Design guidelines that constrain the electric field and the current density to address the lifetime of CMOS devices operated at cryogenic temperatures have been applied to every single transistor (total $\sim$15,000 transistors) in the ASIC design. A picture of the layout of the CMOS ASIC is shown in figure~\ref{fig:figasic}. Test results agree well with simulations and indicate that the analog and the digital circuits (including the digital interface) operate as expected in the cryogenic environment. 

The MicroBooNE \lartpc has  8,256 readout channels and a total of 516 CMOS ASICs are required to fully instrument the detector. The production testing of the CMOS ASICs required two steps: both a warm and cold test. The warm test was performed with a dedicated test board housed in a Faraday box containing a socket to house the ASIC for ease in chip exchange. All programmable parameters (gain, peaking time, baseline, AC/DC coupling etc.) were exercised with the warm test setup for careful screening at room temperature. The yield of the warm testing of the ASICs was 89\%. ASICs must have passed the warm test before going through cold testing. The cold test was performed with a dedicated test board containing 6 sockets to facilitate testing of multiple chips in liquid nitrogen at the same time. A total of 201 ASICs went through cold testing with a yield of 97\%. Based on this high yield, it was decided not to continue the cold screening test on the rest of the production chips, as they were tested cold after being installed on the motherboard. After enough ($\sim$600) ASICs passed the production screening test, they were sent to an assembly house to equip the cold motherboard.

\subsubsection{Cold Motherboards}

\begin{figure}
\centering
\includegraphics[width=0.75\linewidth]{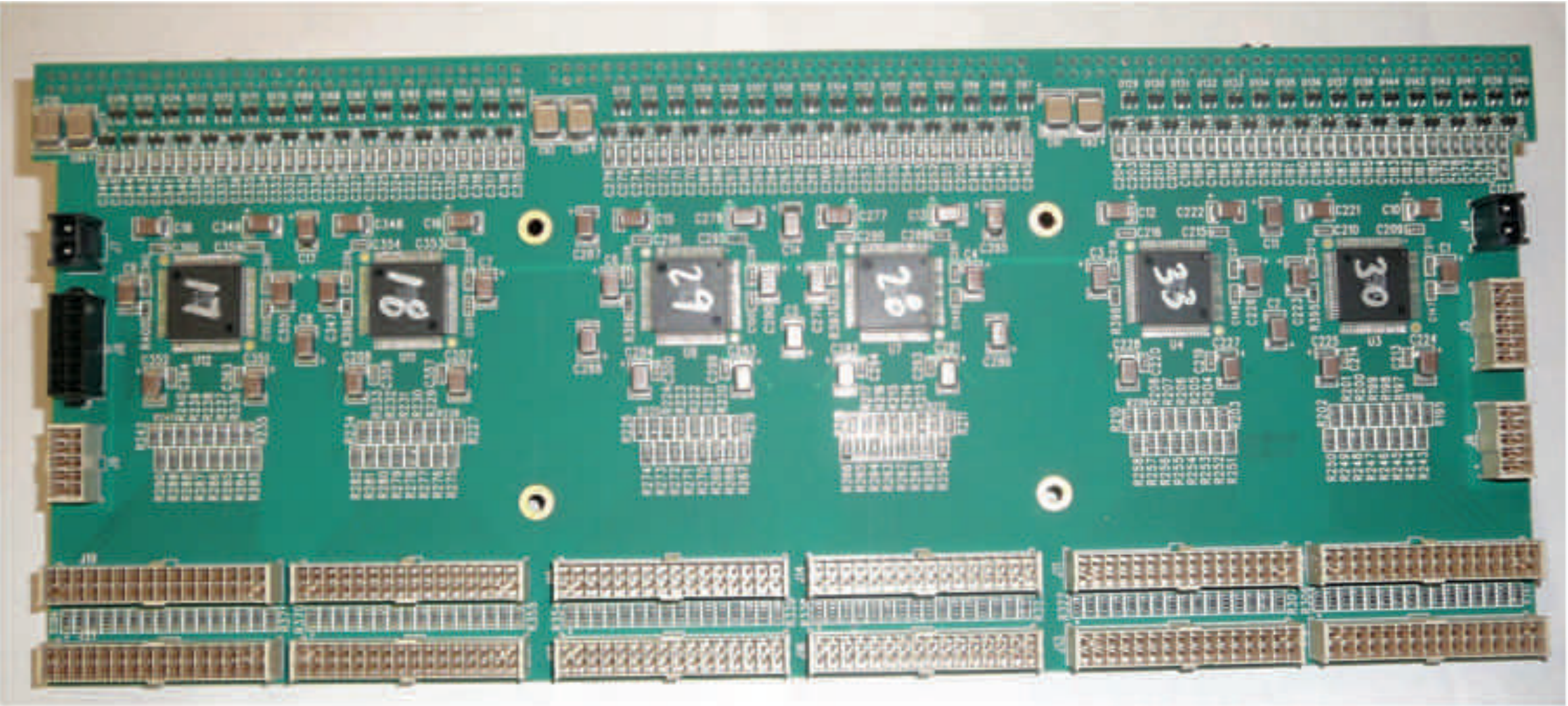}
\caption{\label{fig:figmb} Top version of cold motherboard with 12 ASIC chips, including 6 chips mounted on the top layer and 6 chips on the bottom layer.}
\end{figure}

A cold motherboard was designed to house the MicroBooNE CMOS ASICs. In this capacity, the motherboard provides signal interconnections both between the detector wires and preamplifier inputs as well as between the driver outputs and cold cables to the signal feed-through. The cold motherboard design provides sufficient protection of the ASICs against electrostatic discharge during installation. It also provides a calibration network and bias voltage distribution for the wire planes. Specifically, a calibration signal enters the cryostat via a feed-through and reaches the preamplifiers through the motherboard. Each preamplifier channel in the ASIC has a built-in switch to individually cycle the calibration injection. The bias voltage reaches the \lartpc wires via a two-fold redundant path on the motherboard that allows the detector to operate normally even if one bias voltage channel fails.  As with the PMT bases, the cold motherboards use Rogers 4000 series as the PC-board material due to the cleanliness and thermal properties previously described in section \ref{sec:pmt-bases}.  The different positions of the wire attachments along the top and sides of the \lartpc requires 2 types of cold motherboard. The top version of the motherboard has 192 readout channels that includes 96 Y channels, 48 U channels, and 48 V channels. The side version of the motherboard has 96 readout channels that are either U or V channels. A picture of the top version of a cold motherboard with 12 mounted ASIC chips is shown in figure~\ref{fig:figmb}. 

The MicroBooNE \lartpc required a total of 36 top version motherboards and 14 side version motherboards to instrument the full detector. A test stand was built for testing of the front end electronics. This test stand included a full readout chain from the cold motherboard, cold cable, signal feed-through, warm interface electronics, warm cable and receiver ADC board to a DAQ board based on a Xilinx ML605 FPGA evaluation board which sends acquired data to a PC over a Gigabit Ethernet. The production test of each motherboard also involved both a warm and cold test. Both tests used the same test stand, except the motherboard was placed in a Faraday box for the warm test and submerged in a liquid nitrogen dewar for the cold test. Noise, gain, peaking time, and linearity parameters were measured in both warm and cold to screen the motherboard. Motherboards had to pass both tests  before being installed on the detector.

\subsubsection{Cold Cables}

Cold cables transmit the detector signals from the cold motherboard to an intermediate amplifier on top of the signal feed-through and distribute power to the CMOS ASICs. The cold cable is a custom-built 32-pair twisted pair flat ribbon cable with Teflon FEP insulation and 100 $\Omega~(\pm10\%)$ impedance, using AWG 26 stranded wire with silver-plated copper. Custom designed shells with jack screws used in the cable assembly ensured proper alignment of the insertion on the signal feed-through pin carriers.  Cold cables of two different formats were assembled by an assembly house: signal cables and service cables. Signal cables are used to transmit amplified detector signals while the service cable is used to transmit calibration pulses and slow control/monitoring signals. Signal cables were produced in three different lengths: 203 cm, 254 cm, and 457 cm to accommodate the different lengths between the cold motherboards and the signal feedthroughs, while the service cables were produced in two different lengths: 254 cm and 457 cm. All of the cold cable assemblies were tested with a cable tester in the assembly house before they were shipped out. In addition, 10\% of the cold cables were tested in the test stand at BNL to confirm the quality of the cable assembly. 

\subsubsection{Electronic Calibration}

The MicroBooNE cold electronics include a precision charge calibration system. Through the cold cable and calibration network on the motherboard, a calibration signal enters the cryostat via a feed-through and reaches the preamplifiers. A built-in switch in the ASIC makes it possible to power cycle the calibration injection for every channel individually. The electronics calibration is based on charge injection through known capacitances (180 fF) in the ASIC. This system enables gain (charge sensitivity) calibration, verification of sense wire integrity and noise measurements. The built-in electronics calibration capability is an important tool in testing and characterizing the overall performance of the detector readout system. It was extensively used in the cold electronics production testing and the electronics checkout during installation, commissioning, and data taking.
 
\subsubsection{Performance Tests}

The development of the analog front end ASICs was initiated using 180~nm CMOS technology and 300 K models, though the performance parameters are extracted at 77 K. CMOS was found to function at cryogenic temperatures with increased gain and lower noise. The noise, gain, and pulse shaping were found to be as expected in evaluation tests of the ASICs. Extensive testing of the ASICs mounted on the motherboards was performed; these tests were done in liquid nitrogen rather than at room temperature, since noise levels and characteristics of the ASIC performance in liquid nitrogen are similar to the performance in liquid argon.  Thus, cold tests were performed on all production cold motherboards fully populated with 12 chips. A total of $\sim$2,200 chip-immersions were accumulated in liquid nitrogen without any failures due to thermal contraction or expansion. 

\begin{figure}
\begin{center}
\includegraphics[scale=0.4]{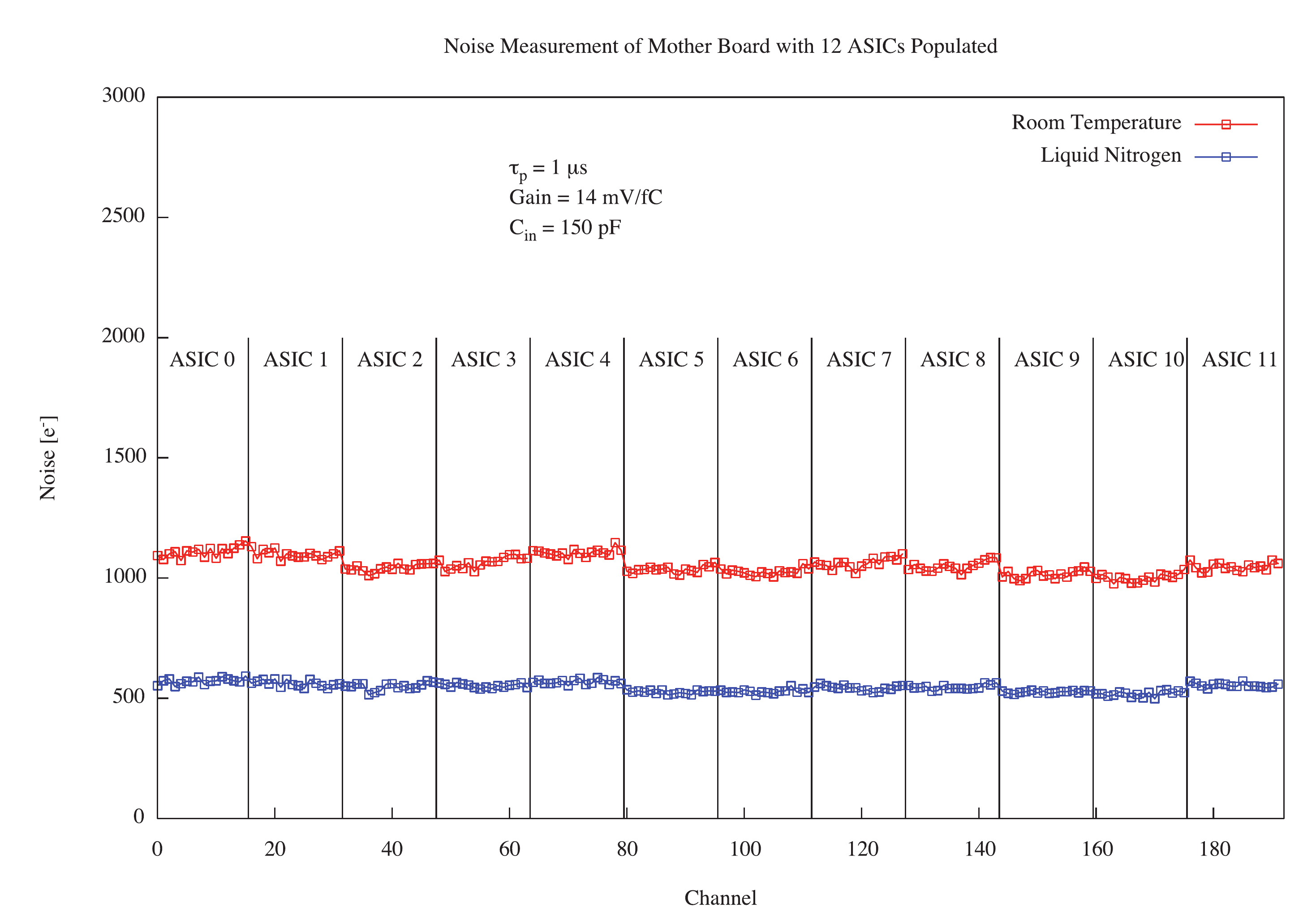}
\end{center}
\caption{\label{fig:fignoise}Plot of noise vs. channel number for 192 channels (12 ASICs) and at two different temperatures. Noise is $\sim$1,200 $e^{-}$ at 293 K, and $\sim$550 $e^{-}$ at 77 K with 150 pF $C_{d}$}
\end{figure}

\begin{figure}
\begin{center}
\includegraphics[scale=0.4]{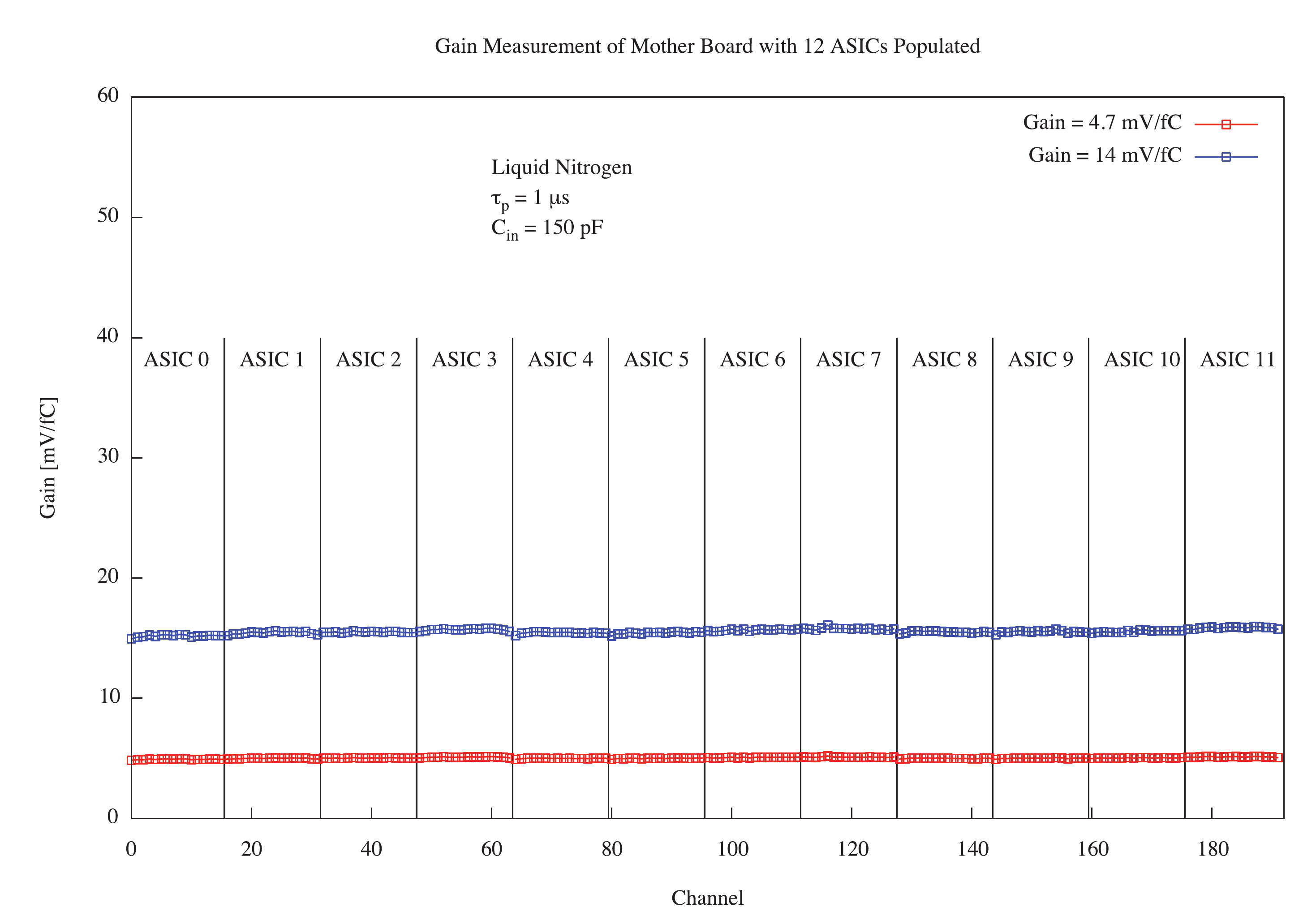}
\end{center}
\caption{\label{fig:figgain}Plot of gain uniformity of 12 ASICs, total 192 channels, at 77~K with two different gain settings.}
\end{figure}

The test results show the noise of the front end readout electronics system decreasing uniformly for all 768 channels from $\sim$1,200 $e^{-}$ at 293 K to less than 600 $e^{-}$ at 77 K with 150 pF detector (sense wire) capacitance. A plot of noise versus temperature of 12 ASICs for a total of 192 channels is shown in figure~\ref{fig:fignoise}. The response of the front end electronics exhibits excellent uniformity at cryogenic temperatures. As shown in figure~\ref{fig:figgain}, the gain variation of a cold motherboard with 12 ASICs is only 7\% peak-to-peak across 192 channels. The spread of the gain variation is only 1\% of the gain setting.

\subsection{Warm Electronic Amplification}
\label{sec:warmelectronics}
\begin{figure}
\begin{center}
\includegraphics[width=0.6\linewidth]{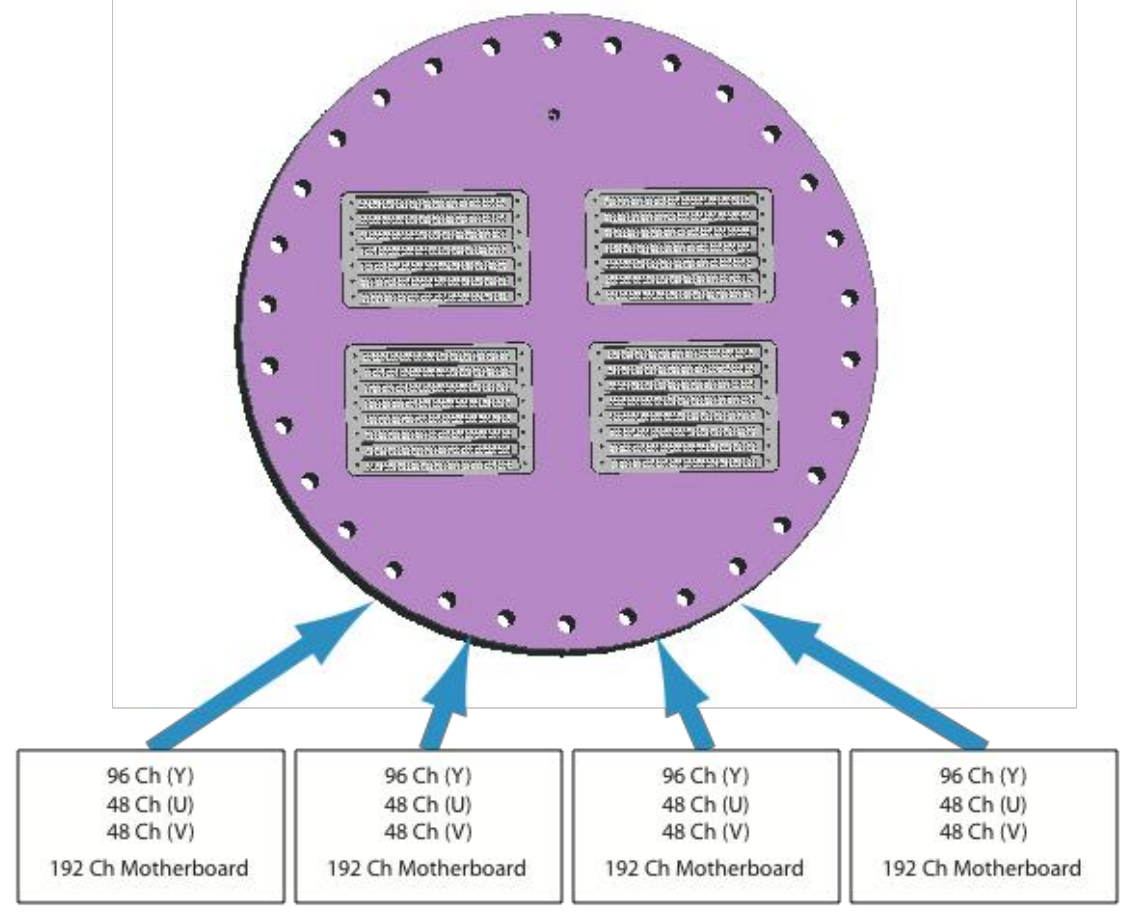}
\end{center}
\caption{\label{fig:feedthroughflange}Signal feedthrough flange consisting of a 356 mm CF flange with two 8-row and two 7-row pin carriers welded in place.}
\end{figure}

Signals from the cold electronics are carried over the cold cables to dedicated feedthroughs mounted on the cryostat.  The cold cables are connected to pin-carriers located on 356 mm outer-diameter CF signal feedthrough flanges that are mounted on nozzles N1A-N1K of the cryostat (see figure \ref{fig:cryostat-drawing}).  The signal feedthrough design must accommodate 100$\%$ hermeticity and high signal density. A design based on the ATLAS pin carrier style was developed for this purpose.  Two 8-row pin carriers and two 7-row pin carriers are welded onto the CF flange, as shown in figure \ref{fig:feedthroughflange}, and create a vacuum-tight seal.  Nine of the 11 signal feedthroughs receive signals from the three \lartpc anode planes (384 Y-plane, 192 U-plane, 192 V-plane), while the remaining two on the extreme ends of the cryostat only receive signals from one of the angled induction planes (672 U-plane on one feedthrough, 672 V-plane on the other).

\begin{figure}
\begin{center}
\includegraphics[width=0.4\linewidth]{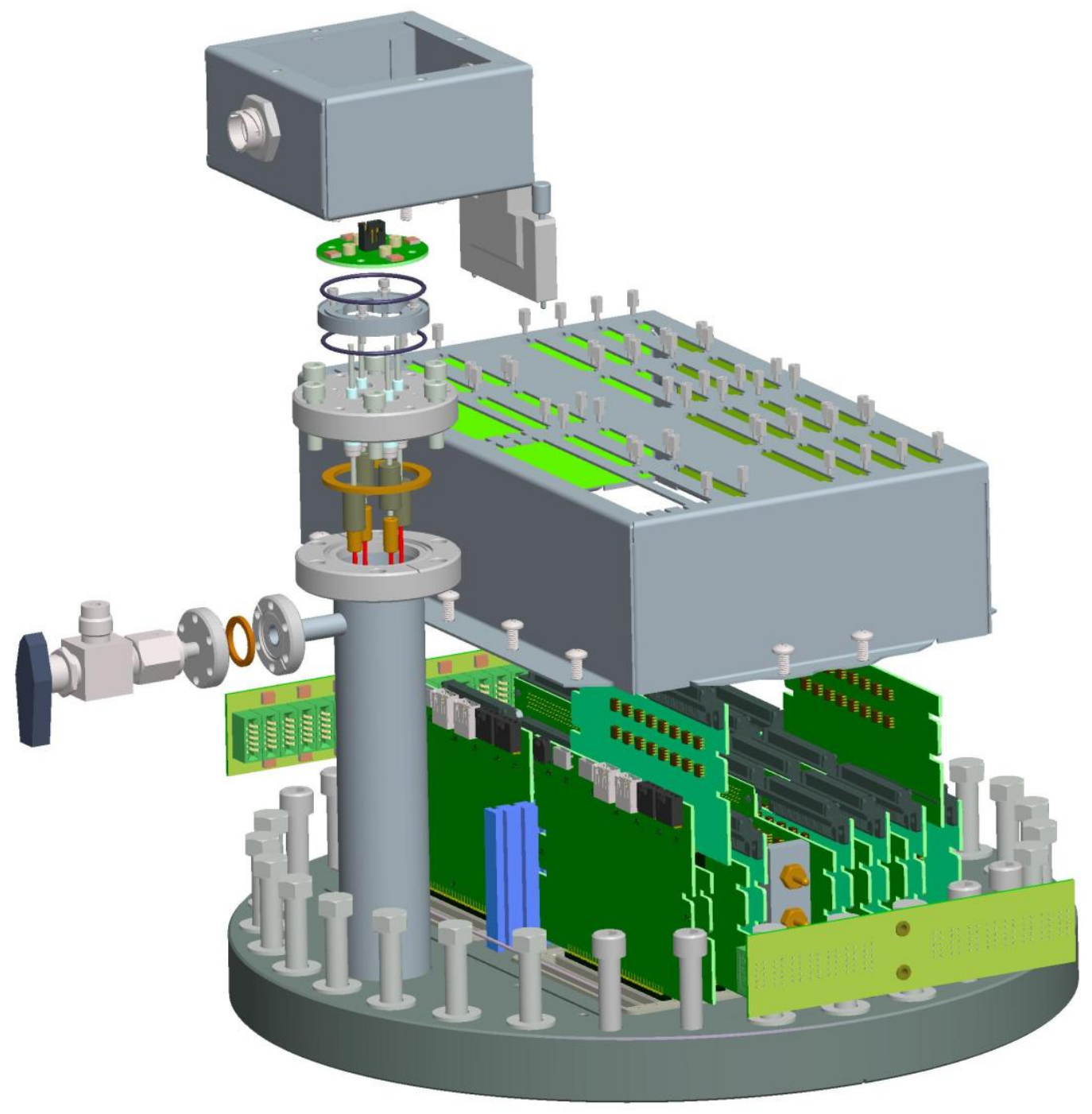}
\includegraphics[width=0.4\linewidth]{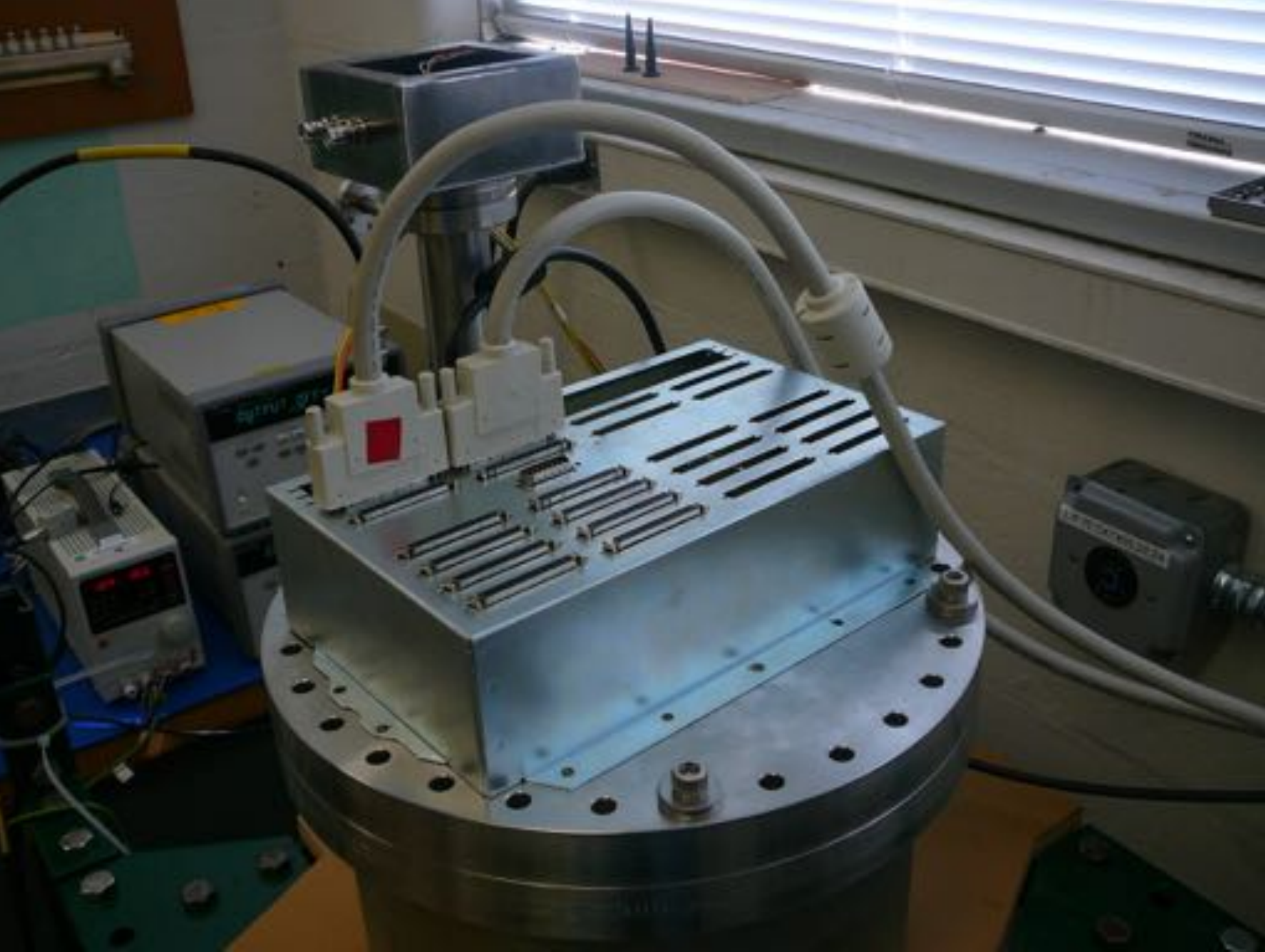}
\end{center}
\caption{\label{fig:feedthroughassembly}Left:  Diagram of the signal feedthrough assembly, which includes intermediate amplifiers, Faraday cage, bias-voltage feedthrough and filtering circuit.  Right: Photograph of one of the feedthrough assemblies, partially constructed and being tested.}
\end{figure}

A Faraday cage is mounted on the external, warm, side of the signal feedthroughs to provide shielding for the intermediate amplifiers located inside.  The bias voltage feedthrough, which supplies anode plane bias voltages into the cryostat, is built onto a small 70 mm CF flange welded onto the signal feedthrough flange. A filter board mounted on the bias voltage flange filters noise and ensures a good ground connection. Figure \ref{fig:feedthroughassembly} shows details of the signal feedthrough assembly with electronics boards, bias voltage feedthrough, and Faraday cage.

\begin{figure}
\begin{center}
\includegraphics[width=0.4\linewidth]{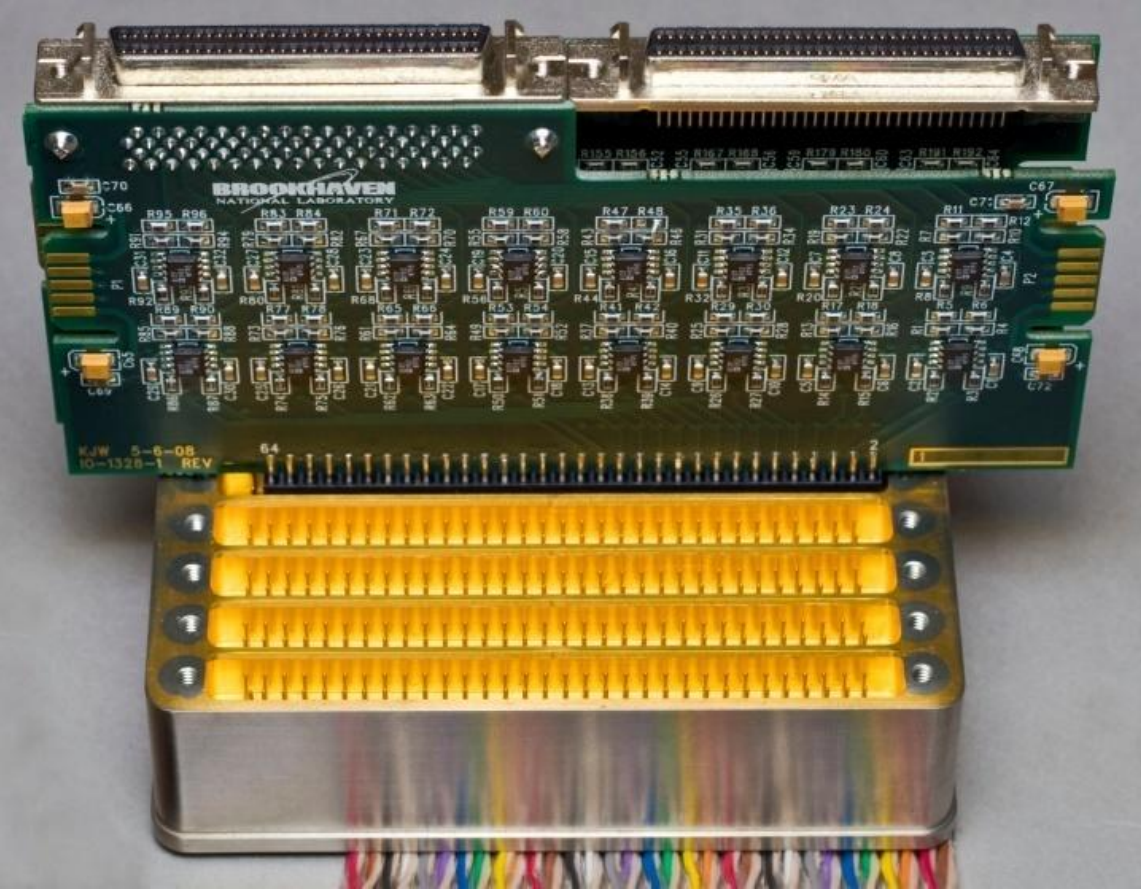}
\includegraphics[width=0.4\linewidth]{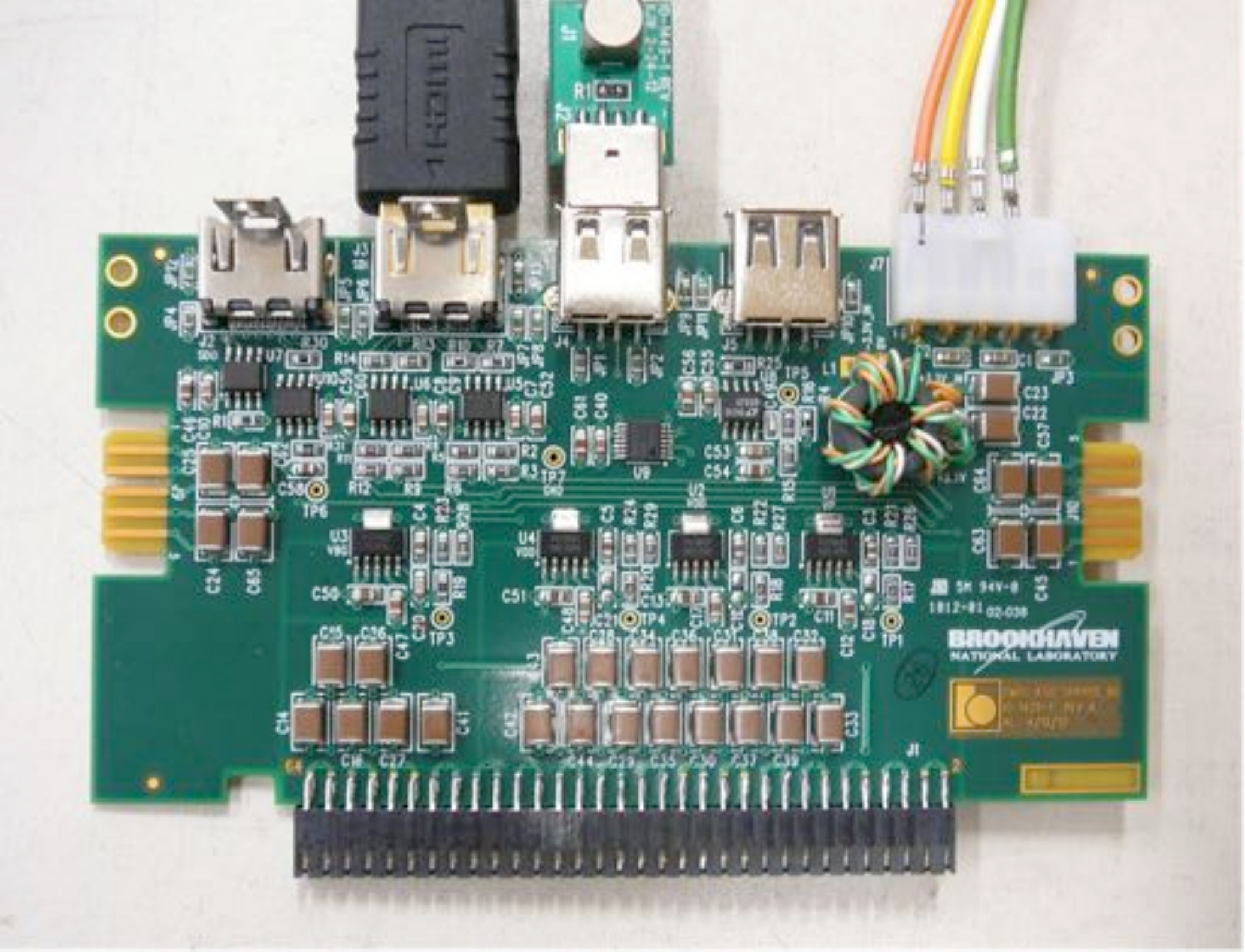}
\end{center}
\caption{\label{fig:intermediateamplifier}Left: Photograph of one of the intermediate amplifier boards plugged into a pin carrier during testing.  Right: Photograph of a prototype service board.}
\end{figure}

The intermediate amplifiers provide $\sim$12 dB gain to the \lartpc signals to make them suitable for transmission over a 20 m long cable to the readout electronics (see section \ref{sec:readoutelectronics}).  Each intermediate amplifier has 32 channels installed on the signal feedthrough flange and housed inside the Faraday cage to provide noise isolation. Figure \ref{fig:intermediateamplifier} shows a picture of a prototype intermediate amplifier plugged on the signal feedthrough pin carrier. The intermediate amplifier uses a 68-pin SCSI-3 connector to drive the 32 channels of signal differentially for better noise immunity. The layout and connector position have been carefully designed to ensure the card can be plugged on the pin carrier in either direction. This efficiently utilizes the limited available space on the top of the feedthrough, which also makes the design of the Faraday cage easier.

In addition to the intermediate amplifiers, there are two service boards mounted on the top of each signal feedthrough. The service board provides regulated low voltage, control and monitoring signals to the analog front end ASICs. It also provides pulse injection to the preamplifiers for precision calibration. The control, monitoring, and calibration signals are provided to the front end electronics with two-fold redundancy. Should one set of signals become defective the detector can still operate normally with the redundant set. Each service board plugs onto a 64-pin carrier row. Figure \ref{fig:intermediateamplifier} shows a picture of a prototype service board.

\subsection{LArTPC Readout Electronics}
\label{sec:readoutelectronics}
The MicroBooNE readout electronics system consists of two subsystems: the \lartpc and PMT readout electronics. The \lartpc readout electronics are responsible for the digitization, processing and readout of the induction and collection wire signals after amplification. The PMT readout electronics are responsible for the amplification, shaping, digitization, and handling of PMT (and lightguide paddle) signals, and can be used to provide a trigger signal for the readout and Data Acquisition (DAQ) systems. While the \lartpc and PMT readout systems share the same back-end design that organizes and packages the data for delivery to the DAQ system, they employ different analog front-end and digitization designs, which are described in this and the following subsection.

\begin{figure}
\begin{center}
\includegraphics[width=0.5\linewidth]{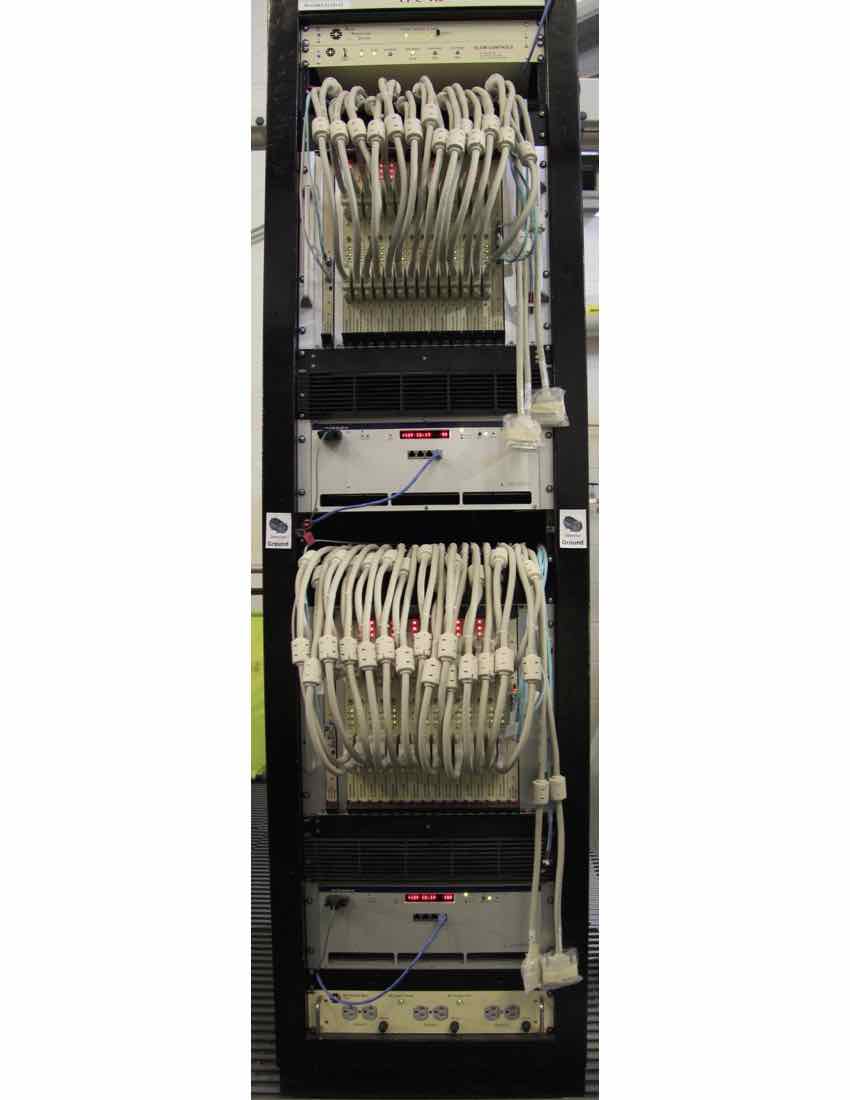}
\end{center}
\caption{\label{fig:rocrates}Rack containing two of the nine MicroBooNE TPC readout crates.}
\end{figure}

The \lartpc readout electronics are responsible for processing the signals from the 8,256 wires in MicroBooNE after pre-amplification and shaping in the cold electronics, as described in section~\ref{sec:coldelectronics}.  The pre-amplified and shaped analog signals from the cold electronics are transmitted to the warm electronics outside the cryostat, as described in section~\ref{sec:warmelectronics}, and then passed to 130 custom-designed ADC and Front End readout Modules (ADC/FEMs) distributed roughly evenly over nine readout crates, as shown in figure~\ref{fig:rocrates}.  The readout modules digitize the analog signals and then process and prepare them for shipping to designated DAQ machines (one DAQ machine per readout crate). These are part of the back-end DAQ system, described in section~\ref{sec:daq}. 

The \lartpc readout crates communicate with the DAQ machines via three duplex 3.125 Gb/s optical links that connect to a crate controller module and data transmitter (XMIT) module on the crate end, and to three PCI Express boards on the DAQ machine end. The controller is responsible for configuration, trigger and run control command distribution as well as the slow monitoring of each readout crate. The controller occupies one of the optical links while the XMIT is responsible for sending two separate streams of readout data to the DAQ machines via the two other optical links. The first XMIT stream contains losslessly compressed \lartpc data associated with event triggers received by the \lartpc readout crates, such as the BNB trigger, and is referred to as the ``NU'' data stream. The second stream is a continuous \lartpc data stream which is compressed with some data loss. The continuous data stream is used for beam-unrelated physics analyses, such as the study of potential supernova neutrino events, and is referred to as the ``SN'' data stream. The compression schemes used in the NU and SN streams are described in section~\ref{sec:tpccomp}.

All readout crates are synchronized to a common 16 MHz clock. The clock sync is provided by a clock fanout board and is sent via coaxial cables to a distribution board which is mounted on each crate backplane. The frame size is set to 1.6 ms, which is equivalent to the time it takes for charge produced on the far end of the \lartpc to drift to the wire planes at the design cathode voltage of -128 kV.

\subsubsection{\label{sssec:tpcdigitization} Data Digitization}

The amplified and shaped analog \lartpc signals are differentially received and digitized in the first section of the ADC/FEM readout modules. Each ADC module holds 8 AD9222 octal-channel 12-bit ADCs and handles signals from 64 wires. The wire signals are grouped in two sets of 32 consecutive wire channels: either 32 induction wires plus 32 collection wires or two sets of 32 induction wires. The induction channel sequence alternates wires between the two induction planes. The ADC module digitizes the signals continuously at 16~MHz. Each channel has a configurable baseline, which is either set low (450 ADC counts) for collection channels or at the middle of the dynamic range (2055 ADC counts) for induction channels, thus ensuring that both the collection plane unipolar differential signals and the induction plane bipolar differential signals can make use of the full ADC analog input range. The requirement to observe a MIP produced at the far end of the \lartpc in the induction plane determines the lower end of the dynamic range, while the requirement to observe a highly-ionizing stopping proton at the close end of the \lartpc without saturation sets the upper end. The digitized outputs from the ADC module are passed directly to a Front End Module (FEM) in the second section of the \lartpc readout module. The FEM houses an FPGA for data processing, data reduction, and preparation for readout by the DAQ system as described in the following section. 

\subsubsection{\label{sssec:tpcFEM} Data Handling}

\begin{figure}
\centering
\includegraphics[width=0.6\linewidth]{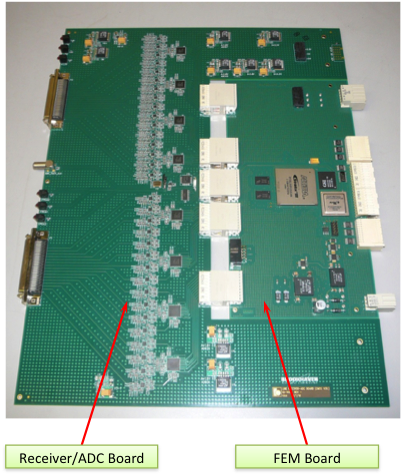}%
\caption{\label{fig:readout_2}Photograph of a MicroBooNE ADC+FEM board.}
\end{figure}

The FEM board consists of a 14-layer printed circuit board which is mechanically integrated with the ADC board as illustrated in figure~\ref{fig:readout_2}. The choice of a smaller board allows for short trace lengths which is beneficial for high speed signals. The full assembly comes together as a standard VME 9U card in height, with a 280 mm depth. Differential outputs from the ADCs connect to the FPGA through HM-Zd connectors that have individual ground shielding on each differential pair. 

\begin{figure}
\centering
\includegraphics[width=0.8\linewidth]{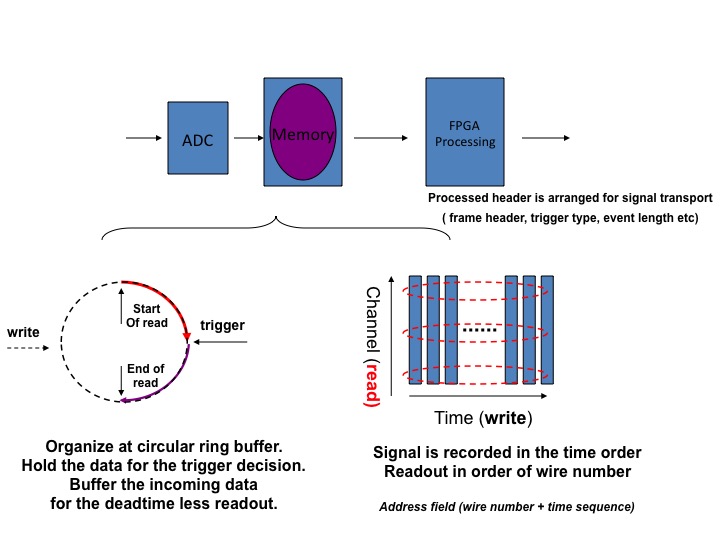}%
\caption{\label{fig:readout_3}MicroBooNE readout sequence in the TPC FEM.}
\end{figure}

The digitized data stream moves from the ADCs to a Stratix III Altera FPGA, which reduces the sampling rate of the ADC from 16~MHz to 2~MHz. The 2~MHz sampling rate is optimized by taking into consideration the expected pulse shape provided by the convolution of the cold electronics, the expected \lartpc field responses, and the O(1$\mu$s) diffusion effects which govern charge drift within the liquid argon. The FPGA stores the data from all 64 wires per board sequentially in time in a 1M $\times$ 36 bit 128 MHz SRAM, grouping two ADC words together in each 36 bit memory word. This requires a data storage rate of (64/2) $\times$ 2 MHz $=$ 64 MHz.  The SRAM chip size and memory access speed allow for continuous readout of the \lartpc data. Since data reduction and compaction algorithms rely on the sequential time information of a given wire, the data readout out from the SRAM takes place in wire order in alternate clock cycles, again at a rate of 64 MHz. This read in/out sequence is illustrated in figure~\ref{fig:readout_3}.

\begin{figure}
\centering
\includegraphics[width=0.8\linewidth]{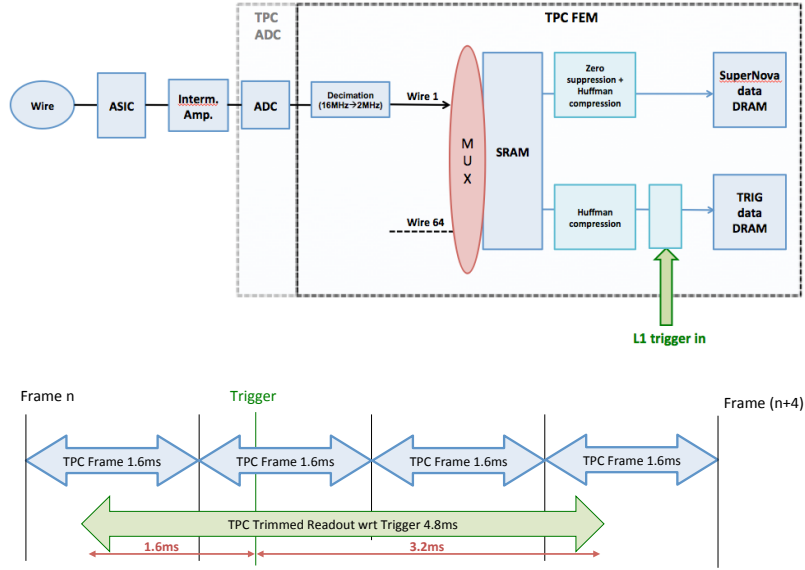}%
\caption{\label{fig:readout_4}MicroBooNE trigger readout.}
\end{figure}

Separate DRAM multi-event buffers on the FEM store the NU and SN data streams. The data divert into the NU readout stream when a trigger is issued and received (as shown in figure~\ref{fig:readout_4}) signaling for example, an accelerator neutrino-induced event. When an event trigger is received, 4.8 ms worth of data, relevant to that event, are packeted per channel and sent to the DAQ through the NU data stream. The 4.8 ms readout size is governed by the maximum drift time and spans three or four frames. In order to reduce the amount of data being transmitted, the FPGA trims the three or four frames to span the exact 4.8 ms required, 1.6 ms before the trigger plus 3.2 ms after the trigger. In parallel, the data is continually sent out through the SN data stream, frame by frame. The compression and data reduction algorithms applied to each of the two streams are described in the following section.
 
After processing by the FPGA, the data passes to the crate backplane dataway on connectors shown in figure~\ref{fig:readout_2}. A token-passing scheme is utilized to transfer data from each FEM board to the data transmitter module (XMIT) in a controlled way, whereby each FEM, in the order of closest to furthest away from the XMIT module, receives a token, transmits its data to the XMIT, and passes the token on to the next FEM in the sequence. For the NU stream, each FEM sends all data associated with a particular trigger number; while for the SN stream, each FEM sends all data associated with a particular frame number. This data transfer is relayed via the otherwise passive crate backplane, and is limited to 512~MB/s. In the XMIT module, the data is buffered temporarily and sent to the DAQ machine through the two streams, SN and NU, which proceed effectively in parallel.


\subsubsection{Compression Schemes}
\label{sec:tpccomp}

\begin{figure}
\centering
\includegraphics[width=0.8\linewidth]{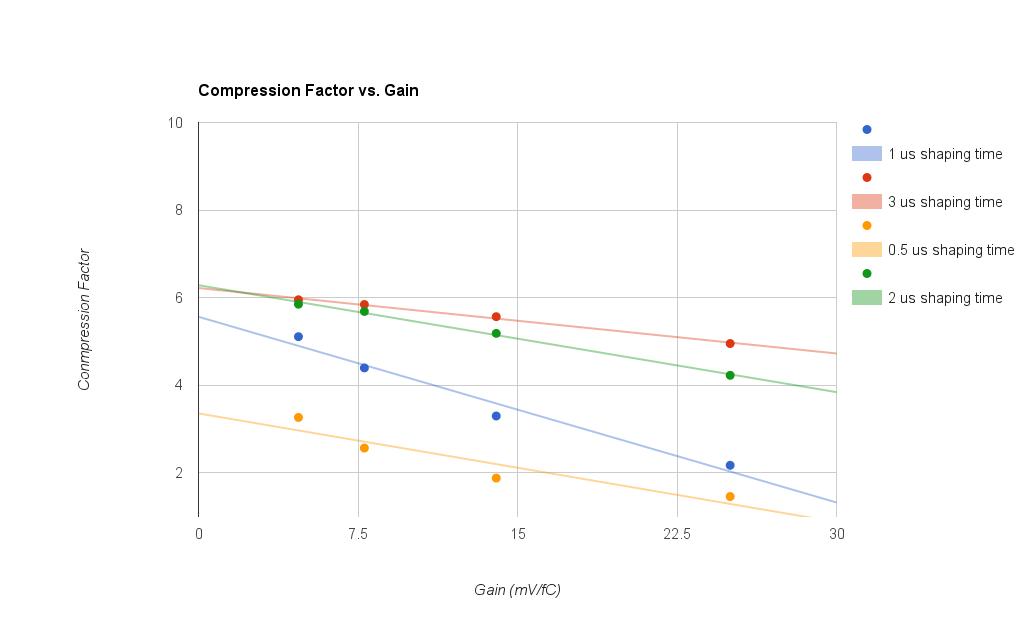}%
\caption{\label{fig:readout_6}Compression factors achieved on ADC data with Huffman compression.  The default shaping time in MicroBooNE is 2~$\mu$s.}
\end{figure}

In the case of the NU data stream, a lossless Huffman coding scheme implemented in the FEM FPGA compresses the data by approximately a factor of 4.5 (dependent on ASIC gain and shaping settings). Further reduction in the overall data rate processed by the DAQ system is achieved by exploiting a PMT trigger in coincidence with the BNB trigger, as described in section~\ref{sec:trigger}. Huffman coding provides for lossless data compression by taking advantage of the slow variation of the waveform TPC data in any given channel. In particular, this compression scheme relies on the fact that successive data samples on any given wire vary relatively slowly in time. As such, when noise levels are low, any two adjacent data samples either coincide or differ by 1 ADC count. The most frequent values for the difference in ADCs between successive data samples are assigned pre-specified bit patterns with the lowest number of bits possible. Those bit patterns are encoded in the 16-bit data words that would otherwise be used for a single 12-bit ADC sample value. As such, data reduction of up to a factor of 14 is theoretically possible (the uppermost two bits are always reserved for header information, in each 16-bit word). In practice, the data reduction is sensitive to noise levels and \lartpc activity, and is also dependent on the gain and shaping time setting. The compression factor achieved during commissioning by MicroBooNE is shown in figure~\ref{fig:readout_6}.

Because of the low trigger rate (the BNB trigger dictates an upper bound on the trigger rate of 15 Hz), lossless Huffman coding compression proves sufficient for the NU data stream. However, for the continuous SN stream, further compression becomes necessary, resulting in unavoidable data loss. A method called ``dynamic decimation'' (DD) handles this case. The DD scheme relies on recognizing regions of interest (ROI) in the data stream that contain waveforms corresponding to drift ionization charges. Portions of the data stream not containing ROI contribute to pedestal determination, and ROI are identified as deviations from the continually-updated pedestal, buffered, and read out to disk. At the time of this writing, the MicroBooNE SN stream compression scheme is being finalized and the SN readout stream is being commissioned.

\subsection{PMT Readout Electronics}

The PMT readout electronics are responsible for processing signals from the 32 PMTs described in section~\ref{sec:light-collection} and identifying light signatures coincident with the BNB and NuMI beam spills. The coincidences generate PMT triggers that can be later mixed with other triggers in the Trigger Board (TB). Signals from the four lightguide paddle guides installed in the \lartpc are also recorded by the PMT readout electronics, but these signals do not participate in the PMT trigger generation. 

\begin{figure}
\centering
\includegraphics[width=0.8\linewidth]{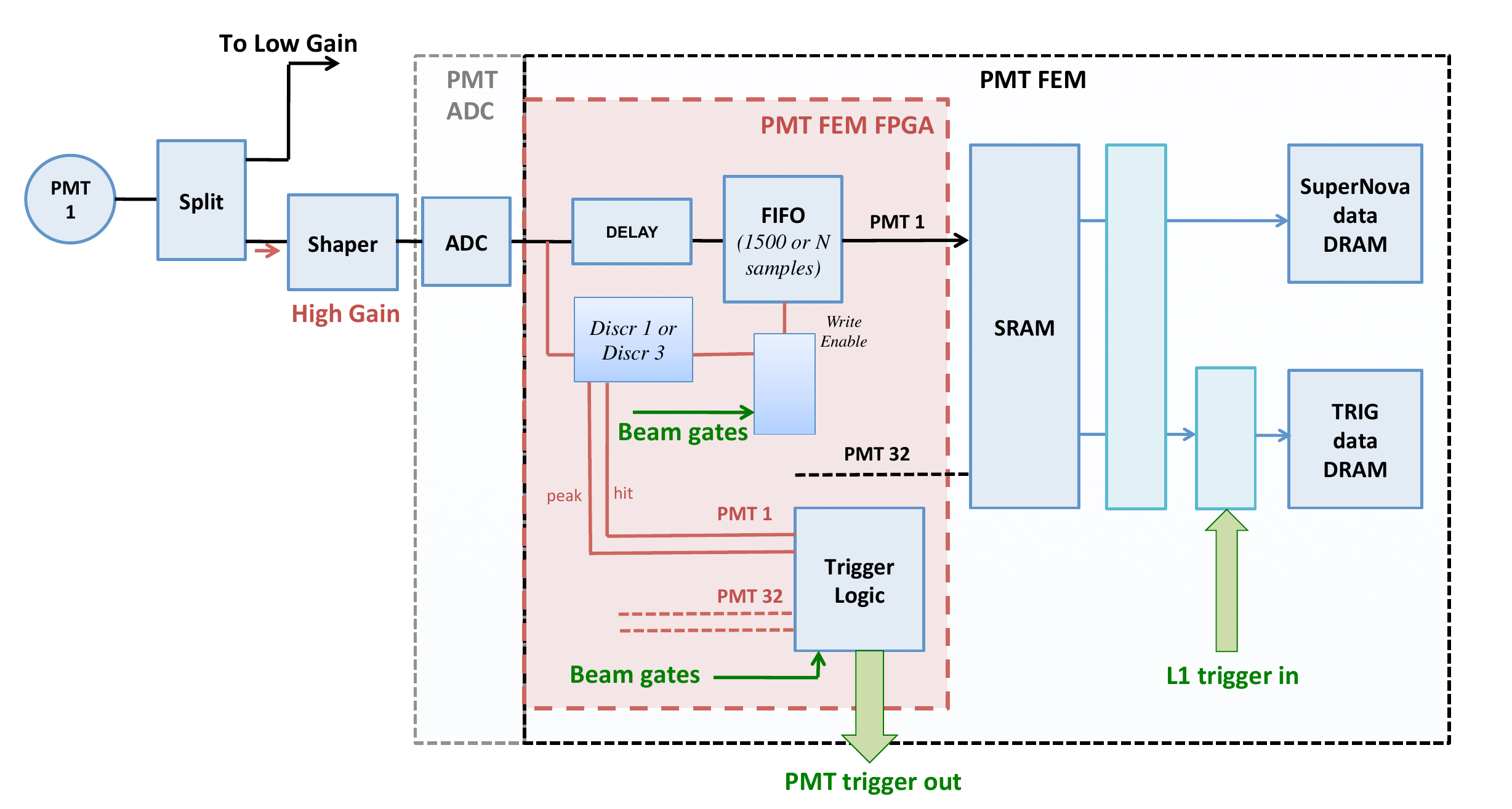}%
\caption{\label{fig:readout_7}PMT signal processing stages, and digital signal processing in the PMT FEM. Each PMT signal is discriminated and gated (in the presence of an external gate, such as a beam gate), and primitives such as hit amplitude and hit multiplicity are used to construct a PMT-based trigger. All discriminated/gated data is transferred continuously to a DRAM for the SN stream. When there is a Level-1 trigger, data corresponding to four frames, including the trigger frame and one/two frames preceding/following the trigger frame, are transferred to a DRAM for the NU data stream for readout by the DAQ.}
\end{figure}

The stages of signal processing are illustrated in figure~\ref{fig:readout_7}. First, each PMT signal (with the exception of the light paddle guide signals) is split into two different gains, as described in section~\ref{LCUnitimplement}, with the HG channel carrying 18$\%$ of the PMT signal and the LG channel carrying 1.8$\%$ of the PMT signal. Each gain is split once again into HG1 and HG2, and LG1 and LG2, in order to allow more flexibility in separately processing beam-related and beam-unrelated PMT signals. All 32$\times$2$\times$2 (PMT) plus 4 (light guide paddle) signals are pre-amplified and shaped in 16-channel pre-amp/shaper boards (section~\ref{sec:PMTamp}). Three PMT readout modules receive the analog shaped signals differentially and digitize them (section~\ref{sec:pmtdigit}) at 64 MHz. The PMT readout modules then process the signals in order to prepare them for shipping to a designated DAQ machine and to form a possible PMT trigger (section~\ref{tpcfem}). 

Each one of the three PMT ADC+FEM readout boards used in the PMT readout system handles one of the following:

\begin{itemize}
\item Readout of and PMT trigger generation using the HG1 PMT signals associated with neutrino beam events. The paddle signals are also readout by this board but they do not participate in the trigger generation. 
\item Readout of and PMT trigger generation using the HG2 PMT signals that are out of beam time (i.e. cosmic rays and other cosmogenic backgrounds). 
\item Readout of the LG1 PMT signals associated with neutrino beam events and signals that are out of beam time.
\end{itemize}

After signal processing, the data is sent to a designated DAQ machine via a transmitter (XMIT) module in the same way as is done for \lartpc data. Two data streams are provided: a NU data stream associated with event triggers and a SN data stream which a continuous version of the NU stream readout.

The PMT readout and trigger electronics share the same 16~MHz clock as the LArTPC readout electronics, and likewise keep track of time since run-start in 1.6~ms-long frames. All PMT readout electronics are housed in a single VME 6U crate.

\subsubsection{\label{sec:PMTamp}PMT Signal Amplification and Shaping}

The preamp/shaper boards read raw PMT signals from the PMT HV/signal splitters and shape them into unipolar signals with a 60 ns rise time. The shaped signals are sent to the \lartpc readout boards differentially via short front-panel cables, in order to minimize noise, where they are digitized at 64 MHz. The 60 ns peaking time allows digitization of two or three samples on the rising edge. This in turn enables an accurate determination of the event $t_0$ needed to determine the $x$ coordinates of ionization signals along the drift direction. An accurate time measurement also helps reject other tracks, such as cosmic rays, that cross the detector during the drift time.

\subsubsection{PMT Data Digitization}
\label{sec:pmtdigit}

\begin{figure}
\centering
\includegraphics[width=0.8\linewidth]{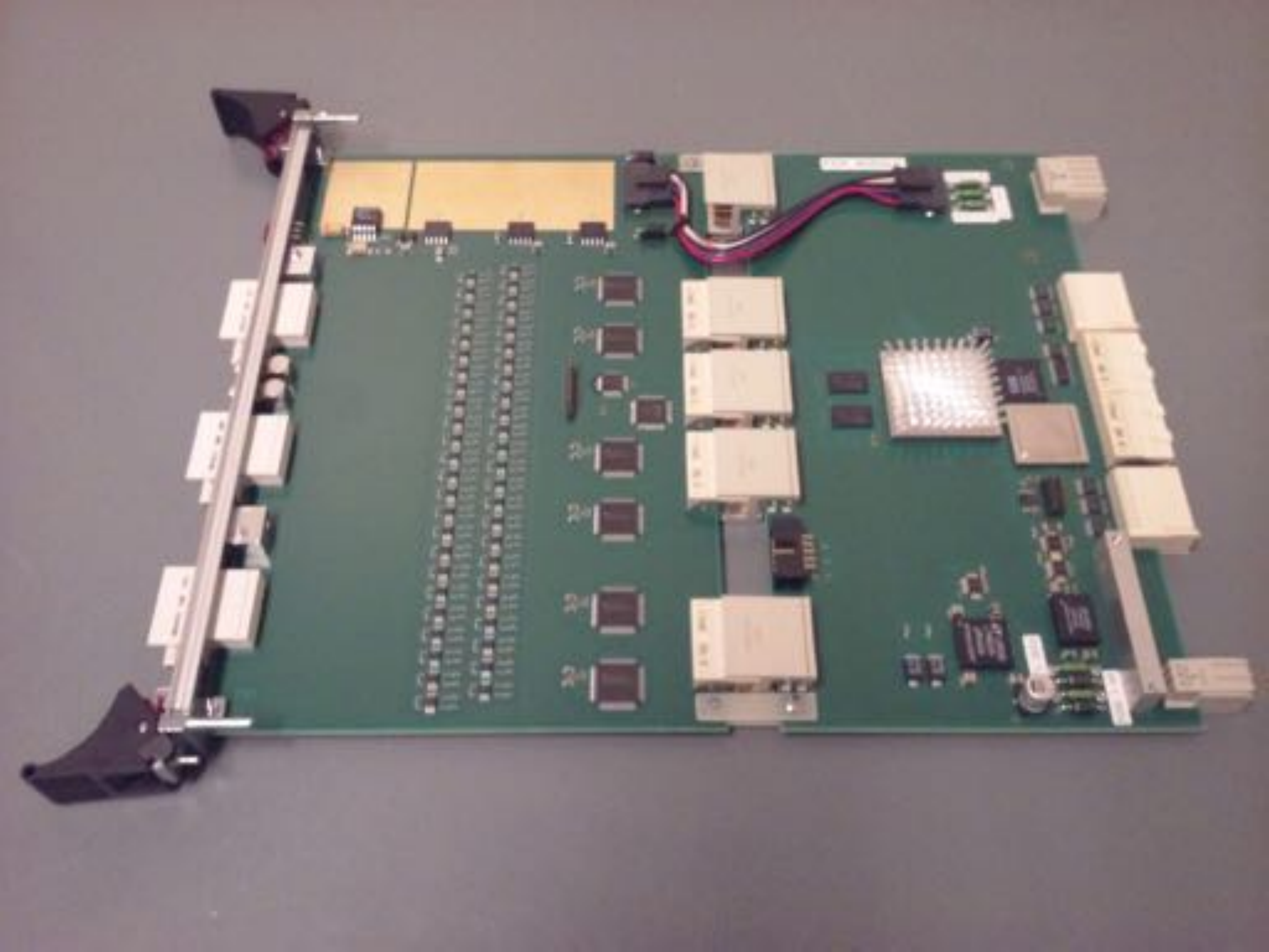}%
\caption{\label{fig:readout_8}The MicroBooNE PMT readout board which digitizes 48 input signals.}
\end{figure}

The ADC (Texas Instruments, ADS5272) module part of the PMT readout board (figure~\ref{fig:readout_8}) is responsible for digitization of up to 48 differentially-driven input signals. The differential signals are digitized at 64~MHz. The 64 MHz clock used by the PMT readout is generated starting from the 16 MHz clock that is common to all readout crates (\lartpc and PMT). 


In addition to HG and LG waveforms from the PMTs, each readout board also receives, digitizes, and processes beam gate signal markers which arrive 4 $\mu$s before the BNB (1.6 $\mu$s) and NuMI (10 $\mu$s) beam gates. These gates are used to (a) specially mark regions of interest where PMT data are read out continuously with no compression and (b) look for coincident PMT light signatures for trigger generation. 

\subsubsection{PMT Data Handling and PMT Trigger Generation}
\label{tpcfem}

PMT information is recorded in the NU data stream for four 1.6 ms frames associated with an event trigger: the frame containing the (asynchronous) trigger, the frame preceding the trigger frame, and two frames following the trigger frame. To avoid the inordinate amount of data that would be generated at a 64 MHz sampling rate, the FEM applies a zero-suppression immediately after digitization, retaining only samples above a given threshold as well as enough information before and after this useful data to establish a local baseline value; this collection of information is referred to as a PMT readout ROI. An exception is formed for beam-related or other likewise-triggered data where, for example,  the 4 $\mu$s-early BNB and NuMI beam gates mentioned in section~\ref{sec:pmtdigit} instruct readout of 1500 consecutive samples (23.4 $\mu$s) surrounding and including the neutrino beam spill period regardless of signal activity.

Two different discriminators are used: one that is active inside the beam spill period(s), and one that is also active outside the beam-spill-surrounding 23.4 $\mu$s. The latter discriminator governs the readout activity due to cosmic rays and other non-beam related activity. The former discriminator enables PMT channels with pulse heights above a configurable threshold (e.g. corresponding to 1 photoelectron)  to participate in trigger multiplicity and pulse height sum conditions, as described in the following section. The thresholds for those two discriminators are set to different levels and configured with different dead times for the HG and LG signals.

\subsection{Level-1 Trigger Generation}
\label{sec:trigger}

The MicroBooNE Trigger Board (TB), which physically resides in the PMT readout crate, issues a ``Level-1'' trigger in order to flag frames that must be treated differently; in the case of the \lartpc readout, the TB flags the 4 frames that must be trimmed and readout through the NU data stream and, in the case of the PMT readout, it flags the 4 frames that must be readout in full through the NU data stream.

The inputs to the TB include a BNB trigger input (maximum rate of 15 Hz), a NuMI trigger input (1.25 Hz), a Fake Beam trigger input (configurable frequency), a PMT trigger input, and two calibration trigger inputs, provided by the laser calibration system and a cosmic ray muon tracker, respectively. The TB also has the ability to receive, via the crate controller, DAQ-issued (via software) calibration triggers, which are used explicitly for cold electronics and PMT calibration. The various input triggers can be independently pre-scaled, masked, and mixed together (OR or AND) to generate an event trigger, referred to as a Level-1 trigger.

The FPGA firmware in the PMT FEM can generate two different types of PMT triggers based on the PMT signals: a cosmic PMT trigger and a beam gate PMT trigger. Beam gate PMT triggers are configured in the same way for the BNB, NuMI, and Fake Beam. The nominal criteria for these triggers are (1) PMT multiplicity $\ge$1 and (2) summed PMT pulse-height $\ge$ 2 PE summed over all 32 HG1 PMT channels. Both criteria must be met during any 100 ns time interval coincident with the beam spill duration (1.6 $\mu$s in the case of the BNB and Fake Beam gates and 10 $\mu$s in the case of the NuMI gate), and only channels enabled by the beam gate discriminator can participate in the active pulse-height and multiplicity sums. The criteria for a cosmic PMT trigger are (1) PMT multiplicity $\ge$1 and (2) summed PMT pulse-height $\ge$ 40 PE summed over any one of 28 preset groups of 5 HG2 PMT channels that are grouped based on their spatial correlation. Again, only channels enabled by the cosmic discriminators can participate in the trigger generation.

In addition, a software-based algorithm has been written to mimic the capabilities of the beam gate PMT trigger performed in the FPGA and provide more flexibility in trigger criteria settings. Details of this higher-level software trigger are described in section~\ref{sec:software-trigger}.

\begin{figure}
\centering
\includegraphics[width=0.8\linewidth]{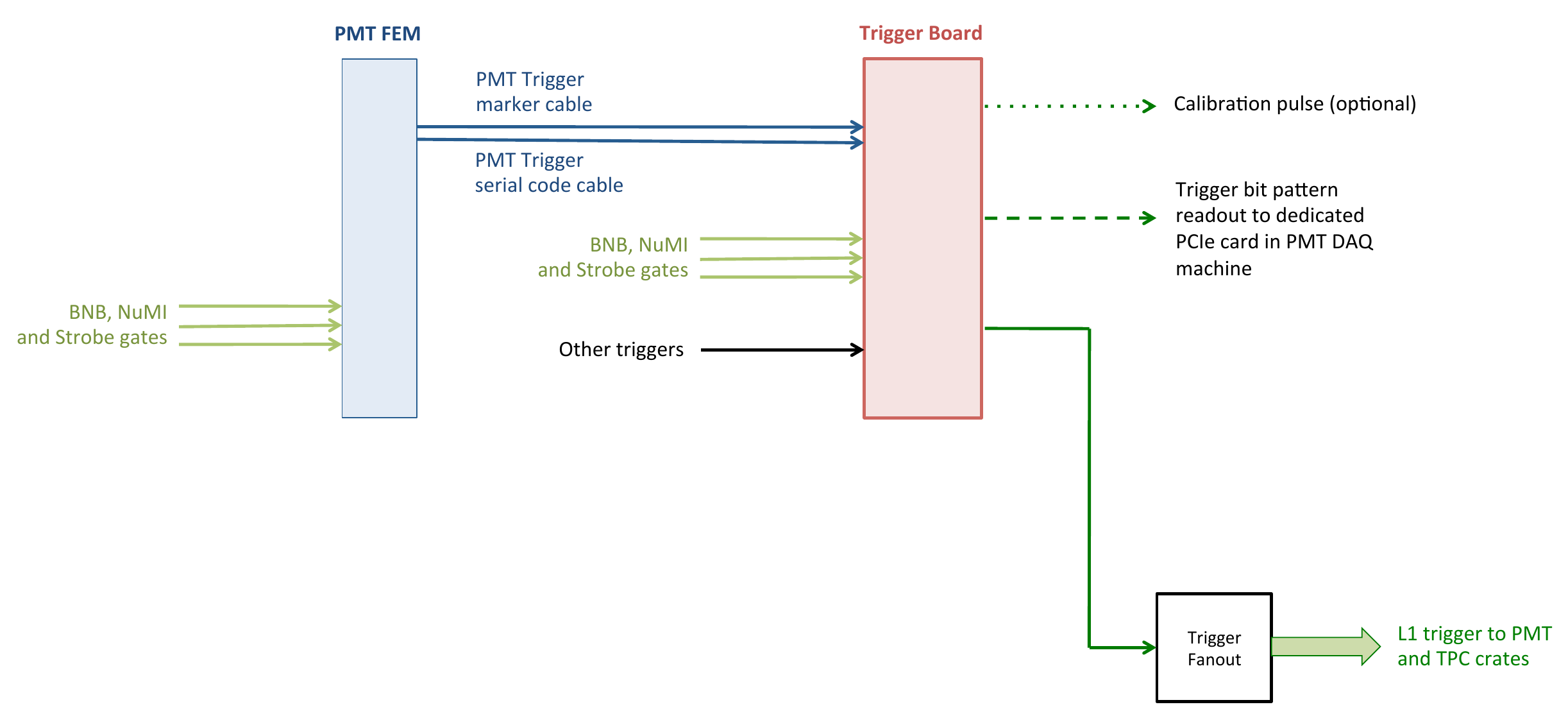}%
\caption{\label{fig:readout_9}PMT readout and trigger logic.}
\end{figure}

Figure~\ref{fig:readout_9} diagrams the PMT readout and trigger logic. Activation or masking of each of the trigger inputs and outputs is DAQ-controlled. The trigger condition and explicit PMT trigger type, if applicable, is available for every event in the NU data stream at both the event-building stage and offline; this information is read out via a dedicated optical data stream, directly from the TB. The trigger number and trigger time are also propagated and available in the NU data streams sent independently from each \lartpc and PMT crate and ``sub-event'' DAQ machine to an ``event assembler'' DAQ machine. They can therefore be used to correctly associate data from the same event.

\begin{figure}
\centering
\includegraphics[width=0.8\linewidth]{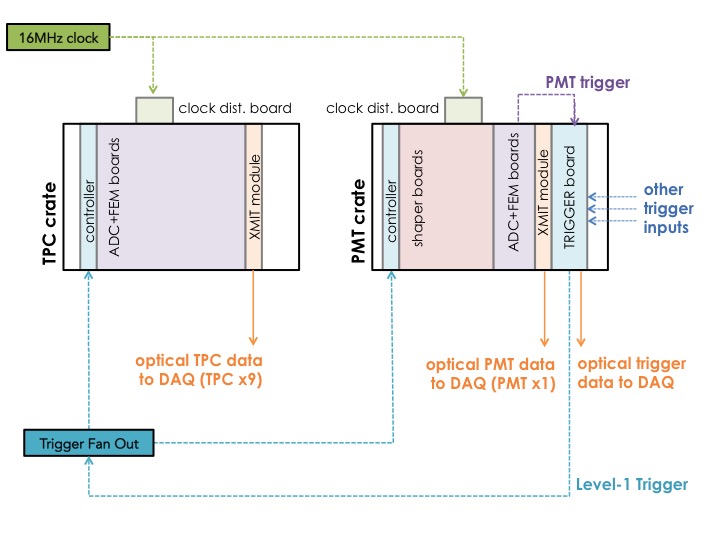}%
\caption{\label{fig:readout_10}Readout control sequence. The MicroBooNE Level-1 trigger is a hardware trigger which consists of the OR between a BNB, NuMI, and Fake Beam (strobe) trigger. Once received, the Level-1 trigger is propagated to all readout crates and instructs PMT and TPC data readout into ten dedicated sub-event buffer DAQ machines. The data across different DAQ machines is correlated at event building stage by the trigger number and corresponding trigger frame and sample numbers recorded in each data stream, per event. All readout crates are synchronized and correlated to the same 16 MHz clock.}
\end{figure}

The overall readout control sequence is illustrated in figure~\ref{fig:readout_10}. When a trigger is generated by the TB it is passed to a fan-out module on a single cable and from there it is distributed to all crate controllers (\lartpc and PMT). Through the crate backplane, the trigger gets propagated to each FEM. An FEM that receives a trigger temporarily inhibits the SN stream with its associated decimation and initiates the loss-less readout scheme to direct the data to the appropriate readout path. SN readout resumes once the XMIT is done sending all NU data associated with an event to the DAQ. 

\subsection{DAQ Design}
\label{sec:daq}
The MicroBooNE DAQ system acquires data from the readout electronics, writes data to local disk before transferring it to long-term storage, configures and controls the readout electronics during data-taking periods, and monitors the data flow and detector conditions. These tasks are performed on a network of commodity servers running both custom and open-source software.

The data from each crate of the backend electronics is sent to a dedicated server (called the sub-event buffer, or SEB) via optical fibers, arriving in dedicated cards on the SEB's PCIe bus. A real-time application places these data in an internal buffer, collects all segments belonging to an event, and creates a sub-event fragment that may be routed to a specified destination.  For the NU data readout stream, in which the data arrives with every Level-1 trigger, these fragments are sent to a single event-building machine (EVB) over an internal network. Full events are checked for consistency and written to local disk on the EVB before being sent offline for further processing. A high-level software trigger, described in section~\ref{sec:software-trigger}, is applied to the data to determine whether events should be written locally or ignored. For the SN stream the data remains on the SEB's where it is written to disk and only sent for offline analysis on explicit requests, e.g.~on receiving a Super Nova Early Warning System (SNEWS) alert \cite{Scholberg:2008fa}.

Data writing to either triggered or SN streams is limited by the RAID6 disk write speeds which are roughly 300 MB/s. This is much less than the network bandwidth bottleneck, which is 10 Gb/s. The 300 MB/s disk write speed therefore sets the maximum aggregate rate at which all SEB fragments can ship data to the EVB without loss of data. With Huffman compression, which gives a data reduction of approximately a factor of five (figure~\ref{fig:readout_6}) and the PMT trigger, which reduces the data rates by another factor of $>70$, this is more than sufficient for MicroBooNE's maximum 15 Hz beam spill rate. MicroBooNE expects a total triggered write rate of around 12 MB/s. The SN stream circular buffers, which are aggressively (non-losslessly) compressed beyond what the NU stream experiences, fill each SEB's 14~TB in on the order of one day, which is ample time to respond to a SNEWS alert.

\begin{figure}
\centering
\includegraphics[width=0.95\textwidth]{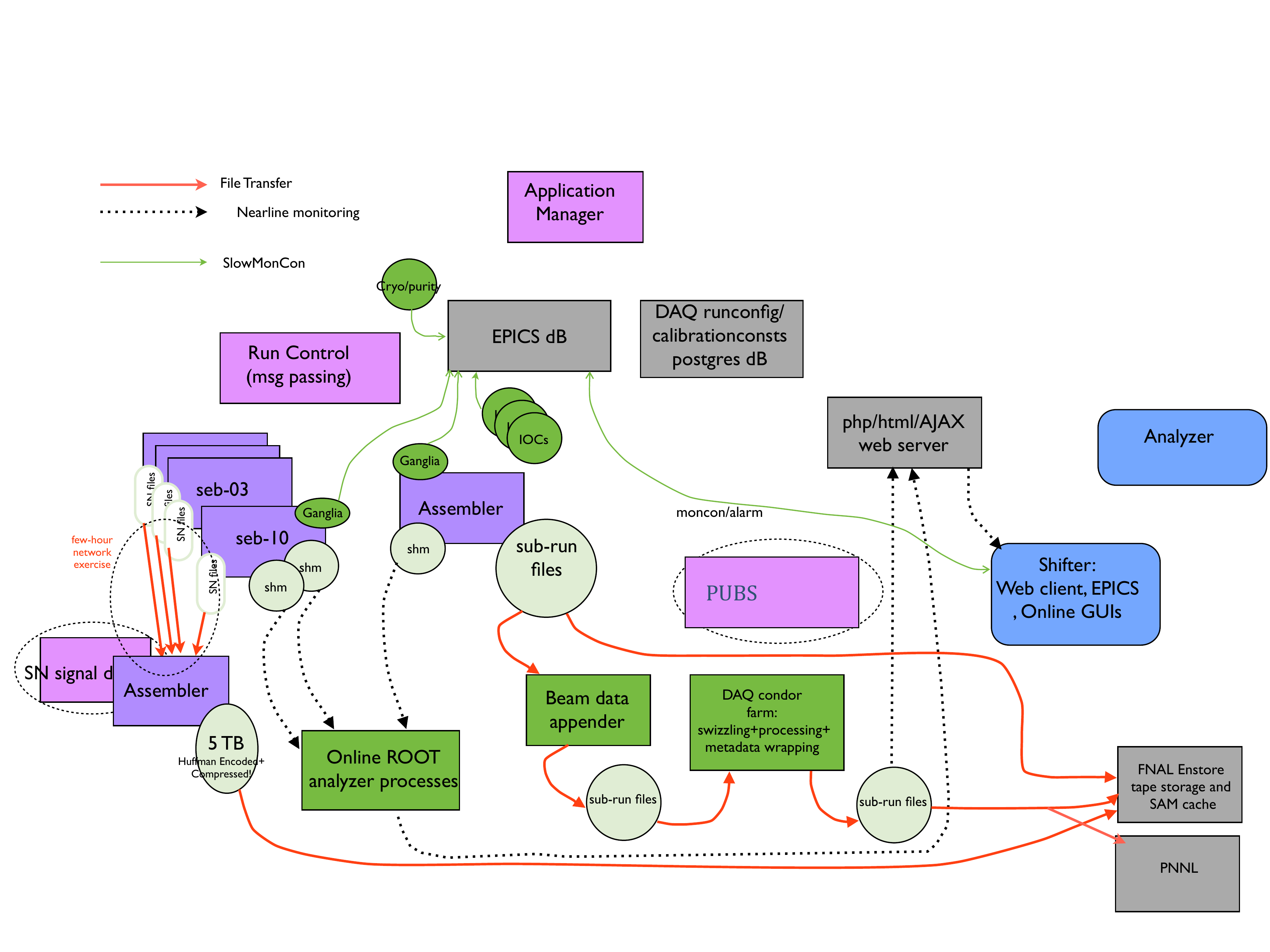}
\caption{MicroBooNE DAQ data flow from raw to processed.}
\label{fig:dataflow}
\end{figure}

After data is written to disk, it is then copied to another server on the internal DAQ network, where the raw data is further compressed, shipped, and queued to be stored on tape and disk cache using the Fermilab central data management system known as SAM.  Offline applications then begin processing the raw data, converting the binary data format into a LArSoft \cite{larsoft} ROOT-based format which can be used as input for reconstruction algorithms.  LArSoft is a common framework of software tools utilized by many \lartpc experiments at Fermilab and elsewhere.  A separate process collects beam data and, during binary to LArSoft conversion, inserts that data into the built events. A duplicate copy of the data is also stored offsite at Pacific Northwest National Laboratory (PNNL). This collection of approximately 15 ``projects''  and the database which holds and monitors the state of the data flow is known as the Python/Postgres for MicroBooNE Scripting system (PUBS), and is patterned after a similar database state machine that the Double Chooz experiment used for data management.  As PUBS pushes the data through this process, the progress of each project is monitored and viewable via GUI. PUBS can also monitor the state of the SN stream data, held locally on the SEBs.  A separate offline PUBS instance controls the processing of the data, including applying newly calculated calibration constants as part of data quality management.  Figure \ref{fig:dataflow} schematically depicts the flow of data throughout the MicroBooNE DAQ system.

Additional software components handle the management of the main DAQ processes moving the data. A run control application issues configuration and state-progressing commands to the SEBs and EVB. Configuration states are stored in a dedicated run configuration database, which allows for the setting and preserving of configuration information for the DAQ, readout, and additional components. This database not only allows for creating the large ($\sim$200 parameters) intricate DAQ run configuration files, but also enforces certain conditions which must hold for consistency. For example, the configuration that initiates the ASICS charge-injection calibration also dials and captures the settings on the external pulser that drives the calibration signal and assures that ASICS gains and peaking time parameters are enforced and recorded. 

\begin{figure}
\centering
\includegraphics[width=0.45\textwidth]{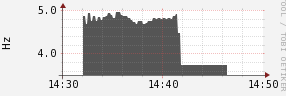}
\includegraphics[width=0.45\textwidth]{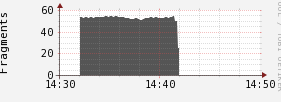}\\
\includegraphics[width=0.45\textwidth]{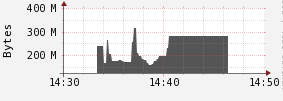}
\includegraphics[width=0.45\textwidth]{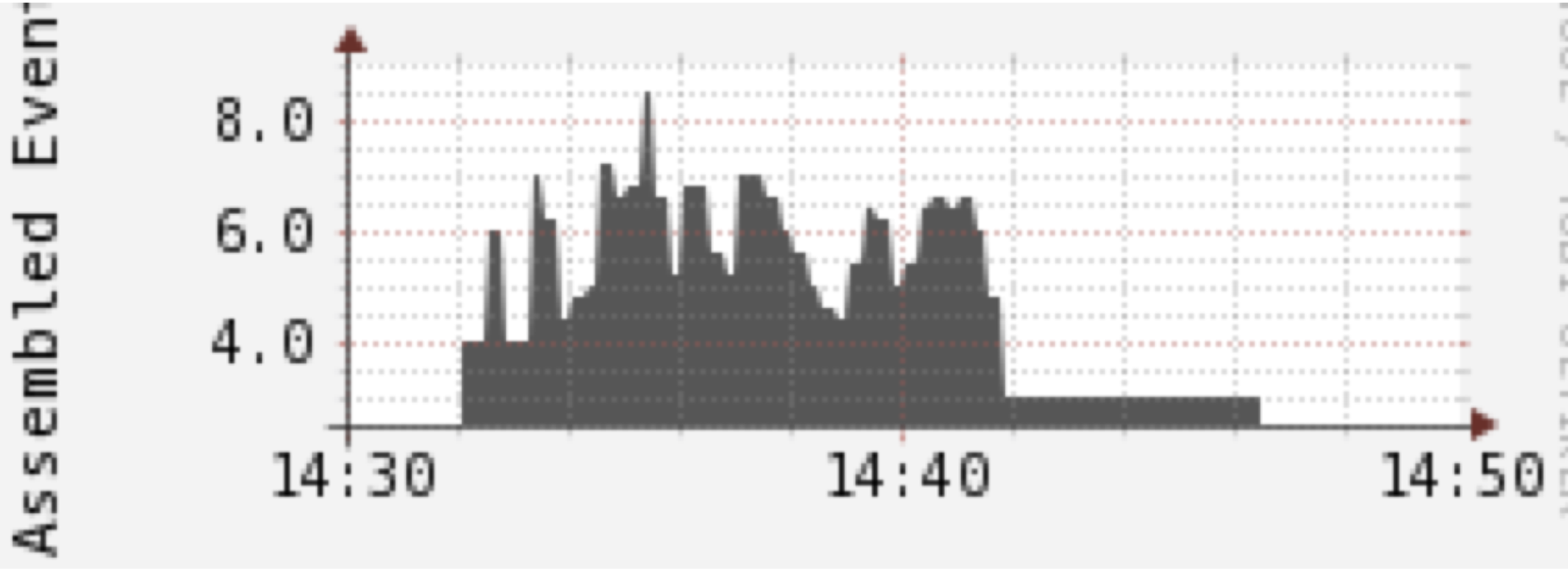}
\caption{Ganglia metrics showing data flow on the EVB machine during a 5 Hz test run. Shown as a function of time are: the external trigger rate (top-left), the received fragment rate (top-right), the disk data write rate (bottom-left), and the number of assembled queued-up events waiting to be written (bottom-right). (11 fragments constitute a complete event in the MicroBooNE DAQ.)}
\label{fig:ganglia}
\end{figure}

Another important aspect of the system is monitoring the health of the DAQ. Monitoring of DAQ components is accomplished through Ganglia which monitors basic system states (such as CPU, memory, and network usage) as well as allows use of custom metrics to monitor the data flow and status of the readout electronics \cite{GangliaBook}. These metrics are sampled and collected by the Experimental Physics and Industrial Control System (EPICS) slow monitoring and control processes, which archives desired quantities and provides alarms when pre-defined thresholds are exceeded \cite{EPICS}. Some examples of Ganglia metrics that are monitored and alarmed in EPICS are the rates of growth of the SEB data buffers, the fragment rates leaving each SEB, and fragment arrival rates at the EVB.  Figure \ref{fig:ganglia} shows examples of Ganglia metrics.  

\begin{figure}
\centering
\includegraphics[width=0.95\textwidth]{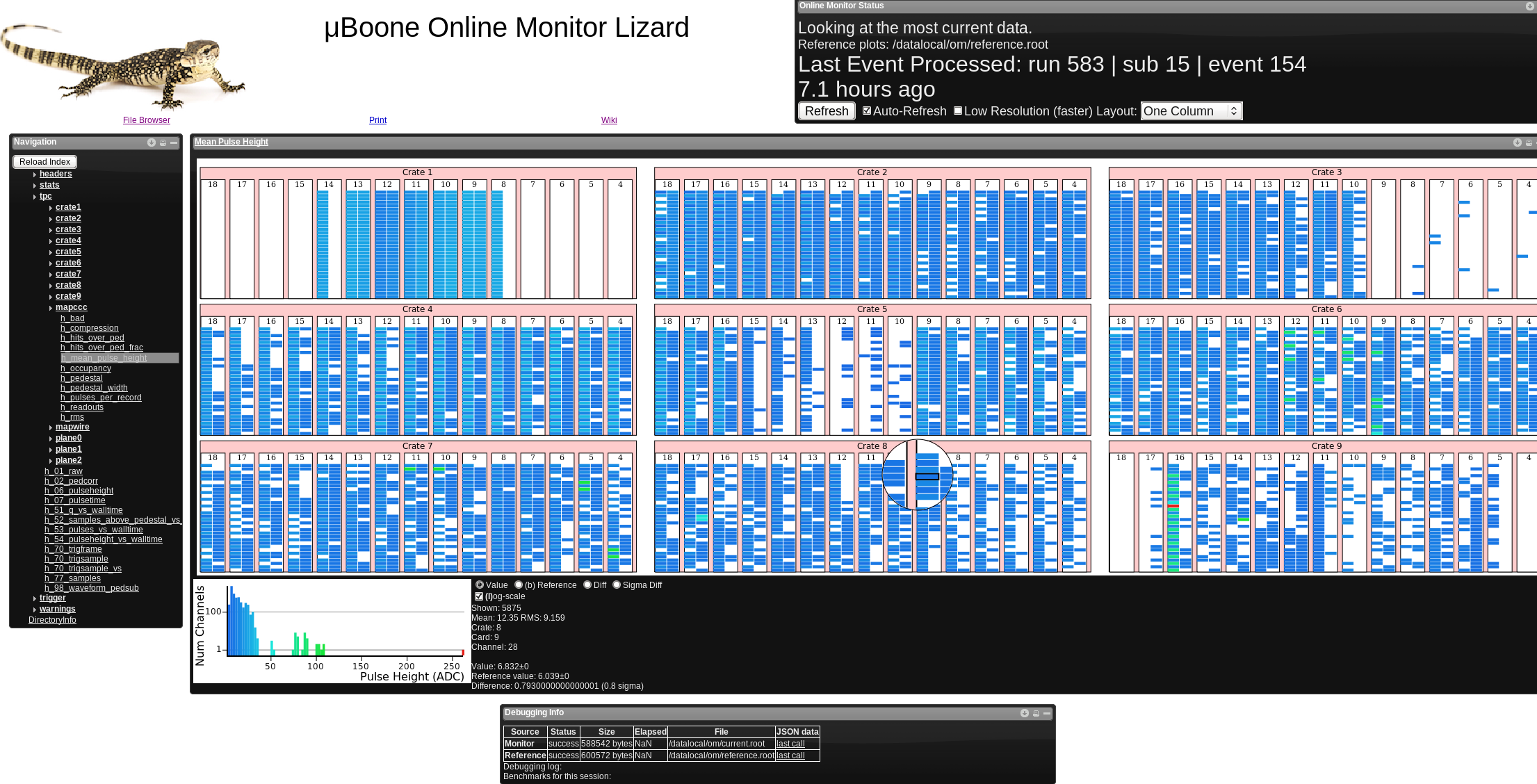}
\caption{The online monitoring GUI (Lizard). The pedestal-subtracted ADC values for all 8256 wires in one minimum bias event are shown. Each box is a channel. Not all crates are fully populated with electronics. }
\label{fig:onlinemonitor}
\end{figure}

Additional online monitoring exists to check data quality in more detail, through both programmed checks and visual checks including a real-time event display. The online monitoring takes snapshots into shared memory segments on the SEBs and the EVB, and thus provides the desired low latency checks of newly-arriving data. It continually walks through these $\sim150$~MB snap-shotted events and outputs histograms of occupancies and rates which are saved in ROOT files \cite{BRUN199781}. The histograms are then displayed in a web-based monitoring system that is easily accessible by the MicroBooNE shift crew. Channels are aggregated in a variety of formats, including the order in which they appear in crates or across the wires and PMTs themselves. In this way, potential problems across connectors or crates, for example, may be more readily identified. Noisy, quiet, and unresponsive channels are easily marked and displayed to the shift crew.  Figure \ref{fig:onlinemonitor} shows an example of available online monitoring information.

\subsection{High-Level Software Trigger}
\label{sec:software-trigger}
After events are collected on the DAQ event-builder server, a suite of trigger algorithms are applied to the data. Currently, these algorithms mimic the PMT readout electronics' FPGA-based beam gate trigger algorithms, described in section~\ref{sec:trigger}. A search over the PMT digitized waveforms from the beam-spill period is performed, and if there is a significant amount of light in coincidence with the expected arrival time of the neutrinos, the event is marked and saved. This selection is performed on data that passes the Level-1 trigger, which typically includes data from the BNB and NuMI beams, and randomly selected off-beam data from an ``external'' (or fake beam) trigger. A fraction of the data from each of these Level-1 trigger input streams is also retained via a random prescale, which provides a selection of data that has not been biased by the trigger. The high-level trigger algorithms take approximately 10 ms to return a result, a latency that is well-below the event-taking rate and so does not impact data-taking performance. The average pass rate for data-events in the PMT beam gate trigger algorithm is roughly 5\%.

\newpage
\section{Infrastructure and Monitoring Systems}
\label{sec:slow-control}

MicroBooNE is housed at LArTF, which is located in the BNB at Fermilab.  Complete knowledge of the electrical and cryogenic systems housed within LArTF is necessary to maintain acceptable operating conditions for the experiment.  Continuous monitoring of the beam being delivered to LArTF is also necessary for subsequent physics analyses.  This section describes the details of infrastructure within LArTF, as well as monitoring of the experiment and beam conditions.

%
%
%
%
%

\subsection{Electronics Infrastructure at LArTF}
\label{sec:lartf}

\begin{figure}
\centering
\includegraphics[width=0.8\textwidth]{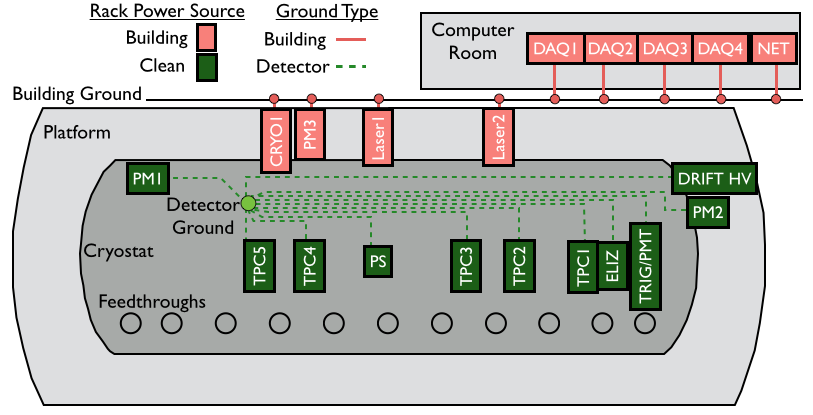}
\includegraphics[width=0.6\textwidth]{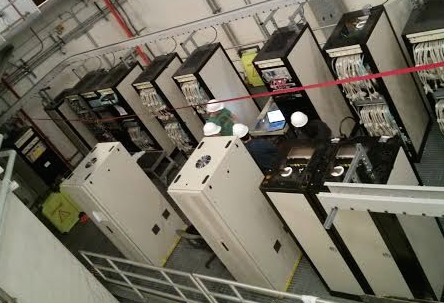}
\includegraphics[width=0.8\textwidth]{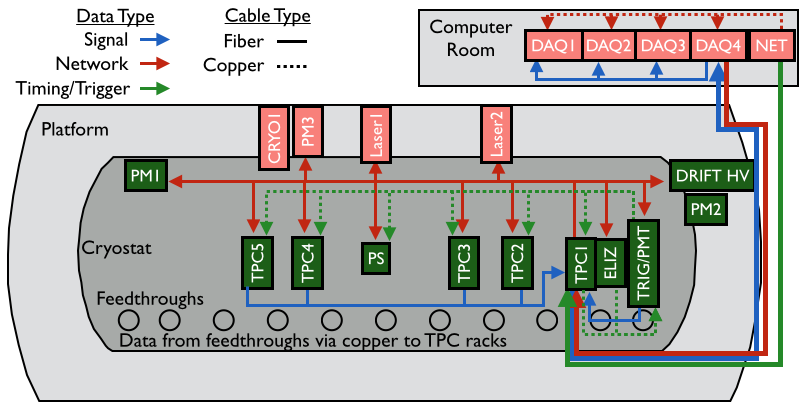}
\caption[]{Top: Diagram illustrating location of deployed electronics racks and separation of detector and building grounds and clean and building power for differing racks.  Middle: Photograph of installed electronics racks on the LArTF platform.  Bottom: A diagram illustrating the general scheme of signal, network, and timing signal cabling in the LArTF computer room and platform.}
\label{fig:ElecInfDiagrams}
\end{figure}

Figure~\ref{fig:ElecInfDiagrams} shows a diagram of the monitoring system at LArTF, depicting racks located on a platform directly above the cryostat that house electronics for: \lartpc control and readout, light collection system, drift high voltage, purity monitors, calibration laser, trigger, and cryogenic control systems.  Additional server racks containing the DAQ servers, beam timing, and external network electronics are located in a separate computer room above and adjacent to the above-cryostat platform.  The distribution of power, data, and network connections to and from all of these racks is also presented in figure~\ref{fig:ElecInfDiagrams}, and is described in more detail in the following sections, along with electronics safety systems and interlocks.

\subsubsection{AC Power Distribution and Grounding for Low-Noise LArTF Data-taking}
As with any large detector operating with a high dynamic range, prevention of electromagnetic interference and its attendant effects on MicroBooNE data is an essential aspect of detector design.  MicroBooNE's strategy for producing a low-noise environment for the \lartpc and associated readout electronics can be largely summarized in a few key points.  AC power distribution-related items will be described here, while cabling, connections, and shielding will be described in a following section.

\begin{itemize}
\item{``Clean power,'' or AC power electrically isolated from AC power for the rest of LArTF (``building power''), is supplied to all sensitive electronics via an isolation transformer.}
\item{Highly sensitive electronics are housed inside the Faraday cage provided by the detector cryostat or inside Faraday cages directly grounded to the cryostat.}
\item{The detector cryostat is grounded to a ``detector ground,'' which is physically and electrically isolated from the ground provided to all other LArTF power circuits, or ``building ground.''}
\item{No direct electrical connections are present between detector ground and building ground.  This is accomplished through the use of insulating platform and cryostat saddle materials, insulating cable trays and cables, and by inserting insulating ``breaks'' (i.e. fiber data links or insulating cryo pipe sections) when connections between sensitive and potentially noisy detector components are necessary.}
\item{Indirect pickup on clean signals through capacitive coupling to adjacent noise sources is minimized through use of detector-grounded shielding and electrically-insulating cable trays.}
\item{Ground loops on detector ground are avoided wherever possible by connecting all electronics racks directly to the cryostat and by minimizing direct electrical connections between racks.}
\item{Direct or capacitive couplings between building and detector ground are constantly monitored during installation and operation with a custom-designed impedance monitor.}
\end{itemize}

\begin{figure}
\centering
\includegraphics[width=0.55\textwidth]{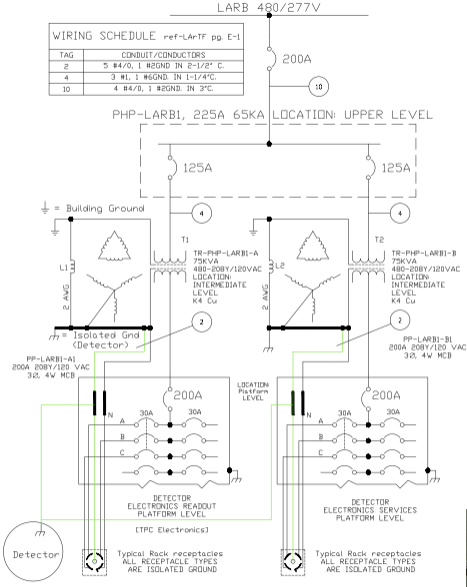}
\caption[]{Line drawing of clean power generation and distribution and connections to detector ground.}
\label{fig:ElecInfAC1}
\end{figure}

A line drawing describing the production of clean power and clean ground is shown in figure~\ref{fig:ElecInfAC1}.  A pair of 200 A clean power circuits produced at isolation transformers are used to power all sensitive racks, which are indicated in figure~\ref{fig:ElecInfDiagrams}.  All racks containing \lartpc readout electronics are placed on one circuit, while all other sensitive equipment is placed on the alternate circuit.  On the platform, all racks utilize clean power with the exception of the calibration laser and in-line purity monitor racks, which either contain noise-producing elements or support building-grounded components.  All racks in the LArTF computer room utilize building power.

\begin{figure}
\centering
\includegraphics[width=0.55\textwidth]{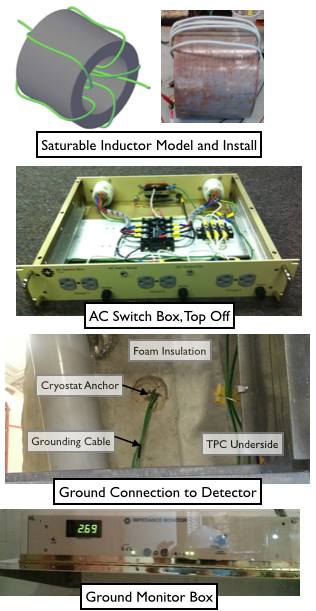}
\caption[]{Photographs of the installed saturable inductor (top), AC switch box (top middle), detector ground strap and connection (bottom middle), and impedance monitor (bottom).}
\label{fig:ElecInfAC2}
\end{figure}

The 208 volt, 3-phase power is distributed to each individual electronics rack.  For racks with significant power requirements or a large number of components, this power is delivered to a Fermilab-designed ``AC switch box,'' which distributes power to an Eaton Power Distribution Unit (PDU) only upon receiving an interlock signal from a smoke detection system in each rack, which will be described in more detail below.  Rack components then receive power from one of the three phases on this PDU.  For racks with fewer requirements, power is supplied to components directly from an interlocked simplified AC switch box or SurgeX SX-1120-RT PDU.  Racks with sensitive electronics are grounded to the cryostat via copper sheeting running throughout insulated cable trays above the cryostat.  Sensitive components within each rack are connected to a tin-plated copper grounding bar electrically connected to the rack bottom and running the height of the rack.  Mechanical attachments to the rack provide grounding for less sensitive rack components.  As mentioned before, any unintentional direct connection between building and detector ground is immediately alerted by the impedance monitor located on the LArTF platform.  Figure~\ref{fig:ElecInfAC2} shows photographs of this equipment.

\subsubsection{DC Power Distribution to the MicroBooNE Detector}
DC power is provided to the \lartpc and readout electronics by power supplies in clean-powered, detector-grounded racks for a variety of purposes including
\begin{itemize}
\item{Holding the \lartpc cathode plane at voltage to produce the desired ionization electron drift speed.}
\item{Holding the two ungrounded anode planes at the proper constant voltage to ensure the planes are transparent to drifting electrons.}
\item{Operating the light collection system PMTs.}
\item{Powering the cold electronics located inside the \lartpc.}
\item{Powering the warm electronics located in the \lartpc and PMT readout electronics racks}
\item{Powering auxiliary systems, such as purity monitors.}
\end{itemize}

Table~\ref{tab:DCPower} summarizes the required voltages or currents for each of these purposes as well as the power supply make and model utilized in each case.  Power supplies are located in the relevant subsystem's electronics rack, with power and grounding connections as dictated by that rack.

\begin{table}[!htb]
	\centering
	  \caption{Overview of MicroBooNE DC power distribution.  Delivered voltages or currents are listed, along with power supply makes and models and whether each supply utilizes clean or building power.}
    \begin{tabular}{llll}
     \hline
      System & Supplied Value & Supply Make/Model & Clean Power? \\ 
      \hline
      Drift High Voltage & 120 kV & Glassman LX150N12 & Y \\ 
      \hline
      Wire Plane Voltage & $\pm$1 kV / 8 mA & Wiener MPOD  & Y \\ 
      \hline
      PMT High Voltage & 2 kV / 3 mA & BIRA T4 & Y \\ 
      \hline
      Cold Electronics Power & 8 V / 10 A & MPOD MPV 800l & Y \\ 
      \hline
      Warm LArTPC/PMT& +/-5 V MDH 2-8 V/25 A& Wiener PL508 & Y\\
       Electronics & +12 V MEH 8-15 V/92 A & & \\ 
      & +3.3 V MEH 2-7 V/115 A & & \\ 
      \hline
       Auxiliary Systems & Various & Various & Laser: N \\
      Power &&& Inline PM: N \\
      &&& Cryostat PM: Y \\ 
      \hline
  \end{tabular}
  \label{tab:DCPower}
\end{table}

DC supply power consumption is minimal in most cases, with the exception of the warm \lartpc electronics PL-508 supplies.  Care was taken to distribute the AC power load by limiting the number of high-draw PC supplies per electronics rack.

\subsubsection{Network, Timing, and Data Distribution for Low-Noise LArTF Data-taking}

Network, timing, and data connections must be made between the detector, building-ground, and detector-ground racks to properly read out MicroBooNE data.  However, as described above, these connections must be made while maintaining strict detector-building electrical isolation.  Deployed interconnections meeting both of these requirements are displayed in figure~\ref{fig:ElecInfDiagrams}.

Timing and LArTF-external network signals are brought into LArTF via electronics in the computer room, where all racks are building grounded and powered.  These signals are distributed and processed in the computer room via copper cable, while network and processed timing signals to be sent to the platform are converted onto fiber cables and aggregated into a central fiber termination box.  A fiber trunk line then delivers these signals to the platform, where another fiber termination box on a detector-ground rack is used to fan out these signals.  Network connections are fanned out via fiber to a network switch in each platform rack, while timing signals are re-converted to copper and further processed for use by the trigger system on a different detector-grounded rack.  All rack-to-rack cables are run in insulating cable trays beneath the platform.

PMT and \lartpc data are transferred from each detector feedthrough to readout crates in detector-ground racks via insulated copper cable whose shield is tied to detector ground.  Digitized crate output is then sent to the aforementioned platform fiber termination box, where these signals are sent via fiber trunk line to the computer room.  In the computer room, these fibers are then fanned out to the appropriate DAQ computer.  Readout crate and cold electronics control commands are transmitted in the opposite direction utilizing a similar scheme, with crate controls delivered directly via fiber, and cold electronics commands delivered via fiber to a copper fanout in a detector-ground rack.

Clock and trigger signals must also be sent from a central trigger rack to all detector-ground LArTPC/PMT readout racks.  These signals are transmitted via copper connections, and represent the only source of ground loops on detector ground.  To further reduce the possible impact of induced noise in these and all copper cables mentioned above, all insulating cable trays beneath the platform are lined with copper sheeting grounded to the detector.  As an additional precaution, all \lartpc signal copper cables are run in separate cable trays from power and auxiliary cabling beneath the platform as well as inside every rack.

All cables between all detector components have been uniquely labelled with serial number, source, and destination to allow for ease of replacement and reconnection.  Ample fiber and copper spares for every major cable type are also installed along with the production cables to allow for quick replacement of any failed cable.

\subsubsection{Interlocks and Safety Systems}
All electronics racks contain smoke-sensing and temperature-monitoring systems, which, when interlocked with AC and DC power transmission in each rack, constitute a rack protection system (RPS) designed to meet Fermilab safety requirements and reduce the risk of fire and related damage in LArTF and to individual rack components.

The RPS principally consists of a smoke sensor connected to a Fermilab-designed rack protection box.  This box produces and outputs a 12~V interlock signal when the rack protection box is on and receiving a ``no-smoke'' signal from the smoke sensor.  This 12~V signal can be sent to the AC distribution box located in each rack, as described above, to allow AC transmission to all rack components only if the RPS is on and not detecting smoke.  A similar 12~V ``RPS Status'' signal is also produced by the RPS box for input into the MicroBooNE slow control box, which will be described in following sections.  Alternate contacts are available on the rear of the RPS box for coupling the status of additional subsystems, such as the DAQ and calibration laser uninterruptible power supply (UPS), to smoke sensor or rack power status.

Temperature sensors deployed in two or three locations in each electronics rack sample air temperature within each rack.  Temperatures at each sensor are read out and recorded in the slow-control database by the slow-control monitoring box.  In addition, the box also produces a 5 V interlock signal if all sampled temperatures are within pre-programmed thresholds.  In electronics racks distributing PMT- or LArTPC-related DC power these temperature interlock signals are input into each relevant power supply, allowing DC power distribution only when this interlock signal is present, for safety purposes.

Additional hardware interlocks ensure the non-simultaneous operation of particular systems.  In particular, the PMT system is disabled when cryogenic system liquid level sensors detect a level below that of the highest PMT bases, or when the UV laser system is active.  The former requirement is enforced with a dry-contact hardware interlock, while the latter is enforced with a software interlock in the MicroBooNE online software.  The UV laser system is also dry-contact hardware interlocked.  Finally, the HV drift power supply is directly interlocked with the cryogenic system controls liquid-level sensor via a dry-contact hardware interlock.

\subsubsection{Performance Measurements}

The proper operation of each production electronics rack's AC and DC distribution and RPS systems has been tested prior to installation at LArTF.  Furthermore, test stands exercising functionality of DAQ, PMT and \lartpc electronics, trigger, and drift HV systems have successfully incorporated and tested various aspects of these same AC and DC distribution and RPS systems.  Impedances between detector and building grounds were recorded throughout the installation of the rack infrastructure at LArTF using the impedance monitor located on the LArTF platform.  

\subsection{Slow Monitoring and Control System}

\begin{figure}
\centering
\includegraphics[width=0.95\textwidth]{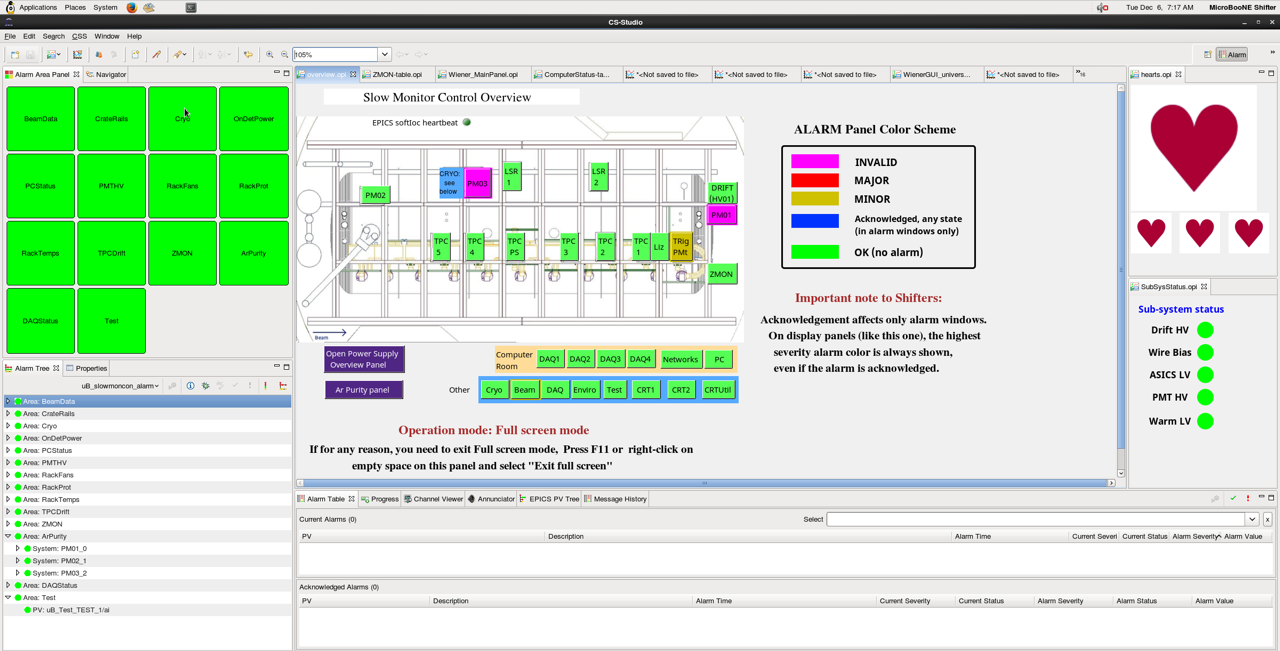}
\caption[]{Standard detector controls and monitoring operator page.  The overview panel (top, middle) provides a ``one-click-away'' access to any sub-system panel of the MicroBooNE detector. The alarm area (top, right), alarm tree (bottom, left) and the alarm table (bottom) panels alert the operator and provide alarm related information at various levels. The 
heartbeats (top, right) show whether programs for important subsystems such as cryogenics, drift high voltage and beam are running. The green LEDs (bottom, right) show the status of important power supplies used in the detector.}
\label{fig:slowmoncon}
\end{figure}

MicroBooNE uses EPICS for controlling and monitoring most devices and conditions important to the experiment.  These include power supply controls, temperatures, fan speeds, rack protection interlock status, and various environmental conditions.  The DAQ, cryogenics systems, and beam data collection systems operate independently of the EPICS slow monitoring, but export data which are imported into EPICS for archiving and status displays.  Applications from the Control System Studio software collection \cite{ControlSystemStudio} are used for providing displays, alarm notifications, and data archiving.  Figure \ref{fig:slowmoncon} is a screenshot of the slow monitoring and controls display.  This system is responsible for monitoring approximately 4500 different variables for MicroBooNE.

\paragraph{EPICS architecture}
An EPICS system consists of any number of server programs implementing the EPICS Channel Access (CA) protocol \cite{EPICS_CAP_Spec} to provide client programs access to any number of process variables, where each process variable represents a quantity being controlled (an output) or measured (an input).  The EPICS base distribution provides a standard type of channel access server called an Input/Output Controller (IOC), which can be extended to support specific hardware as desired.

\paragraph{Power supply controls}
Most power supplies are controllable over the network through the NetSNMP protocol \cite{NetSNMP}.  Several EPICS driver modules are available for SNMP, and MicroBooNE utilizes one written at NSCL \cite{devSNMP}.  An IOC with this SNMP module runs on a central computer and contacts the power supplies over a private network for monitoring and control.  The photomultiplier power supplies are reused from the D\O\  experiment and have custom IOCs running in their own controllers.  The main high voltage power supply has only a simple RS-232 serial interface; control and monitoring for it is provided by a nearby computer running an IOC with the EPICS asynDriver \cite{asynDriver} and StreamDevice \cite{StreamDevice} modules.

\paragraph{Slow controls box}
MicroBooNE has a number of racks in various positions above the detector and in an adjacent server room.  Each is equipped with a rack-protection system and multiple digital temperature sensors, and most contain one or two fan packs, each containing 6 fans.  To monitor and control these devices, each rack has an 1U rack-mount enclosure containing an ARM-based single-board computer (SBC) running Debian GNU/Linux 7 and a custom interface board, collectively known as a ``slow controls box''.  An off-the-shelf GESBC-9G20 from Glomation Inc.~\cite{GlomationInc} is utilized for the SBC.  The custom interface board \cite{HuffmanFanInterfacePage} connects the SBC to front panel LEDs, temperature probes, fan packs, and rack-protection-status input. The temperature sensors are DS1621 chips, controlled and read out over an I2C bus by the SBC's I2C controller.  The DS1621 also has a thermostat output with programmable trip and reset temperatures, which are connected via the interface board to outputs that can be used to interlock devices in the racks, such as power supplies.  The fans provide pulse-per-rotation outputs, which are monitored by a 12-channel tachometer implemented via a PIC16F887 microcontroller, and also read out by the I2C bus.  An EPICS IOC runs in each SBC, with custom device drivers for reading all status information and controlling the heartbeat LED and temperature sensor trip and reset points.

\paragraph{External data sources}
Data are imported into EPICS channels from a number of external sources. The primary reason for duplicating these data in EPICS is to integrate displays and warnings into one system for the experiment operators, and to provide integrated archiving for sampled data in the archived database. An IOC running on a central computer provides ``soft'' process-variables channels for these data.  The data acquisition system provides many metrics describing its operation via the Ganglia system\cite{GangliaBook,GangliaHomePage}, which makes the data available in an XML format easily read by a Python script, which in turn writes to EPICS using the PyEPICS module \cite{PyEPICS}.  The hardware and system status of the DAQ computers is monitored through the industry standard Intelligent Platform Management Interface (IPMI); rather than writing a script to import data from IPMI directly into EPICS, a IPMI-to-Ganglia interface provided by the FreeIPMI's ``ipmi-sensors'' package \cite{FreeIPMI} is used, allowing data to be imported via the same mechanism used for the DAQ metrics.  Separate Python scripts periodically retrieve data about outside weather conditions from various sources, cryogenics system data from a file retrieved non-intrusively from the IFIX cryogenics control system, and beam data from Fermilab's Intensity Frontier Beam Database (IFDB) \cite{IFBeamDB}. 

\subsection{Beam Monitoring}
\label{sec:beam-monitoring}

The primary source of neutrinos for the MicroBooNE experiment is the BNB.  NuMI beam data is also recorded on a spill-by-spill basis.  The primary beamline is lined with instrumentation including toroids which indicate beam intensity, ``multiwires'' showing beam profile in the horizontal and vertical planes, and beam position monitors measuring the mean beam position. Data from these monitors are stored on a spill-by-spill basis in the IFDB. Many of MicroBooNE's physics analyses require that beam data are recorded for each spill and matched to detector events.

Primary beam monitoring in MicroBooNE is done using a ``dashboard'' interface to IFDB. By using the IFBD instead of the accelerator control system, the experiment can also verify that data are being acquired by the IFBD. The dashboard is accessible over the network using a web browser. The final monitoring step includes a post-data-merge check, ensuring that beam data are successfully matched with detector data for all beam spills. This is done once the detector DAQ binary data file is closed. 

\begin{figure}
\centering	
\includegraphics[width=0.95\linewidth]{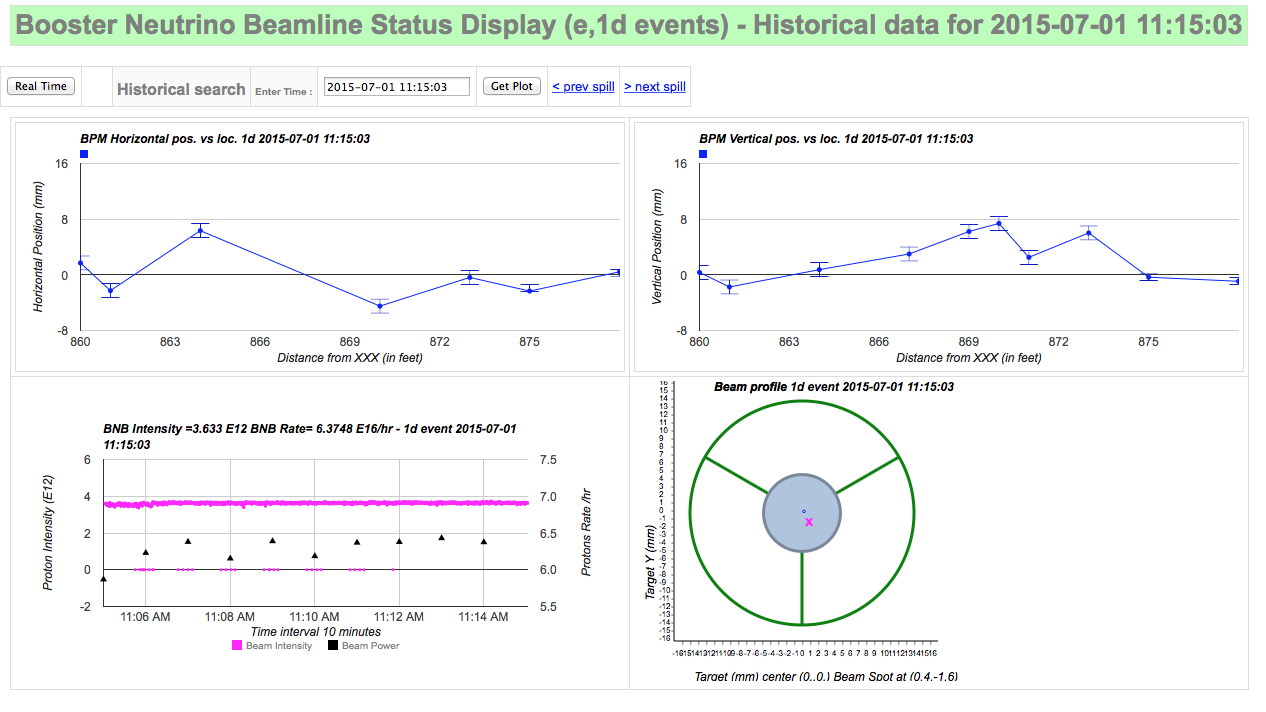}
\caption{The BNB dashboard, showing graphical representation of beam instrumentation data, is used to monitor the beam. The top box shows the timestamp of the beam spill and indicates if data is stale by changing the color. The two top plots show the primary proton beam position along the BNB. The bottom left plot shows the recent beam spill intensity and rate. The bottom right plot shows the beam as projected onto the Beryllium target (the grey circle in the middle with radius of 0.5~cm). The dashboard is accessible via web page providing both real time updates and the review of past data. The page can be easily extended to monitor additional beam devices.}
\label{fig:beammonitor}
\end{figure}

The dashboard presents a graphic representation of the data, allowing for easy error identification, as shown in figure \ref{fig:beammonitor}.  The experiment monitors two toroids, which indicate beam intensity; three multiwires, each of which shows beam profile in each plane; and beam position monitors along the beamline, which show the vertical and the horizontal position.  Parameters pertaining to the target and horn, such as cooling air temperature and horn current, can also be monitored.  The dashboard allows the experiment to easily add additional devices if experience demonstrates the need for their monitoring.

Data are monitored in near real-time.  A reasonable history is also kept so that changes are easily identified.  The accelerator control system provides detailed diagnostics tools to experts and can be used in case of any problems.


\newpage
\section{UV Laser System}
\label{sec:laser}

\begin{figure}
\centering	
\includegraphics[width=0.8\textwidth]{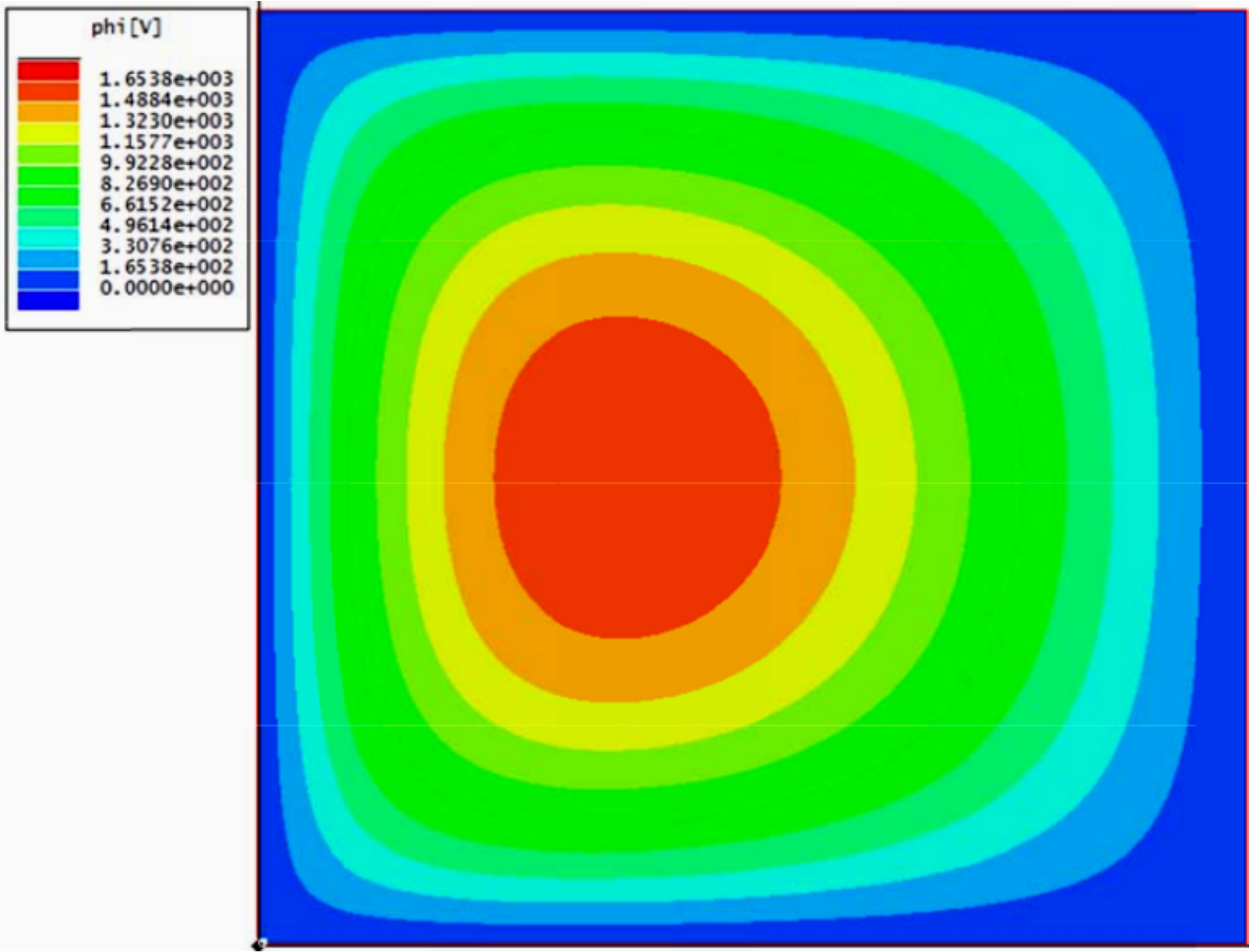}
\caption{Simulation of distorting potential distribution due to positive space charge in equilibrium in the MicroBooNE detector.}
\label{Ions}
\end{figure}

\begin{figure}
\centering	
\includegraphics[width=0.8\textwidth]{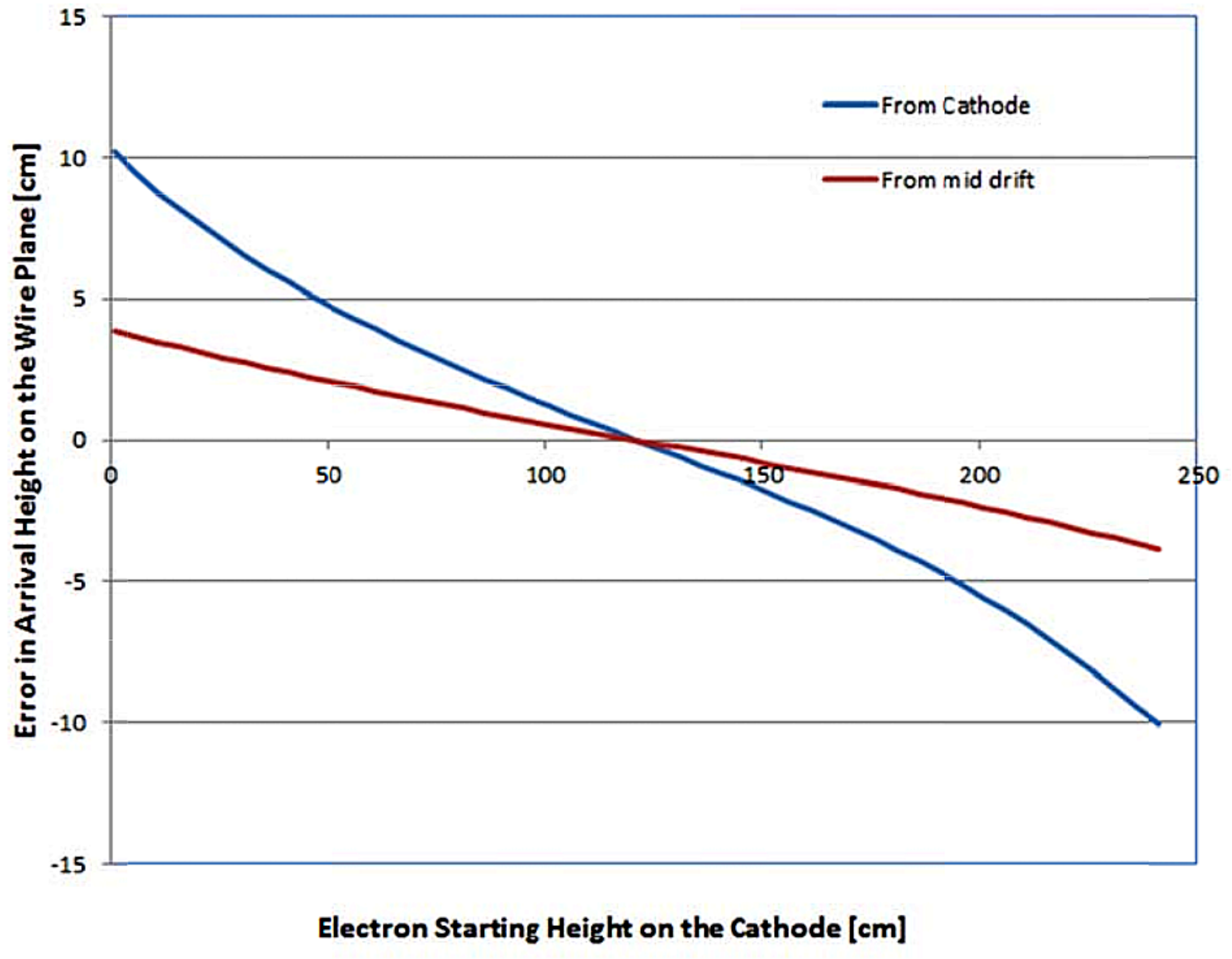}
\caption{Deviation of a crossing track from its true coordinates due to a simulated positive ion space charge in MicroBooNE.}
\label{Coordinates}
\end{figure}

\begin{figure}
\centering	
\includegraphics[width=0.8\textwidth]{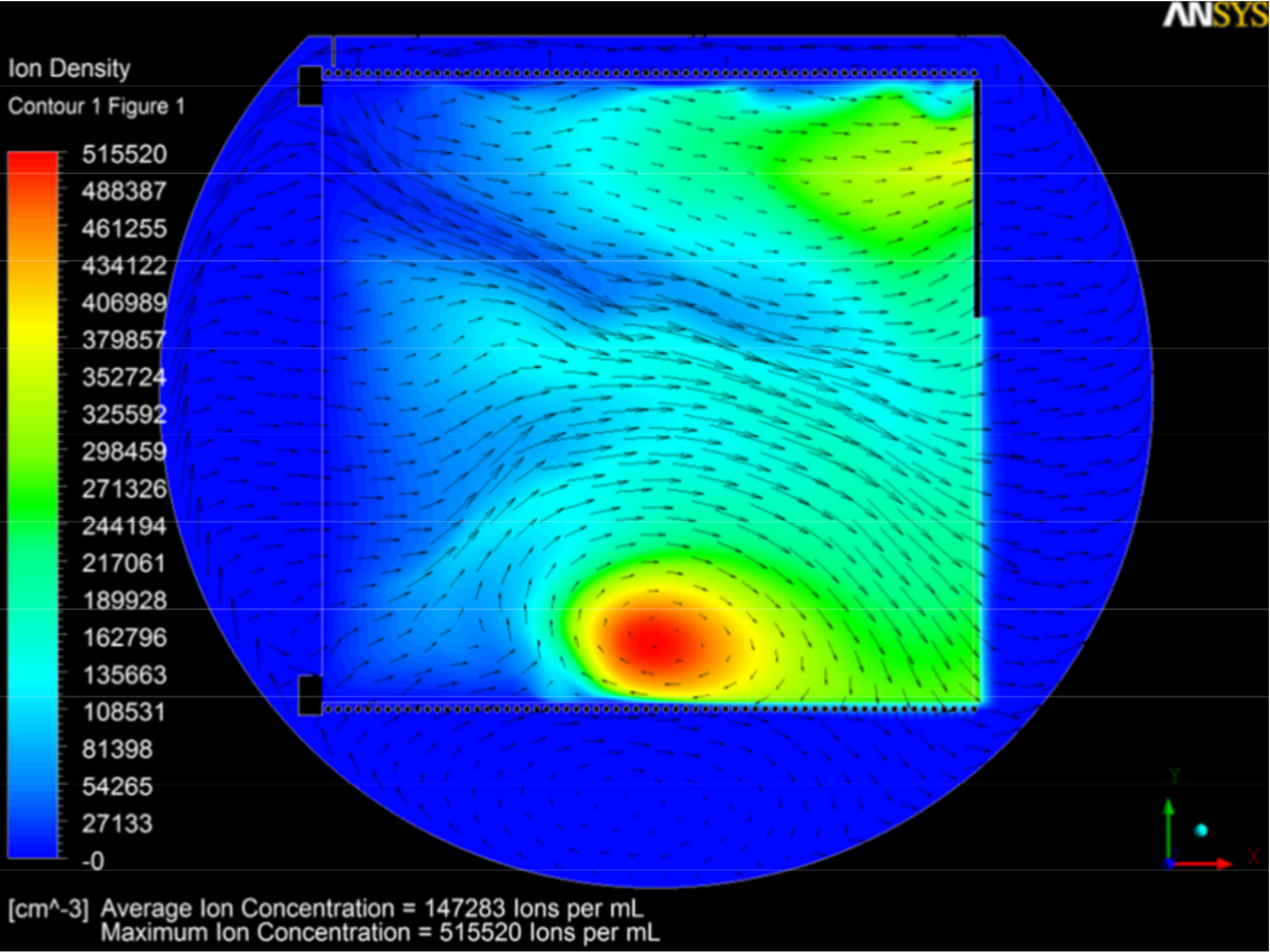}
\caption{Simulation of distribution of the positive space charge in presence of argon circulation.}
\label{Voirin}
\end{figure}

Knowledge of the electric field inside the drift volume of a \lartpc is a necessary aspect for performing subsequent event reconstruction. Distortions of particle tracks due to field non-uniformities affect the accuracy of the particle momentum reconstruction based on multiple scattering.  Deviations from a uniform drift field may arise mainly due to accumulation of positive argon ions in the drift volume. These ions are produced by ionizing particles from neutrino interactions, as well as by cosmic rays. While electrons produced by ionizing particles are quickly (within few milliseconds) swept towards the readout system, ions have significantly lower mobility. Their drift velocity in the MicroBooNE detector at nominal drift field is of the order of 0.8 cm/s. The rate of cosmic muons in the \lartpc volume is calculated to be 11,000 muons/s within the active volume (assuming no overburden, and a cosmic rate of 200 muons/m$^2$/s through a horizontal plane at the earth's surface and 63 muons/m$^2$/s through a vertical plane), traversing a summed total length of 1.9$\times$10$^4$ m through the liquid argon \cite{mcdonald1,mcdonald2}.  Assuming that cosmic muons are minimum-ionizing (2.1 MeV/cm) and produce 23.6 eV per ion pair, positive ion charge is produced at a rate of 2.8$\times$10$^{-8}$ C/s in the MicroBooNE LArTPC. These ions are continuously neutralized at the cathode. The resulting positive space charge distribution distorts the nominal \lartpc electric field potential, as shown in figure \ref{Ions}. Such accumulated space charge leads to noticeable distortion of the drift field and, consequently, to deviations of reconstructed track coordinates by up to 10 cm (see figure \ref{Coordinates}).  The ion drift velocity is comparable to local argon flow velocities, produced by global argon recirculation flow and thermal convection. Therefore the distribution of positive space charge inside the drift volume may be not only nonuniform (figure \ref{Voirin}), but also time varying.

A nonuniform drift field in the \lartpc leads to the apparent bending of truly straight tracks of high-momentum ionizing particles. In principle, a set of events from such particles allows for the reconstruction of the field in any small region of the \lartpc drift volume, using the systematic apparent curvature of tracks at different angles passing through that region. In practice, the rate of such events from cosmic muons is too low to acquire sufficient statistics in reasonable time. A method to generate straight ionization tracks at a defined location in liquid argon is described in \cite{Badhrees:2010}. A collimated photon beam from a pulsed UV laser with $\lambda$=266~nm can ionize liquid argon via multi-photon absorption. The resulting ionization track is straight, characterized by low electron density, therefore featuring little charge recombination loss, unlike cosmic muon tracks. Laser tracks are also free from $\delta$-electrons, which complicate track reconstruction in the case of muons. The method was successfully exploited in the Argontube long drift \lartpc \cite{Badhrees:2012-argontube,Ereditato:2013-argontube,Zeller:2013-argontube} to derive the non-uniformity of the electric field along its 5~m long drift volume \cite{Ereditato:2014-argontubedrift}. 

The MicroBooNE \lartpc requires a set of such tracks in order to cover the whole sensitive volume to reconstruct field distortions. This is the purpose of the laser calibration system, which features two UV lasers, one on either end of the LArTPC.  Tracks are generated by steering a pulsed laser beam, introduced to the cryostat via a custom-designed opto-mechanical feedthrough \cite{Ereditato:2014-laser}, through the LArTPC active volume.   

\subsection{UV Laser Calibration}
In order to unambiguously reconstruct a drift field vector at any point within the detector fiducial volume, a minimum of two ionization tracks are required to cross in the region of interest. The total number of crossing points is determined by the required reconstruction granularity. In MicroBooNE the initial scenario is to acquire 100 tracks from each direction, producing a reconstructed 3-D map with voxels that are approximately 10x10x50 cm$^3$ in volume. This map provides a rough picture of the space charge distribution. Depending on the results of this measurement, areas of interest can be studied in more detail. Repetitive study of small volumes may reveal dynamics in the space charge distribution due to turbulent circulation, and should further inform an optimized scenario for a standard UV laser calibration procedure.  

\begin{figure}
\centering	
\includegraphics[width=0.98\textwidth]{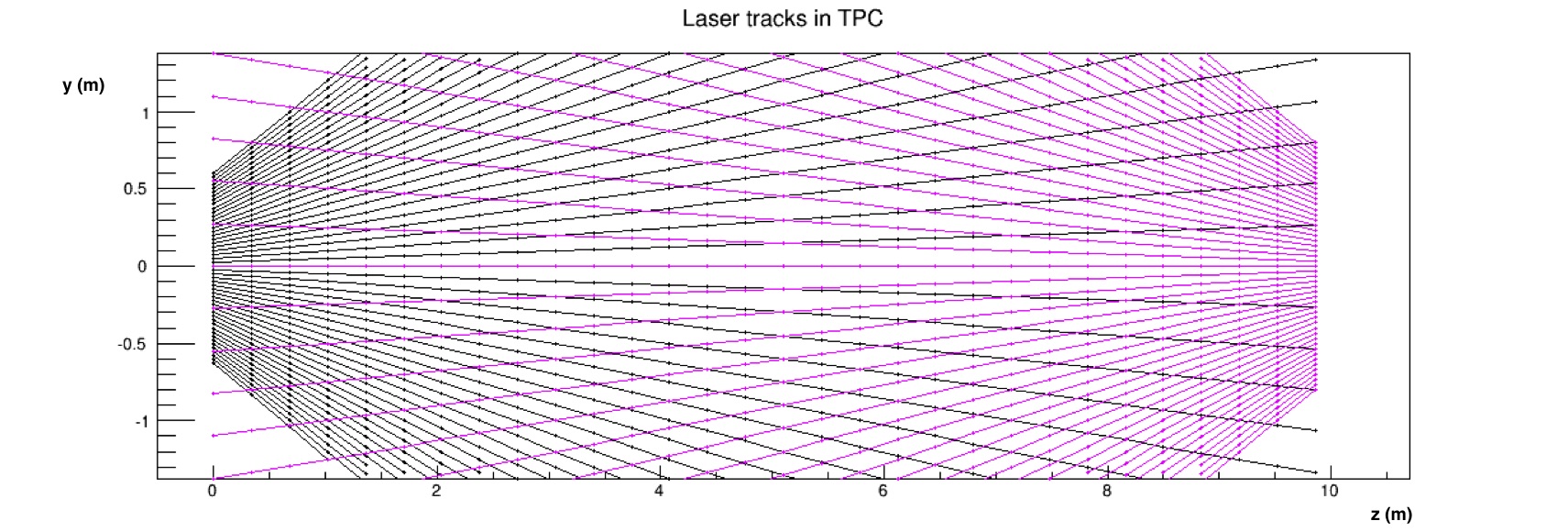}
\includegraphics[width=0.98\textwidth]{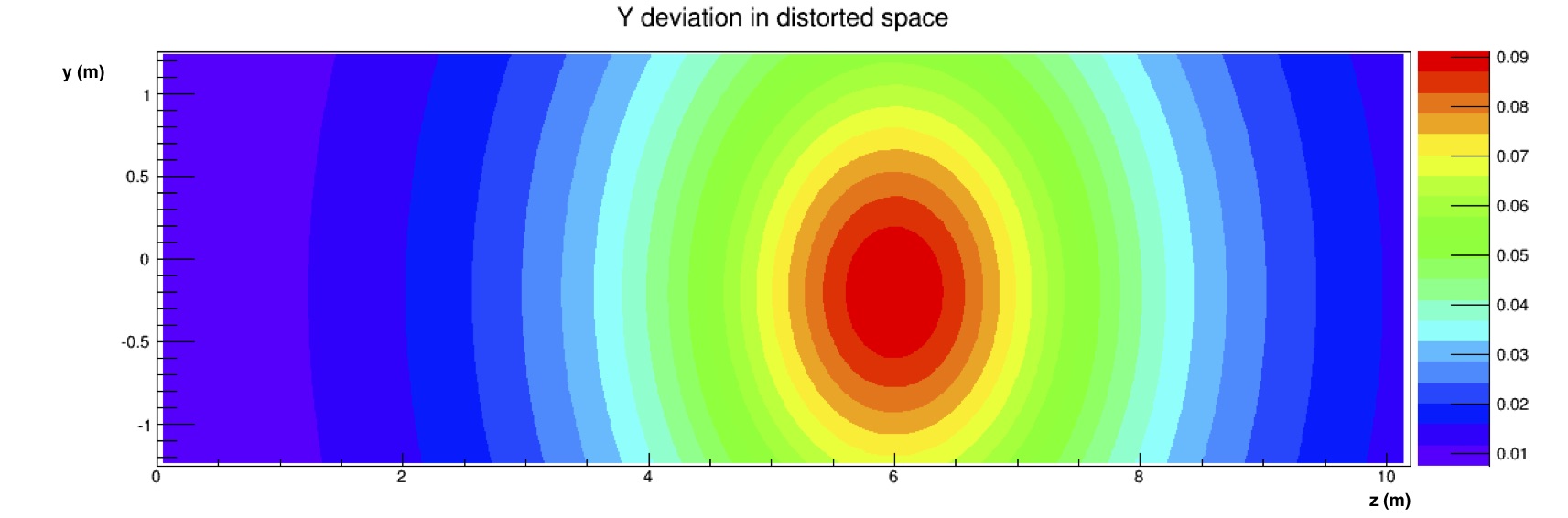}
\includegraphics[width=0.98\textwidth]{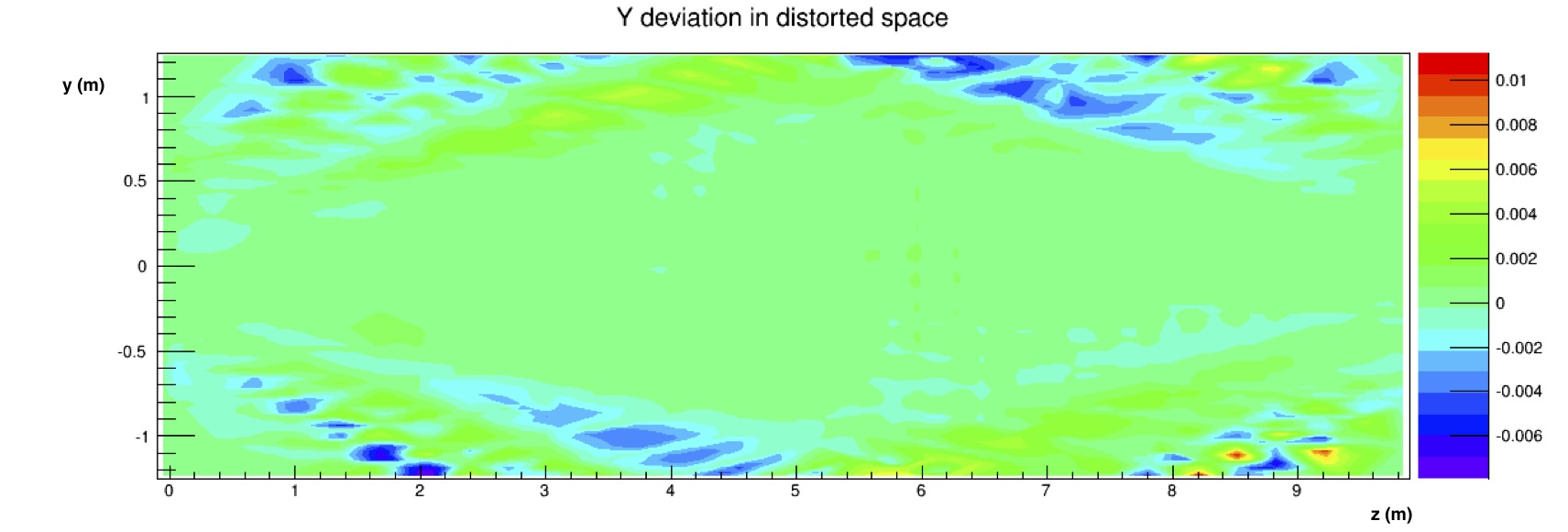}
\caption{Top: Simulated true laser beam trajectories in the detector; Middle: A map of the discrepancy between reconstructed and true track Y coordinate under influence of non-uniform electric field, as a function of the true Z coordinate of the track; Bottom: Map of the residual (true - reconstructed) track Y-coordinates, as a function of the true Z coordinate of the track, after applying the UV laser electric field calibration methods.} 
\label{Reco}
\end{figure}

An algorithm of drift field calibration utilizes an input array of detector events with one straight ionization track in each element of the array. The result of the calibration is a coordinate correction map, to be applied to each track, which converts apparently curved track images back to the true coordinate system where they are straight. The algorithm is iterative with an optimizable iteration step and required accuracy. An example of simulated reconstruction in 2-D space is depicted in figure \ref{Reco}, showing that the magnitude of the field distortions can potentially be reduced from 10~cm down to several millimeters in 99\% of the detector volume.

\subsection{Laser Source and Optics}
A Nd:YAG laser emitting light at a wavelength of 1024~nm is used as the primary light source \cite{continuum}. Inside the laser head, nonlinear crystals are installed in the beam line for frequency doubling and summing, resulting in a wavelength of 266~nm needed for ionization of liquid argon. For this wavelength, the Nd:YAG laser is specified to produce an output energy of 60~mJ for each 4 to 6~ns long pulse and a horizontal polarization. The maximal repetition rate is 10~Hz; the beam has a divergence of 0.5~mrad.

\begin{figure}%
    \centering
    \includegraphics[width=.6\textwidth]{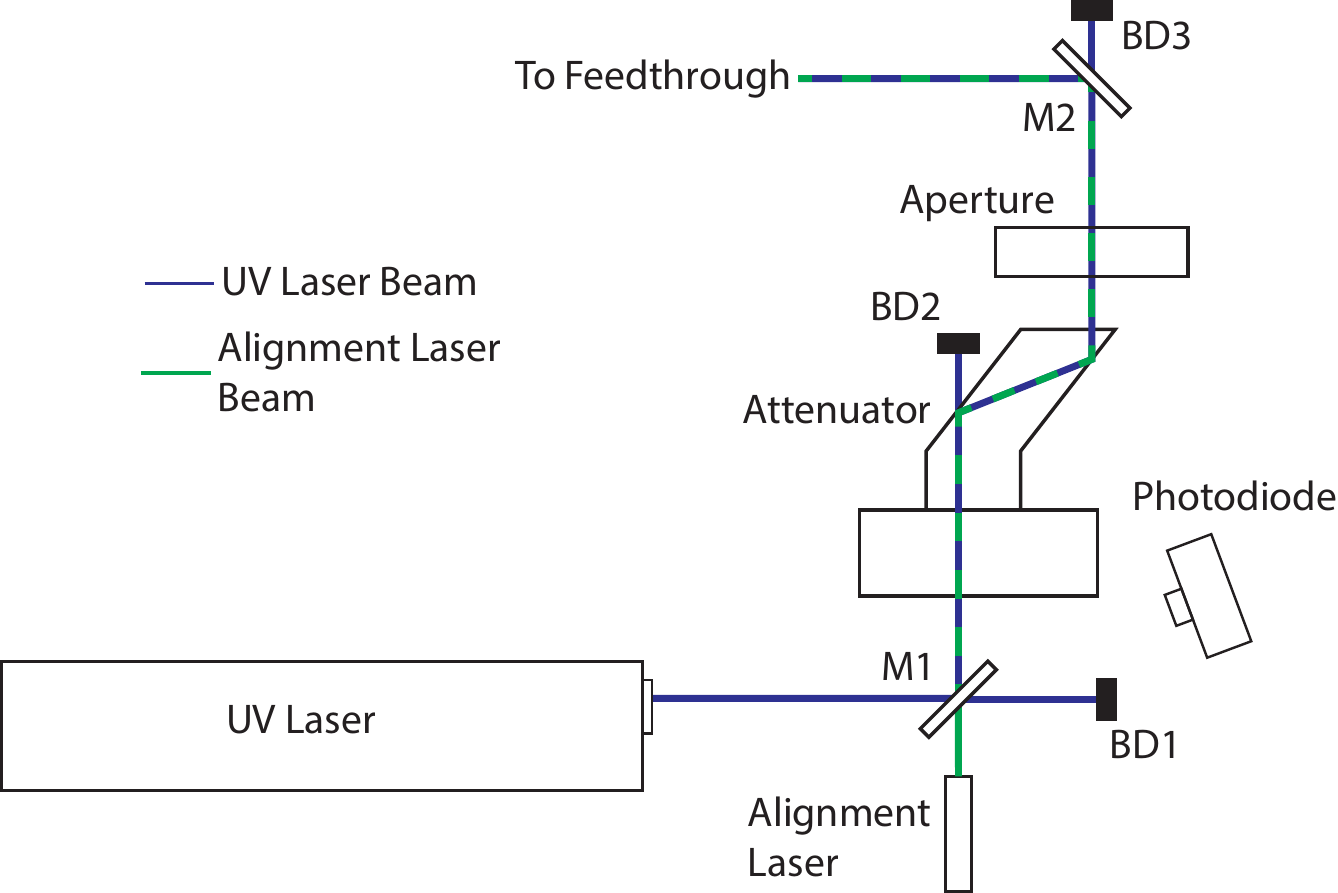}
    \qquad
    \includegraphics[width=.2\textwidth]{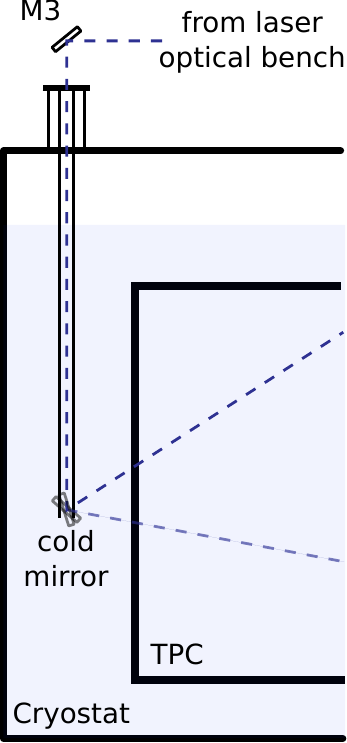} 
    \caption{Left: A schematic drawing (not to scale) of the components used for laser beam configuration.  An alignment laser (visible light) is introduced along the UV-laser path at the first dichroic mirror (M1), such that the paths overlap. In the attenuator the UV-laser beam intensity can be adjusted to the desired level, and the diameter of the beam is controlled by an aperture. A motorized mirror (M2) deflects the light into the direction of the feedthrough. Beam dumps (BD) are installed behind all mirrors to absorb the non-reflected laser light. Right: Side view of the cryostat indicating the mirror support structure with respect to the LArTPC.}%
    \label{fig:optics}%
\end{figure}

Figure \ref{fig:optics} depicts the arrangement of the laser system in MicroBooNE.  An optical table, mounted on the cryostat, was developed to accommodate the necessary parts in a stable and compact environment. The components were chosen to be accessible remotely where necessary. The emitted laser beam contains not only ultraviolet light but also all other harmonics generated in the crystal and the primary light of 1024~nm. Dichroic mirrors optimized to reflect only wavelengths in the UV region are used to filter higher wavelengths out. To absorb the transmitted wavelength behind the mirrors, glass-ceramic plates are installed. The beam leaving the laser head is reflected by the first $45^{\circ}$-mirror into an attenuator. For optical adjustment and verification of the non-visible UV-beam, a green alignment laser is placed behind this mirror and adjusted such that its path is coincident with the UV-laser beam. In the attenuator (Altechna Wattpilot) a turnable $\lambda/2$-plate enables rotation of the orientation of the laser beam polarization. Behind the attenuator two parallel plates are installed such that the angle of the incident beam matches the Brewster angle of the reflector. Modulating the polarization of the beam adjusts the intensity of the reflected beam. An aperture is placed in the optical path of the beam after the attenuator to control the beam diameter. The last part in the beam line is a remotely-controllable mirror mount (Zaber T-OMG), which directs the beam to the laser feedthrough on the cryostat. A photodiode (Thorlabs DET10A/M), which is sensitive in the ultraviolet region, detects the scattered light when a laser pulse is fired.  Its signal is then used as a trigger for data taking. Both the UV-laser head and the optical table are mounted on a 15~mm thick aluminum plate.

\subsection{Steering System}
\begin{figure}
\centering
     \includegraphics[height=0.6\textwidth]{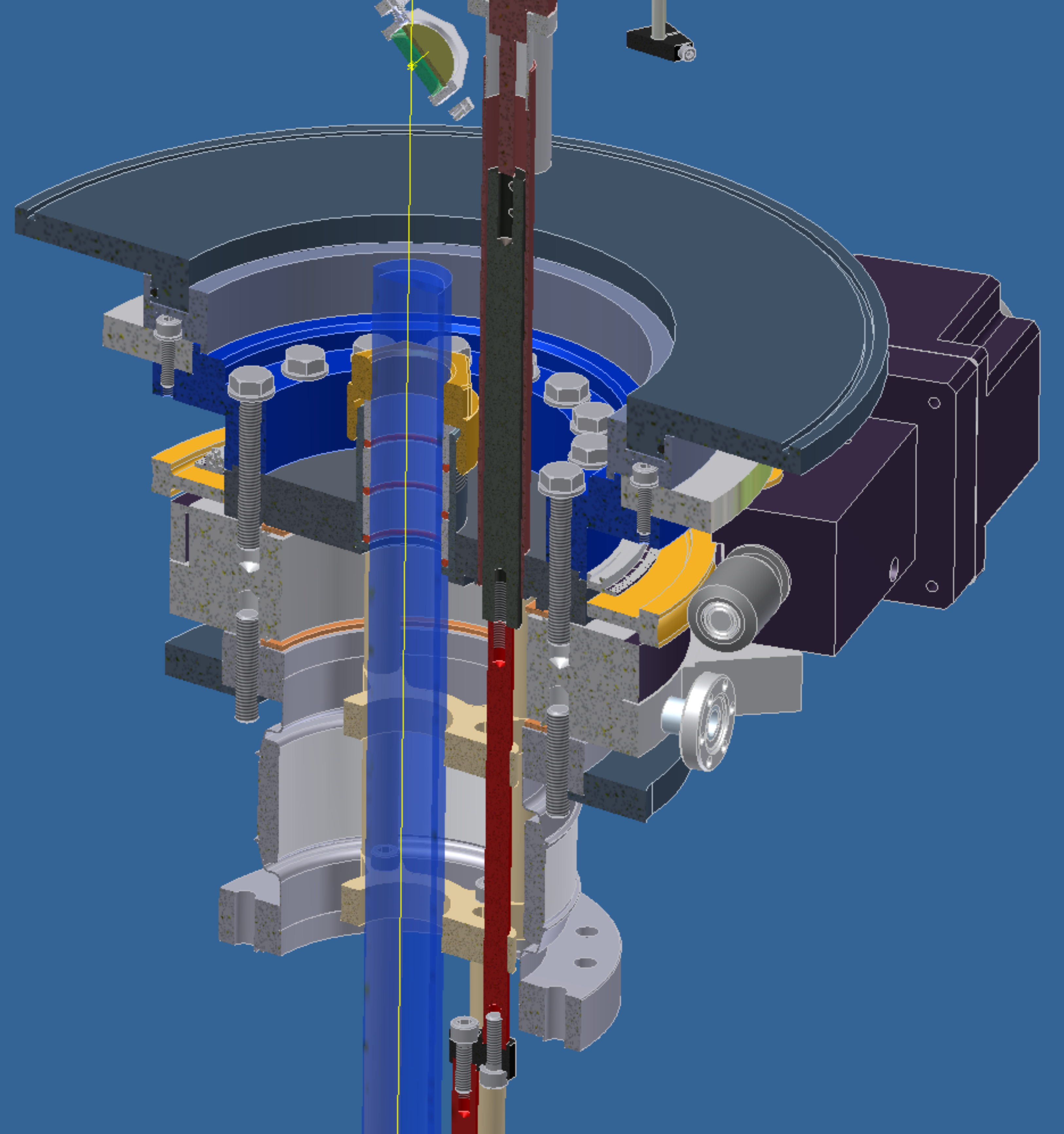} \includegraphics[height=0.6\textwidth]{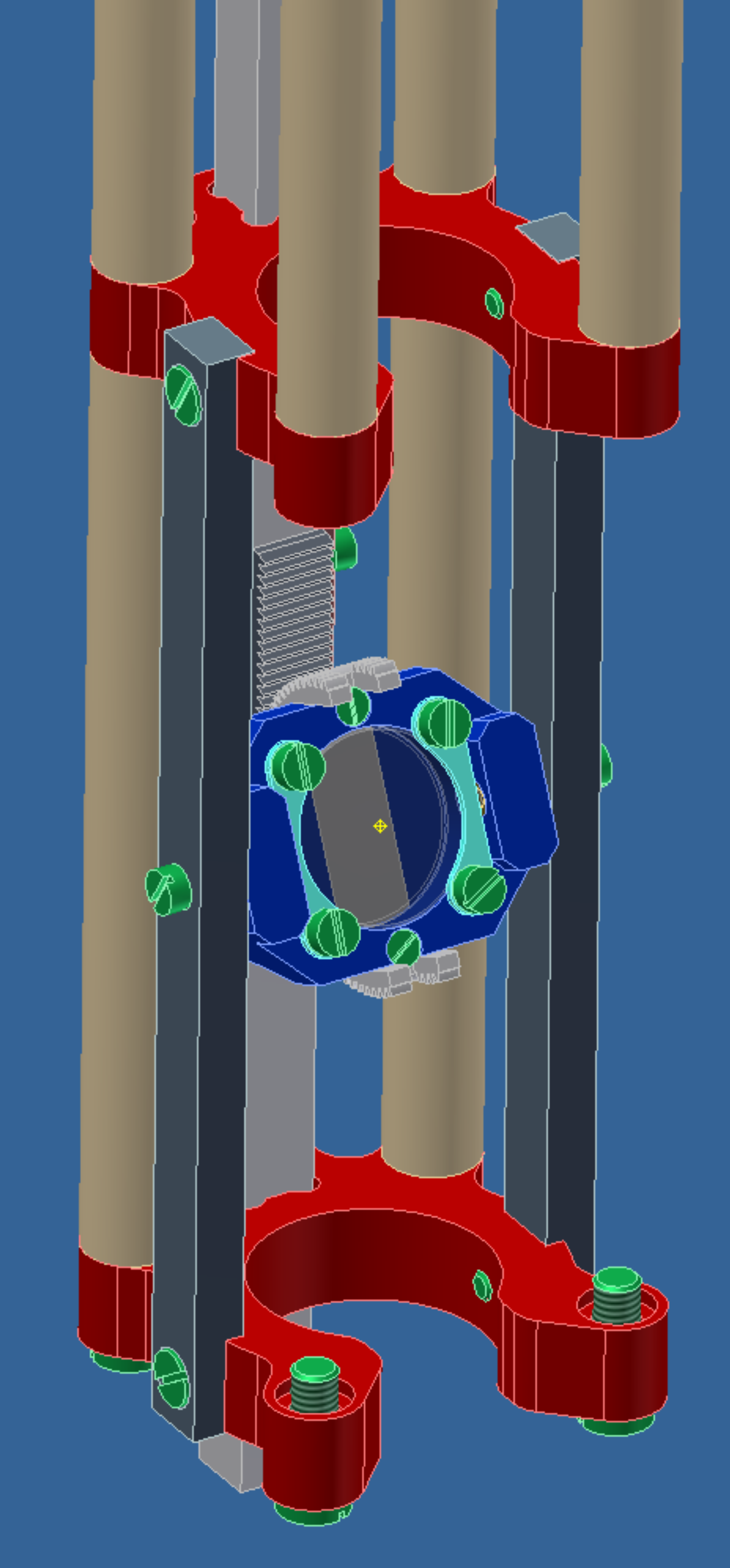}
     \caption[CAD drawing of the feedthrough and the mirror support structure.]{Left: CAD cutaway drawing of the feedthrough construction is shown. The yellow line indicates the laser path. Right, CAD drawing of the cold mirror including the support structure.}
  \label{fig:feedthrough}
\end{figure}

One of the main challenges of the laser calibration system is the introduction of a steerable laser beam into the detector. Earlier, an evacuated quartz-glass \cite{BernLaser} was utilized to introduce a laser beam into liquid argon, however this beam had a fixed path through the detector. For the purpose of scanning the full detector a fully steerable mirror in liquid argon is necessary. In the MicroBooNE detector, this is achieved by mounting a mirror on a horizontally-rotatable support structure. A rack and pinion construction, where the mirror is mounted on the frontside of a half gear (pinion), provides the necessary freedom for the vertical movement (see figure \ref{fig:feedthrough} right). To steer the horizontal movement from outside the cryostat, a commercial differentially-pumped rotational feedthrough is deployed (see figure \ref{fig:feedthrough} left). The rack and pinion construction is attached to a linear feedthrough. Both feedthroughs are motorized to allow for remote control and automation of the mirror movement. The mirror support structure was fabricated out of polyamide-imide (Duratron T4301 PAI), which has a very low outgassing rate and low thermal expansion coefficient, and is certified for operation at 87~K. To minimize the probability of discharges due to the close location of the feedthrough to the field cage structure in MicroBooNE, no conductive parts were used in the support structure. The support structure has a total length of 2.5~m in MicroBooNE.
Both feedthroughs are equipped with high precision position encoders from Heidenhain. The accuracy of the encoders is chosen such that a position accuracy of 2~mm for the laser beam spot over 10~m distance is achieved.  An external interface box controls the encoders and records a position reading upon receiving a trigger signal from the photodiode. The DAQ computer accesses the position information over an ethernet connection. The same computer is also used for steering the two motors via a motor driver system (over a RS232 interface).

\subsection{Performance Tests and Initial Operation}

A full performance test of the laser calibration system identical to the one installed in the MicroBooNE \lartpc was performed prior to the final installation. Apart from the general proof-of-principle of the laser calibration system, several operationally relevant parameters were identified. These include scanning speed, positioning accuracy, positioning limits, optimal laser beam intensity, beam diameter, and the minimal achievable field distortion which can be resolved. The test system consists of a \lartpc equipped with 64 readout channels and an active area of about 400~cm$^2$, with a drift distance of 40~cm (see \cite{lasertest} for further details).

Several tests of the motorized feedthrough were performed under warm conditions before cold tests were conducted. One crucial parameter for the quality of the electric field calibration is the resolution at which laser tracks can be aimed in the detector. For the rotational axis this angle is directly measured on a circular scale. For the vertical movement the linear displacement of the bellow is translated into a rotation inside the cryostat, as can be seen in the CAD drawing in figure \ref{fig:feedthrough}. This construction introduces uncertainties to the measurement position and backlash. The backlash can be compensated by always approaching positions from the same direction. For the translation of the linear movement $\Delta L$ into a rotation $\Delta \phi$ the translation ratio $s$ according to $\Delta L = s \cdot \Delta \phi$ was measured with a Bosch GPL3 laser alignment device.  The obtained ratio is $s=$0.3499$\pm$0.0002~mm/$^{\circ}$. The dominant uncertainty in the vertical position measurement is the accuracy of the encoder $\sigma_{\mathrm{linear}} = \pm 1~\mu m$, which translates into a vertical rotation measurement accuracy $\sigma_{\mathrm{vertical}} =$ 0.050~mrad. 

Horizontal movement limitations arise from the construction of the feedthrough system, namely the warm mirror support structure. This limitation originates in the manner the laser table is mounted relative to the feedthrough on the MicroBooNE cryostat. Vertically the mirror can be rotated more than 45$^{\circ}$ relative to the horizon in both directions. In an upward looking configuration, no limitations arise which would affect the coverage of the detector with the beam.  When the mirror faces the opposite downward direction and the laser is properly aligned onto the center of the mirror, only the laser diameter and the size of the mirror limit the achievable coverage. However slight misalignment will affect this, since the beam will not be in the optimal spot anymore. In warm tests a maximal downward angle of the beam of 52.5$^{\circ}$ with respect to the horizon was achieved. During the cold tests the horizontal and vertical movement speeds were set to 2.6~{$^{\circ}$/s} and 1~{$^{\circ}$/s}, respectively, and horizontal and vertical angles of 81$^{\circ}$ and 22$^{\circ}$, respectively, were covered. Warm tests of the fully expanded setup showed vibrations if the chosen movement speed was too large. The vibrations are expected to be dampened with a more stable installation on the detector, and with the immersion of the setup in liquid argon.

Modulation of the beam energy with respect to the vertical alignment of the cold mirror was found to be crucial for obtaining sufficient ionization in the detector. Investigations showed that the reflectivity of the selected dielectric mirrors, which were optimized for 45$^{\circ}$ in air, are very sensitive to the angle of incidence in liquid argon. Therefore during a calibration run, the beam energy has to be controlled. The emitted UV-laser beam has a diameter of 6~mm and will spatially diverge during propagation.  A beam with this diameter will produce an ionization signal larger than the wire spacing, which will limit the capabilities of the full system. With the aperture a small as possible diameter of the laser was selected to enter the detector. Measurements of the diameter were performed with thermal paper (used for thermal printing) on which the selected beam spot burns in. The minimal achieved diameter was 1~mm.

\newpage
\section{Conclusion}
\label{sec:conclusion}

The MicroBooNE detector is the culmination of several years of development and construction.  Innovations such as custom cold (in-liquid) electronics, non-evacuated cryostat, 2.5~m drift, and a UV laser calibration system represent major technological advances that future experiments will build upon.  Operations of MicroBooNE began in the summer of 2015, and a cosmic ray tagger system was added in the fall of 2016.  Future publications will be dedicated to describing the performance of the detector systems introduced in this paper.

\begin{acknowledgments}
This material is based upon work supported by the following: the United States Department of Energy, Office of Science, Offices of High Energy Physics (OHEP) and Nuclear Science, Intensity Frontier Fellowships and Neutrino Physics Center Fellowships through OHEP, the U.S. National Science Foundation, the Swiss National Science Foundation, the Science and Technology Facilities Council of the United Kingdom, and The Royal Society (United Kingdom). Additional support for the laser calibration system and cosmic ray tagger was provided by the Albert Einstein Center for Fundamental Physics. Fermilab is operated by Fermi Research Alliance, LLC under Contract No. DE-AC02-07CH11359 with the United States Department of Energy.
\end{acknowledgments}

\bibliographystyle{JHEP}
\bibliography{references}

\end{document}